\documentclass[12pt]{article}

\usepackage{times}

\usepackage{url}
\usepackage{float}
\usepackage{amsfonts} 
\usepackage{color}
\usepackage{amsmath}
\usepackage[table]{xcolor}
\usepackage{bm}
\usepackage{bbm}
 
\DeclareMathOperator*{\argmax}{arg\,max}

\usepackage{float}
\usepackage{lipsum}
\usepackage{cleveref}
\usepackage{graphicx}
\usepackage{dsfont}
\usepackage{subcaption}
\usepackage{braket,amsfonts,amsopn} 
\usepackage{algorithm}
\usepackage{algorithmic}
\newcommand{\mat}[1]{\boldsymbol{#1}}
\newcommand{\myvec}[1]{\boldsymbol{#1}}
\newcommand{\indicator}[2]{\mathds{1}_{#1}(#2)}
\newcommand{\ans}[1]{\textcolor{black}{#1}}

\DeclareMathOperator*{\card}{\#}

\newenvironment{sciabstract}{%
	\begin{quote} \bf}
	{\end{quote}}

\newcounter{lastnote}

\title{Bayesian 3D Reconstruction of Complex Scenes from Single-Photon Lidar Data}

\author
{Juli\'an Tachella$^{1}$, Yoann Altmann$^{1}$, Ximing Ren$^{1}$, Aongus McCarthy$^{1}$, Gerald S. Buller$^{1}$, \\ Jean-Yves Tourneret$^{2}$ and Steve McLaughlin$^{1}$
	\\
	\normalsize{$^{1}$School of Engineering and Physical Sciences, Heriot-Watt University, Edinburgh, UK.}\\
	\normalsize{$^{2}$ENSEEIHT-IRIT-TeSA, University of Toulouse, Toulouse, France.}\\
	\\
}

\date{}

\begin{document}

\maketitle

\begin{sciabstract}
 Light detection and ranging (Lidar) data can be used to capture the depth and intensity profile of a 3D scene. This modality relies on constructing, for each pixel, a histogram of time delays between emitted light pulses and detected photon arrivals. In a general setting, more than one surface can be observed in a single pixel. The problem of estimating the number of surfaces, their reflectivity and position becomes very challenging in the low-photon regime (which equates to short acquisition times) or relatively high background levels (i.e., strong ambient illumination). This paper presents a new approach to 3D reconstruction using single-photon, single-wavelength Lidar data, which is capable of identifying multiple surfaces in each pixel. Adopting a Bayesian approach, the 3D structure to be recovered is modelled as a marked point process and reversible jump Markov chain Monte Carlo (RJ-MCMC) moves are proposed to sample the posterior distribution of interest. In order to promote spatial correlation between points belonging to the same surface, we propose a prior that combines an area interaction process and a Strauss process. New RJ-MCMC dilation and erosion updates are presented to achieve an efficient exploration of the configuration space. To further reduce the computational load, we adopt a multiresolution approach, processing the data from a coarse to the finest scale. The experiments performed with synthetic and real data show that the algorithm obtains better reconstructions than other recently published optimization algorithms for lower execution times.
\end{sciabstract}
\pagebreak
\section{Introduction}
Reconstruction and analysis of 3D scenes have a variety of applications, spanning earth monitoring \cite{hakala2012full,nilsson1996estimation,mallet2009full}, underwater imaging \cite{maccarone2015underwater,halimiwater}, automotive \cite{ogawa2006lane,schwarz2010lidar} and defence \cite{gao2011research}. Single-photon light detection and ranging devices acquire range measurements by illuminating a three-dimensional (3D) scene with a train of laser pulses and recording the time-of-flight (TOF) of the photons reflected from the objects in the illuminated scene. 
Using a time correlated single-photon counting (TCSPC) system, a histogram of time delays between emitted and reflected pulses is constructed for each pixel. For a given pixel, the presence of an object is associated with a characteristic distribution of photon counts in the histogram. The position and number of counts provide depth and reflectivity information respectively. In scenarios where the light goes through a semitransparent material (e.g, windows or camouflage) or when the laser beam is wide enough with respect to the object size (e.g., distant objects), it is possible to record two or more surfaces in a single pixel. The recovery of multiple objects per pixel is thus very important in many applications, such as tree layer analysis \cite{wallace2014design} or detection of hidden targets behind camouflage \cite{halimi2017}. 

In order to reconstruct the 3D scene from single-photon Lidar data, it is necessary to discriminate the photon counts associated with each surface from the ones linked to the background illumination. When the background level can be neglected, the traditional approach consists, under the single-peak assumption, of log-match filtering the Lidar waveforms and finding the maximum of the filtered data for each pixel \cite{snyder2012random}, which is the maximum likelihood (ML) solution for a Poisson noise assumption (a matched filter is used for Gaussian noise). While this method obtains good results for high photon counts, it gives poor estimates when the background illumination is high or the number of recorded photons is low. Several studies have focused on improving the maximum likelihood (ML) estimates in the single-depth estimation problem. Altmann \emph{et al.} \cite{altmann2016lidar} proposed a Bayesian approach, whereas Shin \emph{et al.} \cite{shin2015photon}, Halimi \emph{et al.} \cite{halimi2016restoration} and Rapp \emph{et al.} \cite{rappfew} suggested three different optimization alternatives. The method introduced in \cite{shin2015photon} estimates the reflectivity and depth information independently, considering a rank-ordered mean censoring of background photons as a preprocessing step. The optimization method in \cite{halimi2016restoration}, assumes a negligible background and estimates the depth and reflectivity jointly using an alternating direction method of multipliers (ADMM) algorithm. The algorithm proposed in \cite{rappfew} uses an adaptive superpixel approach to censor background photons and improve depth and reflectivity estimates. In the multiple surface per pixel configuration, Hernandez-Marin \emph{et al.} \cite{hernandez2007bayesian} proposed a pixelwise reversible jump Markov chain Monte Carlo (RJ-MCMC) algorithm. While this approach is able to find an \emph{a priori} unknown number of surfaces and compute associated uncertainty intervals, it involves a prohibitive computation time. Moreover, it performs poorly when photon counts are relatively low, as it does not account for spatial correlation between neighbouring pixels. In later work, Hernandez-Marin \emph{et al.} \cite{hernandez2008multilayered} proposed an extension to the latter algorithm, where a Potts model was used to regularize spatially the number of surfaces per pixel. However, the computational load of their algorithm was prohibitive for large images and the correlation between the amplitude and position of each object was not modelled \emph{a priori}. There have been other attempts to derive statistical models for Lidar waveforms with an unknown number of objects per pixel, such as Mallet \emph{et al.} with full waveform topographic Lidar \cite{mallet2010marked}, where a marked point process was considered for each pixel separately. While they defined interactions between pulses in the same pixel, no spatial interaction between points of neighbouring pixels was considered. Recently, new optimization approaches have been proposed to tackle the multiple object per pixel problem: Shin \emph{et al.} introduced an $\ell_1$ norm regularization for the recovered peak positions, followed by a post processing of the 3D point cloud \cite{shin2016computational}. Halimi \emph{et al.} improved it by considering a TV operator and the $\ell_{21}$ norm \cite{halimi2017}.

\begin{figure}[h!]
	\centering
	\begin{subfigure}{0.25\textwidth}
		\centering
		\includegraphics[width=1\textwidth]{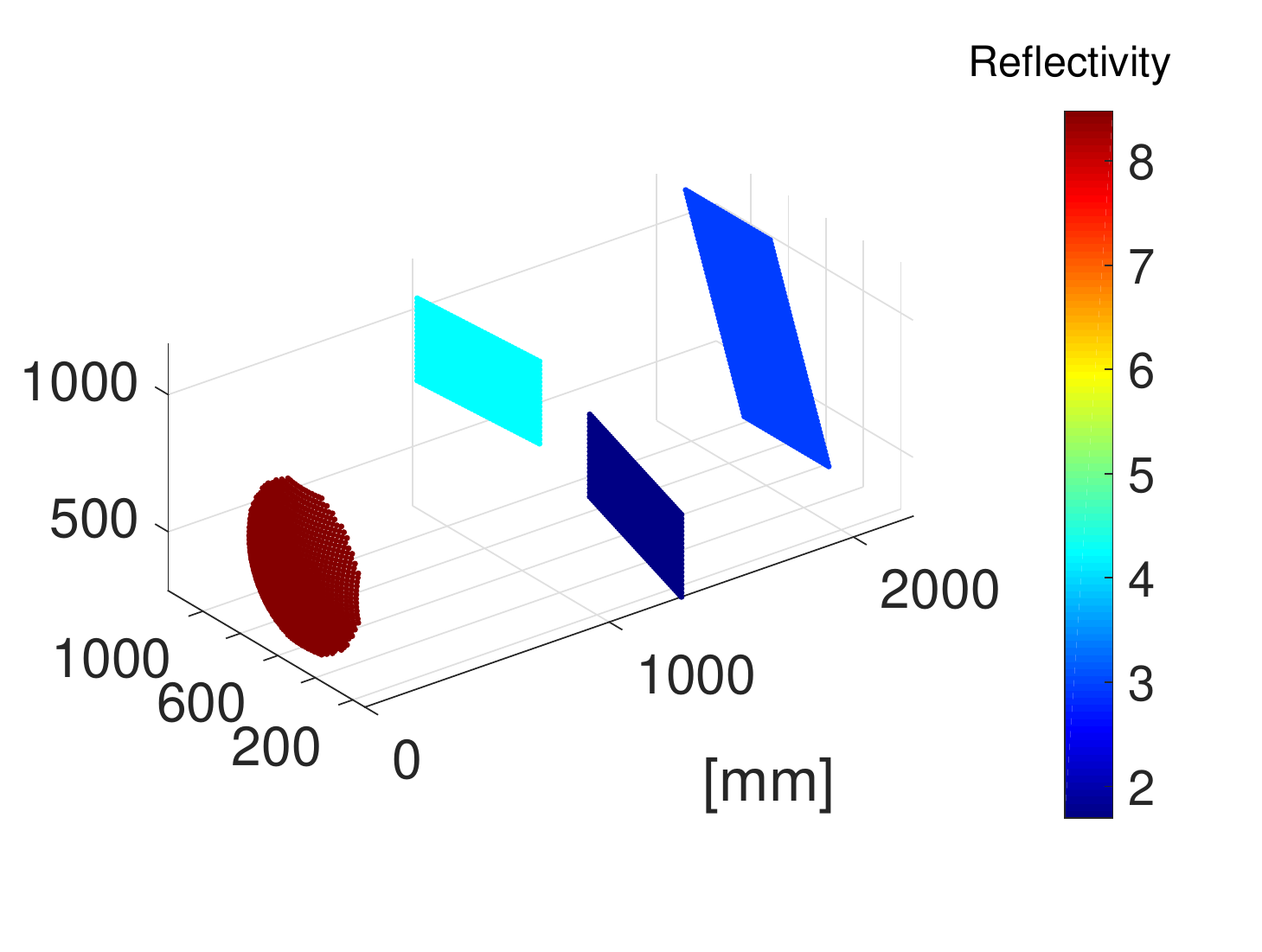}
		\caption{}
	\end{subfigure}%
	\begin{subfigure}{0.25\textwidth}
		\centering
		\includegraphics[width=1\textwidth]{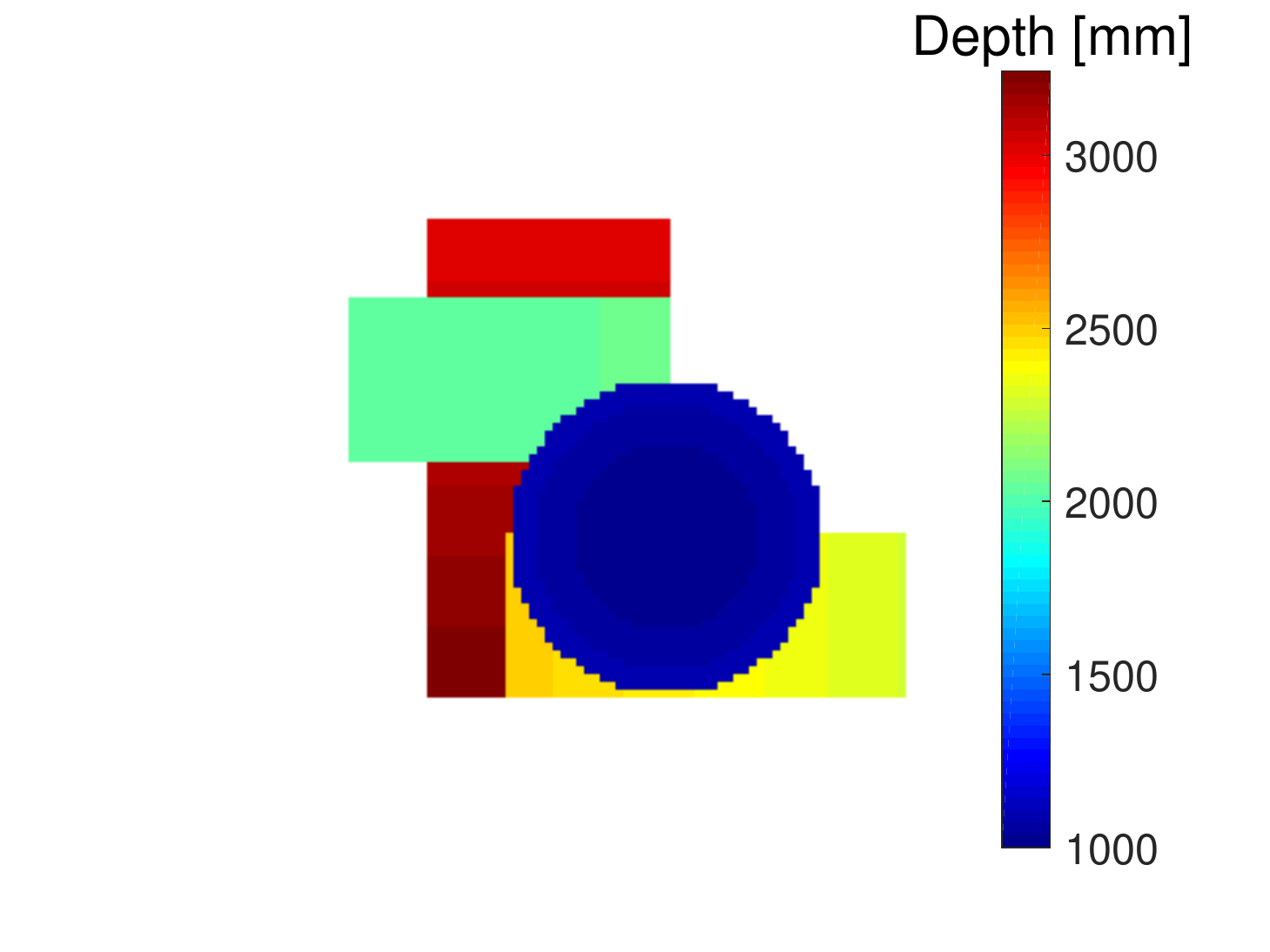}
		\caption{}
	\end{subfigure}%
	\begin{subfigure}{0.25\textwidth}
		\centering
		\includegraphics[width=1\textwidth]{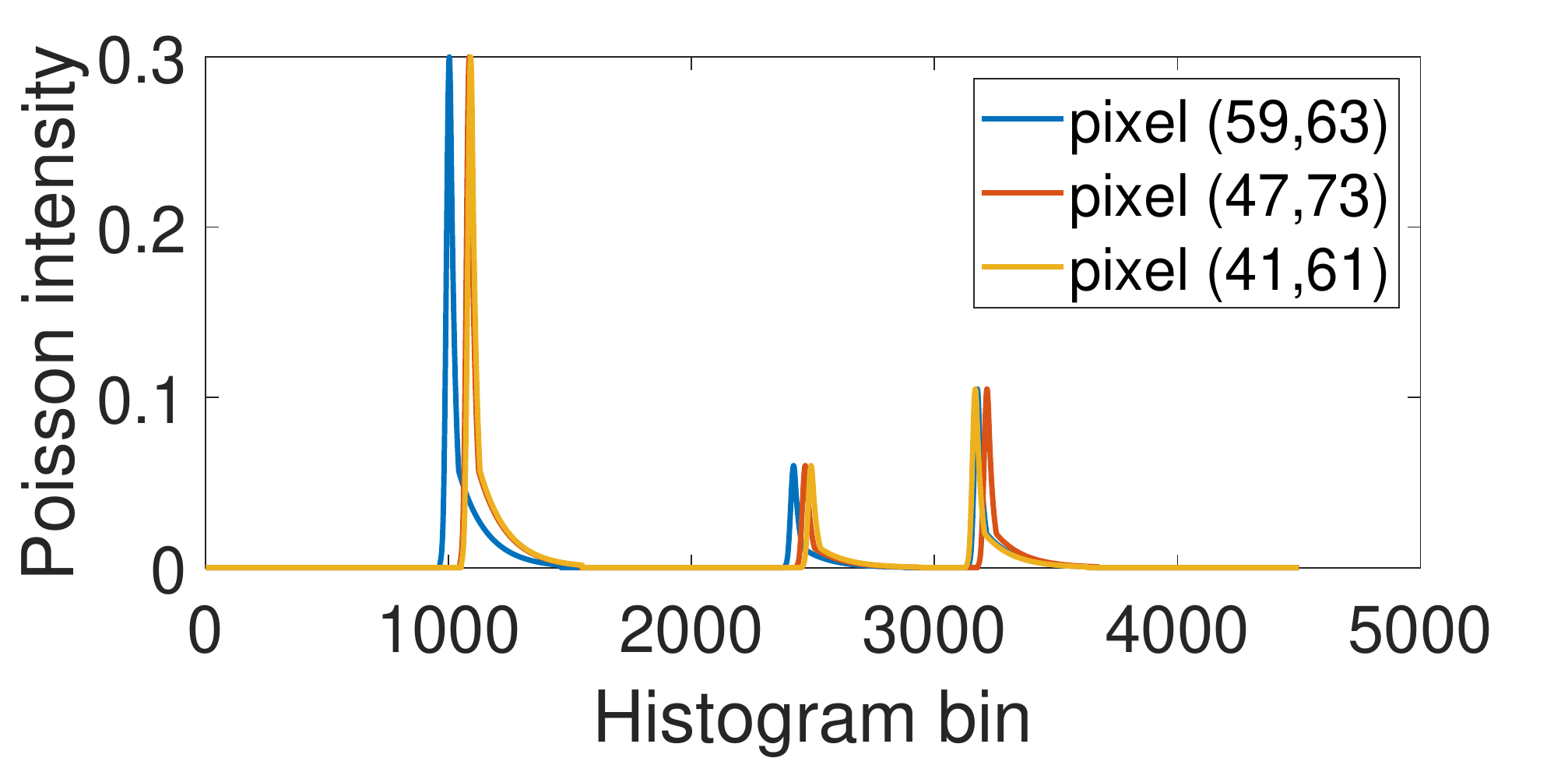}
		\caption{}
	\end{subfigure}%
	\begin{subfigure}{0.25\textwidth}
		\centering
		\includegraphics[width=1\textwidth]{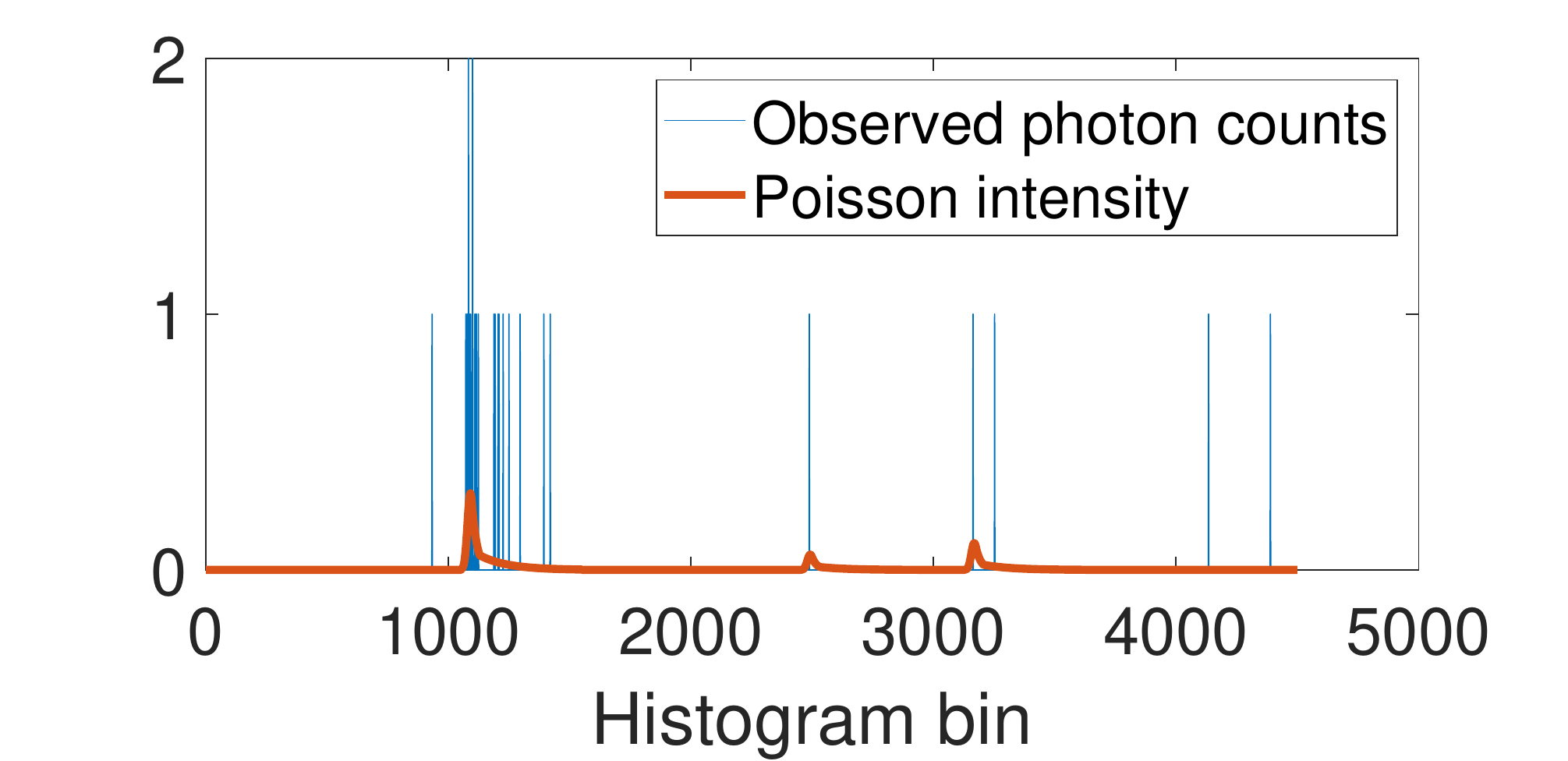}
		\caption{}
		\label{FIG:noisy_hist}
	\end{subfigure}
	\caption{(a) depicts a synthetic 3D point cloud with $N_r=99$ rows, $N_c=99$ columns and $T=4500$ bins. The scene consists of 3 plates with different sizes and orientations and one ball shaped object. (b) illustrates the depth of the first object for each pixel. (c) shows the intensity of three different pixels. The observed photon counts and underlying Poisson intensity of a pixel with 3 surfaces is shown in sub-figure (d).}.
	\label{FIG:example}
\end{figure}

In this work, we introduce a new spatial point process within a Bayesian framework for modelling single-photon Lidar data. This novel approach considers interactions between points at a pixel level and also at an inter-pixel level, in a variable dimension configuration. Here, we consider each \ans{surface within a pixel} as a point in the 3D space, which has a mark that indicates its \ans{intensity}. Natural Lidar point clouds exhibit strong spatial clustering, as points belonging to the same surface tend to be close in range. Conversely, points in a given pixel tend to be separated as they correspond to different surfaces. \Cref{FIG:example} shows an example of a synthetic Lidar 3D point cloud to illustrate this phenomenon. This prior information is added to our model using spatial point processes: Repulsion between points at a pixel level is achieved with a hard constraint Strauss process and attraction among points in neighbouring pixels is attained by an area interaction process, as defined in \cite{van2000markov}. Moreover, the combination of these two processes implicitly defines a connected-surface structure that is used to efficiently sample the posterior distribution. To promote smoothness between reflectivities of points in the same surface, we define a nearest neighbour Gaussian Markov random field (GMRF) prior model, similar to the one proposed in \cite{mccool2016robust}. Inference about the posterior distribution of points, their marks and the background level is done by an RJ-MCMC algorithm \cite[Chapter~9]{brooks2011handbook}, with carefully tailored moves to obtain high acceptance rates, ensuring better mixing and faster convergence rate. In addition to traditional birth/death, split/merge, shift and mark moves, new dilation/erosion moves are introduced, which add and remove new points by extending or shrinking a connected surface respectively. These moves lead to a much higher acceptance rate than those obtained for birth and death updates, as they propose moves to and within regions of high posterior probability. To further reduce the transient regime of the Markov chains and reduce the computational time of the algorithm, we consider a multiresolution approach, where the original Lidar 3D data is binned into a coarser resolution data cube with higher signal power, lower number of points and same data statistics. An initial estimate obtained from the downsampled data is used as the initial configuration for the finer scale, thus reducing the number of burn-in iterations needed for the Markov chains to convergence. We assess the quality of reconstruction and the computational complexity in several experiments based on synthetic Lidar data and two real Lidar datasets. The algorithm leads to new efficient 3D reconstructions with similar processing times to other existing optimization-based methods. This method can be successfully applied to scenes where there is only one \ans{object} per pixel, thus generalizing other \ans{single-depth} algorithms \cite{shin2015photon,altmann2016lidar,halimi2017,rappfew}. \ans{Moreover, the proposed algorithm can also be applied in scenes where each pixel has at most one surface and it generalizes other target detection methods \cite{altmanndetect2016,altmann2016target}.} We refer to the proposed method as ManiPoP, as it aims to representing 2D manifolds with a 3D point process.
In summary, the main contributions of this paper are
\begin{enumerate}
	\item A new Bayesian model based on a marked point process prior for modelling spatially correlated 3D point clouds.
	\item New reversible-jump moves proposed for sampling the posterior distribution more efficiently.
	\item A multiresolution processing approach to improve the convergence rate, which also allows a rapid information extraction using only the coarser scales.
\end{enumerate}

The remainder of this work is organized as follows. \Cref{SEC:model} presents the Bayesian model considered for the analysis of \ans{multiple-depth} Lidar data.  \Cref{SEC:estimation_strategy} details the sampling strategy using an RJ-MCMC algorithm.  \Cref{SEC: implementation} discusses the proposed multiresolution approach and other implementation details to reduce the computational load of the algorithm. Results of experiments conducted on synthetic and real data are presented in \cref{SEC:experiments}. \Cref{SEC:Conclusions} finally summarizes our conclusions and discusses future work.

\section{Proposed Bayesian model}\label{SEC:model}
Recovering the position and \ans{intensity} of the \ans{objects} from the raw Lidar data is an ill-posed problem, as the solution is not uniquely identified given the data (e.g., the histogram of  \cref{FIG:noisy_hist}). This problem can be tackled in a Bayesian framework, where the data generation mechanism is modelled through a set of parameters $\myvec{\theta}$ that can be inferred using the available data $\mat{Z}$. The probability of observing a Lidar cube $\mat{Z}$ is given by the likelihood $p(\mat{Z}|\myvec{\theta})$. The \emph{a priori} knowledge of the unknown parameters $\myvec{\theta}$ is embedded in the prior distribution $p(\myvec{\theta}|\mat{\Psi})$ given a set of hyperparameters $\mat{\Psi}$. Following Bayes theorem, the posterior distribution of the model parameters is 
\begin{equation}
p(\myvec{\theta}|\mat{Z},\mat{\Psi})=\frac{p(\mat{Z}|\myvec{\theta})p(\myvec{\theta}|\mat{\Psi})}{\int p(\mat{Z}|\myvec{\theta})p(\myvec{\theta}|\mat{\Psi})d\myvec{\theta}}.
\end{equation}
\subsection{Likelihood}
A 3D point cloud is represented by an unordered set of points
\begin{equation}\label{EQ:set_of_3D_points}
\mat{\Phi}=\{(\myvec{c}_n,r_n), n=1,\dots,N_\Phi\} 
\end{equation}
where $N_\Phi$ is the total number of points, $\myvec{c}_n = (x_n,y_n,t_n)^T \in \mathbb{R}^3$ is a coordinate vector and $r_n \in \mathbb{R}^+$ is the \ans{intensity}\footnote{\ans{The reflectivity of the point, limited to $(0,1]$, can be obtained as $\max\{1,r_n/(\eta N_\textrm{rep}\sum_{t}h(t))\}$, where $\eta\in[0,1]$ is the quantum efficiency of the detector and $N_\textrm{rep}$ is the number of laser pulses sent per pixel.}} of the $n$th point. For clarity in the notation, we will also denote the set of point coordinates as $
\mat{\Phi}_c=\{\myvec{c}_n, n=1,\dots,N_\Phi\}$ and the set of \ans{intensity} values as $\mat{\Phi}_r=\{r_n, n=1,\dots,N_\Phi\}$.

According to \cite{hernandez2007bayesian}, in the presence of distributed objects, the observed photon count in bin $t$ and pixel $(i,j)$ follows a Poisson distribution, whose intensity is a mixture of the pixel background level $b_{i,j}$ and the responses of the surfaces present in that pixel, i.e.,
\begin{equation}
\label{EQ:likelihood}
z_{i,j,t} | (\mat{\Phi} ,b_{i,j})\sim \mathcal{P} \left(\sum_{\substack{n : (x_n,y_n)=(i,j)}}g_{i,j} r_{n}h(t-t_{n})+g_{i,j} b_{i,j} \right) 
\end{equation}
where $t\in\{1,...,T\}$, $T$ is the number of histogram bins, $h(\cdot)$ is the known temporal instrumental response and $g_{i,j}$ is a scaling factor that represents the gain/sensitivity of the detector in pixel $(i,j)$. Assuming mutual independence between the noise realizations in different time bins and pixels, the full likelihood can be written as 
\begin{equation}
\label{EQ:full_likelihood}
p(\mat{Z} | \mat{\Phi},\mat{B}) = \prod_{i=1}^{N_c} \prod_{j=1}^{N_r} \prod_{t=1}^{T} p(z_{i,j,t} | \mat{\Phi},b_{i,j})
\end{equation}
where $\mat{Z}$ is the full Lidar cube with $[\mat{Z}]_{i,j,t} = z_{i,j,t}$, $\mat{B}$ is the background 2D image and $N_r$ and $N_c$ are the numbers of pixels in the vertical and horizontal axes respectively. Note that $p(z_{i,j,t} | \mat{\Phi},b_{i,j})$ in \cref{EQ:full_likelihood} is the Poisson distribution associated with \cref{EQ:likelihood}.

\subsection{Markov marked point process}\label{SUBSEC:point process model}
\ans{The set of points $\mat{\Phi}$ is defined inside the 3D space $\mathcal{T}=[0,N_r]\times[0,N_c]\times[0,T]$. 
Interactions between points can be characterized by defining densities with respect to the Poisson reference measure, i.e.,}
\begin{equation*}
f(\mat{\Phi}_c) \propto f_1(\mat{\Phi}_c)\dots f_r(\mat{\Phi}_c),
\end{equation*}
\ans{where $\propto$ means "proportional to". A more detailed definition of the point process theory can be found in \ref{APP:point_process_theory}.} In this work, we only consider Markovian interactions between points. The benefits of this property are twofold: a) Markovian interactions are well suited to describe the spatial correlations in natural 3D scenes \cite{moller2007modern} and b) inference is performed using only local updates, which leads to a low computational complexity. We can constrain the minimum distance between two different surfaces in the same pixel using the hard object process with density  
\begin{equation}
\label{EQ:strauss}
f_1(\mat{\Phi}_c) \propto \left\{
\begin{array}{ll}
0  & \text{if } \exists~ n \ne n' : x_n=x_{n'}, y_n=y_{n'} \\
&   \text{and } |t_{n}-t_{n'}| < d_{\min} \\
1 & \mbox{otherwise}
\end{array}
\right. 
\end{equation}
which is a special case of the repulsive Strauss process \cite{van2000markov}, where $d_{\min}$ is the minimum distance between two points in the same pixel. Attraction between points of the same surface in neighbouring pixels cannot be modelled with another Strauss process, due to a phase transition of extremely clustered realizations, as explained in \cite{van2000markov,moller2007modern}. However, a smoother transition into clustered configurations can be achieved by the area interaction process, introduced by Baddeley and Van Lieshout in \cite{baddeley1995area}. In this case, the density is defined as
\begin{equation}
\label{EQ:area_interaction_density}
f_2(\mat{\Phi}_c|\gamma_{a},\lambda_{a})=k_1\lambda_{a}^{N_\Phi}\gamma_{a}^{-m\left( \bigcup_{n=1}^{N_\Phi}S(\myvec{c}_{n}) \right)}
\end{equation}
where $\lambda_{a}$ is a positive parameter that controls the total number of points, $\gamma_{a}\ge1$ is a parameter adjusting the attraction between points and $k_1$ is an intractable normalizing constant. The exponent of $\gamma_{a}$ in \cref{EQ:area_interaction_density} is the measure $m(\cdot)$ over the union of convex sets $S(\myvec{c}_{n})\subseteq \mathcal{T}$ centered around each point $\myvec{c}_{n}$. In this way, the density is bigger when the intersection of the convex sets around two interacting points is closer to the union of them, i.e., if the points are clustered together. The special case $\gamma_{a}=1$ corresponds to a Poisson point process (without considering a Strauss process) with an intensity proportional to $\lambda_{a}\lambda(\cdot)$ (see \cref{APP:lambda_ai_equal_1} for details). In the rest of this work, we fix $\lambda(\mathcal{T})=1$ and control the number of points with the parameter $\lambda_{a}$. 
The set $S(\myvec{c}_n)$ is defined as a cuboid with a face of $N_{p}\times N_{p}$ squared pixels and a depth of $2N_{b}+1$ histogram bins and $m(\cdot)$ is the Lebesgue measure on $\mathcal{T}$. This set determines a cuboid of influence around each point, allowing interactions up to a distance of $N_{p}-1$ pixels and $N_{b}$ bins.   As two points in the same pixel generally correspond to different surfaces, we set $d_{\min}>2N_{b}$, thus constraining the minimum distance between two surfaces in the same pixel. The combination of the Strauss process and the area interaction process implicitly defines a connected-surface structure, where each point belongs. 

\begin{figure}[!h]
	\centering
	\begin{subfigure}{0.3\textwidth}
		\centering
		\includegraphics[width=1\textwidth]{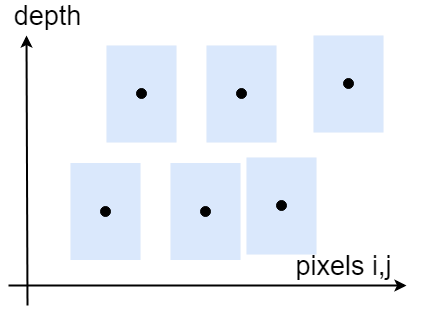}
		\caption{}
	\end{subfigure}%
	\begin{subfigure}{0.3\textwidth}
		\centering
		\includegraphics[width=1\textwidth]{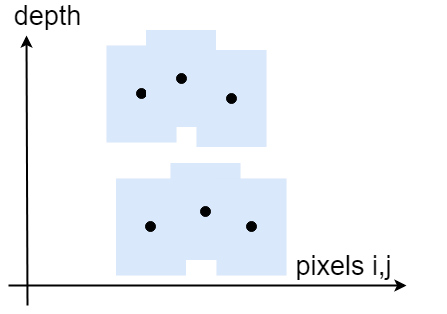}
		\caption{}
	\end{subfigure}%
	\begin{subfigure}{0.3\textwidth}
		\centering
		\includegraphics[width=1\textwidth]{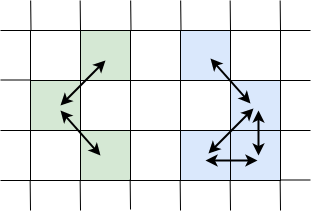}
		\caption{}
	\end{subfigure}
	\caption{Figures (a) and (b) show two different point configurations. Each point $\myvec{c}_n$ is denoted by a black dot and the corresponding blue rectangle depicts the area of the convex set $S(\myvec{c}_n)$. The configuration shown in (a) has a lower prior probability than the one shown in (b), as the union of all sets $S(\myvec{c}_n)$ is smaller in (b) with respect to the Lebesgue measure. Figure (c) shows the connectivity at an inter-pixel level when $N_{p}=3$. The green and blue squares correspond to pixels with points associated with two different surfaces. The white squares denote pixels without points. For simplicity, in this example all points are considered to be present at the same depth. Note that each pixel can be connected with at most 8 neighbours.}
	\label{FIG:connected_surfaces}
\end{figure}

\begin{figure}[t!]
	\centering
	\begin{subfigure}{0.35\textwidth}
		\centering
		\includegraphics[width=1\textwidth]{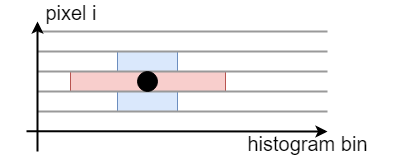}
		\caption{}
	\end{subfigure}%
	\begin{subfigure}{0.35\textwidth}
		\centering
		\includegraphics[width=1\textwidth]{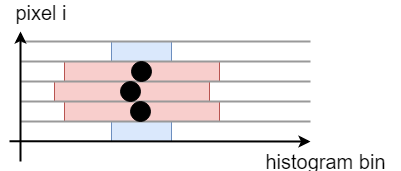}
		\caption{}
	\end{subfigure}%
	\caption{In both figures, the red colour denotes the space where no other points can be found (Strauss process) whereas the blue colour denotes the volume where other points are likely to appear (area interaction process). (a) Example of configuration with $1$ point. (b) Example of configuration with $3$ points.}
	\label{FIG:area_int+strauss}
\end{figure}

\Cref{FIG:connected_surfaces,FIG:area_int+strauss} illustrate the connected-surface structure via several examples.
The hyperparameters  $\gamma_{a}$ and $\lambda_{a}$ of the area interaction process are difficult to estimate, as there is an intractable normalizing constant in the density of \cref{EQ:area_interaction_density} and standard MCMC methods cannot be directly applied. Although there exist ways of bypassing this problem (e.g., \cite{murray2012mcmc}), we fixed these hyperparameters in all our experiments to ensure a reasonable computational complexity.  

After defining the spatial priors, the marked point process is constructed by adding the \ans{intensity} marks $r_n$ to the set $\mat{\Phi}_c$ with the density detailed in the next section. An illustration of the proposed prior can be found in \cref{APP:sampling_prior}.

\subsection{Intensity prior model}\label{SEC:intenreg}
In natural scenes, the \ans{intensity} values of points within a same surface exhibit strong spatial correlation. Following the Bayesian paradigm, this prior knowledge can be integrated into our model by defining a prior distribution over the point marks.  Gaussian processes are classically used in spatial statistics. However, the underlying covariance structure needs to consider too many neighbouring points to attain sufficient smoothing, which involves a prohibitive computational load. In order to obtain similar results with a lower computational burden, we propose to exploit the connected-surface structure to define a nearest neighbour Gaussian Markov random field (GMRF), similar to the one used by McCool \emph{et al.} in \cite{mccool2016robust}. First, we alleviate the difficulties induced by the positivity constraint of the \ans{intensity} values by introducing the following change of variables, which is a standard choice in spatial statistics dealing with Poisson noise (see \cite[Chapter~4]{rue2005gaussian})
\begin{equation}
m_n = \log(r_n) \quad n=1,\dots,N_{\Phi_c}.
\end{equation}
Second, the spatial correlation is promoted by defining the conditional distribution of the log-intensities, i.e.,
\begin{equation} \label{EQ:a_n_cond}
p(m_{n}|\mathcal{M}_{pp}(\myvec{c_{n}}),\sigma^2,\beta) \propto  \exp\left(-\frac{1}{2\sigma^2}\left(\sum_{n'\in\mathcal{M}_{pp}(\myvec{c}_n)}\frac{(m_{n}-m_{n'})^2}{d(\myvec{c}_{n};\myvec{c}_{n'})}+m^2_{n}\beta\right)\right)
\end{equation}
where $\mathcal{M}_{pp}(\myvec{c}_{n})$ is the set of neighbours of $\myvec{c}_{n}$, $d(\myvec{c}_{n};\myvec{c}_{n'})$ denotes the distance between the points  $\myvec{c}_{n}$ and $\myvec{c}_{n'}$, and $\beta$ and $\sigma^2$ are two positive hyperparameters. The set of neighbours $\mathcal{M}_{pp}(\myvec{c}_{n})$ is obtained using the connected-surface structure, where each point can have at most $8$ neighbours, as illustrated in  \cref{FIG:connected_surfaces}. The distance between two points is computed according to
\begin{equation*}
d(\myvec{c}_{n};\myvec{c}_{n'})= \sqrt{(y_n-y_{n'})^2+(x_n-x_{n'})^2+\left(\frac{t_n-t_{n'}}{l_z} \right)^2}
\end{equation*}
with $l_z=\Delta_{p}/\Delta_{b}$, which normalizes the distance to have a physical meaning, where $\Delta_{p}$ and $\Delta_{b}$ are the approximate spatial resolutions of one pixel and one histogram bin respectively. \ans{This prior promotes a linear interpolation between neighbouring\footnote{\ans{The combination of a local Euclidean distance with a nearest neighbours definition can be seen to approximate the  manifold metrics \cite{Tenenbaum2319}.}} intensity values, as explained in \cite{rue2005gaussian}.} \ans{In this work, we assume that $\Delta_{p}$ is constant throughout the scene. If the scene presents significant distortion, i.e., objects separated by a significant distance in depth, $\Delta_{p}$ should depend on the position by computing the projective transformation between world coordinates and Lidar coordinates (a detailed explanation can be found in  \cite{Ferstl_2013_ICCV,hartley2003multiple}).} Following the Hammersley and Clifford theorem \cite{hammersley1971markov}, the joint intensity distribution is given by the multivariate Gaussian distribution
\begin{equation} \label{EQ:a_joint}
\myvec{m}|\sigma^2,\beta, \mat{\Phi}_c \sim \mathcal{N}(\myvec{0},\sigma^2\mat{P}^{-1})
\end{equation}
where $\mat{P}$ is the unscaled precision matrix of size $N_\Phi\times N_\Phi$ with the following elements
\begin{equation} \label{EQ:P}
[\mat{P}]_{n,n'}=
\begin{cases}
\beta + \sum_{\tilde{n}\in\mathcal{M}_{pp}(\myvec{c}_n)} \frac{1}{d(\myvec{c}_n;\myvec{c}_{\tilde{n}} )} & \text{if $n=n'$} \\
-\frac{1}{d(\myvec{c}_n;\myvec{c}_{n'} )} & \text{if $\myvec{c}_n\in\mathcal{M}_{pp}(\myvec{c}_{n'})$} \\
0 & \text{otherwise}
\end{cases}
\end{equation}
The parameter $\sigma^2$ controls the surface \ans{intensity} smoothness and $\frac{\beta}{\sigma^2}$ is related to the \ans{intensity} variance of a point without any neighbour. In addition, the parameter $\beta$ ensures a proper joint prior distribution, as $\mat{P}$ is diagonally dominant, thus full rank \cite{rue2005gaussian}.

\subsection{Background prior model}\label{SEC:backreg}
\begin{figure}[!h]
	
	\centering
	\begin{subfigure}[t]{0.35\textwidth}
		\centering
		\includegraphics[width=1\textwidth]{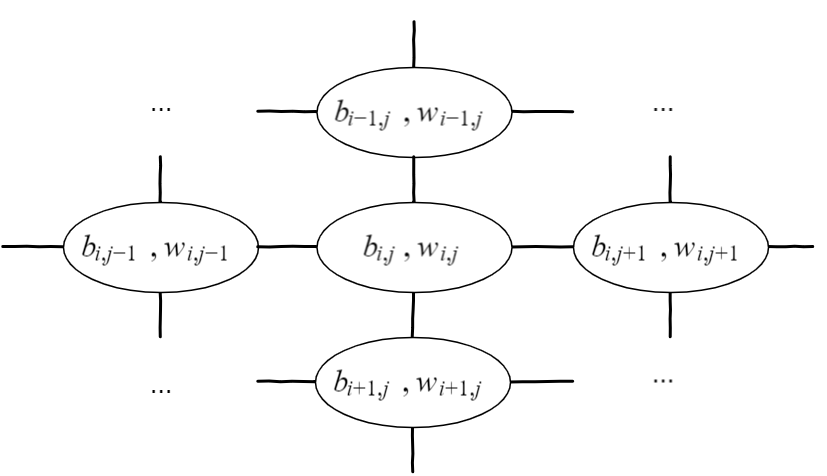}
		\caption{}
	\end{subfigure}
	\begin{subfigure}[t]{0.35\textwidth}
	\centering
	\includegraphics[width=1\textwidth]{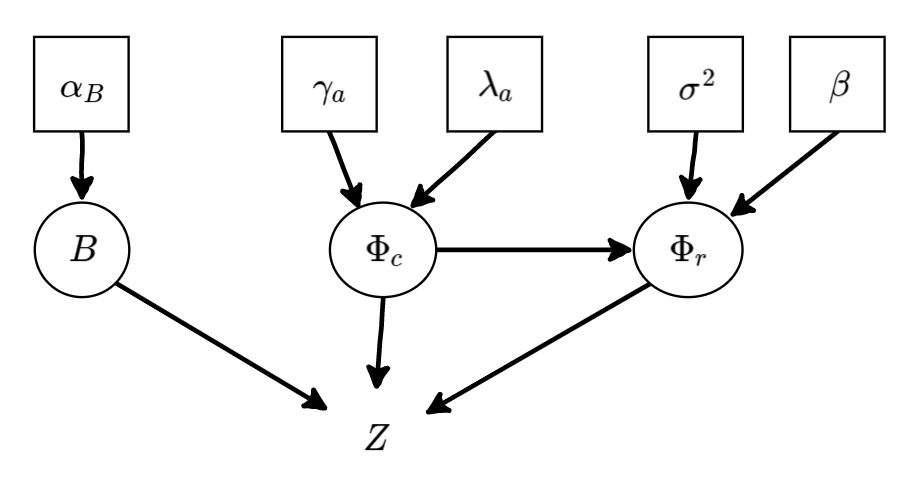}
	\caption{}
	\end{subfigure}
	\caption{(a) illustrates the Gamma Markov random field neighbouring structure $\mathcal{M}_B$. Each ${b}_{i,j}$ is connected to 5 auxiliary variables $w_{i',j'}$ as depicted by the continuous lines , including the one with the same subscript. Similarly, each $w_{i,j}$ is also connected to other $5$ variables ${b}_{i',j'}$ as indicated by the continuous lines. (b) shows the directed acyclic graph (DAG) of the proposed hierarchical Bayesian model. The variables inside squares are fixed, whereas the variables inside circles are estimated.}
	\label{FIG:gamma mrf}
\end{figure}

Non-coherent illumination sources, such as the solar illumination in outdoor scenes or room lights in the indoor case, are related to arrivals of photons at random times (uniformly distributed in time) to the single-photon detector. The level of these spurious detections is modelled as a 2D image of mean intensities $b_{i,j}$ with $i=1,\dots,N_r$ and $j=1,\dots,N_c$. \ans{If the transceiver system of the Lidar is mono-static\footnote{The transceiver system is mono-static when the transmit and receive channels are co-axial and thus share the same objective lens aperture.} (e.g., the system described in \cite{McCarthy:09}), the background image is usually similar to the objects present in the scene and exhibits spatial correlation, as background photons generally arise from the ambient light reflecting from parts of the targets and being collected by the system}. Hence, we use a hidden gamma Markov random field prior distribution for $\mat{B}$ that takes into account the background positivity and spatial correlation. This prior was introduced by Dikmen and Cemgil in \cite{dikmen2010gamma} and applied in many image processing applications with Poisson likelihood \cite{altmann2017bayesian,altmann2017robust}. In \cite{dikmen2010gamma}, the distribution of $b_{i,j}$ is defined via auxiliary variables $[\mat{W}]_{i,j}=w_{i,j}$  such that
\begin{align}
b_{i,j} | \mathcal{M}_B(b_{i,j}), \alpha_{B} &\sim \mathcal{G} \left(\alpha_{B},\frac{\overline{b}_{i,j}}{\alpha_{B}} \right) \\
w_{i,j} | \mathcal{M}_B(w_{i,j}), \alpha_{B} &\sim \mathcal{IG}(\alpha_{B},\alpha_{B}\overline{w}_{i,j})
\end{align}
where  $\mathcal{M}_B$ denotes the set of 5 neighbours as shown in  \cref{FIG:gamma mrf}, $\mathcal{G}$ and $ \mathcal{IG}$ indicate gamma and inverse gamma distributions, $\alpha_{B}$ is a hyperparameter controlling the spatial regularization and 
\begin{align}
\label{EQ:b_bar}
\overline{b}_{i,j}=\left(\frac{1}{4}\sum_{(i',j')\in \mathcal{M}_B(b_{i,j})}w_{i',j'}^{-1}\right)^{-1} \\
\label{EQ:u_bar}
\overline{w}_{i,j}=\frac{1}{4}\sum_{(i',j')\in \mathcal{M}_B(w_{i,j})}^{}b_{(i',j')}
\end{align} 
We are interested in the marginal distribution of the gamma Markov random field $p(\mat{B} |\alpha_{B})$ that integrates over all possible realizations of the auxiliary variables $w_{i,j}$. The expression of this marginal density can be obtained analytically (as detailed in  \cref{APP: GMRF marginal}) as 
\begin{align}
\label{EQ:marginalized GMRF}
p(\mat{B} |\alpha_{B}) &\propto \int p(\mat{B},\mat{W}|\alpha_{B})d\mat{W}\\
&\propto \prod_{i=1}^{N_c} \prod_{j=1}^{N_r} \frac {b_{i,j}^{\alpha_{B}-1}} { \left(\sum_{(i',j')\in \mathcal{M}_B(w_{i,j})} b_{i',j'}\right)^{\alpha_{B}}}.
\end{align}
In this work, we fix the value of $\alpha_{B}$, even if it could also be estimated using a stochastic gradient procedure as explained in \cite{pereyra2014maximum}, at the expense of an increase in the computational load. \ans{If the system is not mono-static, i.e., there is no prior assumption of smoothness in the background image, the value of $\alpha_{B}$ is set to 1.}

\subsection{Posterior distribution}
The joint posterior distribution of the model parameters is given by
\begin{multline}
\label{EQ:posterior}
p(\mat{\Phi}_c,\mat{\Phi}_r,\mat{B}|\mat{Z},\Psi) \propto p(\mat{Z}|\mat{\Phi}_c,\mat{\Phi}_r,\mat{B})  p(\mat{\Phi}_r|\mat{\Phi}_c,\sigma^2,\beta) \\ \times f_1(\mat{\Phi}_c|\gamma_{a},\lambda_{a}) f_2(\mat{\Phi}_c|\gamma_{st})\pi(\mat{\Phi}_c)p(\mat{B}|\alpha_{B})
\end{multline}
where $\Psi$ denotes the set of hyperparameters  $\Psi=\{\gamma_{a},\lambda_{a},\gamma_{s},\sigma^2,\beta,\alpha_{B}\}$. \Cref{FIG:gamma mrf} shows the directed acyclic graph associated with the proposed hierarchical Bayesian model.
\section{Estimation strategy}\label{SEC:estimation_strategy}
Bayesian estimators associated with the full posterior in \cref{EQ:posterior} are analytically intractable. Moreover, standard optimization techniques cannot be applied due to the highly multimodality of the posterior distribution. However, we can obtain numerical estimates using samples generated by a Monte Carlo method denoted as
\begin{equation}
\{\mat{\Phi}^{(s)} ,\mat{B}^{(s)}   \quad \forall s=0,1,\dots, N_{i}-1 \}.
\end{equation}
where $N_{i}$ is the total number of samples.
In this work, we will focus on the maximum-a-posteriori (MAP) estimator of the point cloud positions and \ans{intensity} values, i.e.,
\begin{equation}
\mat{\hat{\Phi}}=\argmax_{\mat{\Phi}} p(\mat{\Phi},\mat{B}|\mat{Z},\Psi),
\end{equation}
which is approximated by
\begin{equation}
\mat{\hat{\Phi}} \approx \argmax_{s=0,...,N_i-1} p(\mat{\Phi}^{(s)} , \mat{B}^{(s)} |\mat{Z},\Psi). 
\end{equation}
In our experiments, we found that the minimum mean squared error estimator (MMSE) of $\mat{B}$, i.e.,
\begin{equation}
\mat{\hat{B}}=\mathbb{E}\{\mat{B}|\mat{Z},\Psi\}
\end{equation}
achieves better background estimates than the MAP estimator. This estimator can be approximated by the empirical mean of the posterior samples of $\mat{B}$, that is
\begin{equation}
\label{EQ:B_estimator}
\mat{\hat{B}} \approx \frac{1}{N_{i}}\sum_{s=N_{\textrm{bi}}+1}^{N_{i}}\mat{B}^{(s)} .
\end{equation}
where $N_{\textrm{bi}}=N_{i}/2$ is the number of burn-in iterations.
In many applications, assessing the presence or absence of a target at a pixel level can be of special interest (e.g., \cite{halimi2017,altmanndetect2016}). Here, we can use the Monte Carlo samples to estimate the probability of having $k$ objects present in pixel $(i,j)$, as
\begin{equation}\label{EQ:prob_k_returns}
P(\text{$k$ returns in $(i,j)$} | \mat{Z}, \Psi) = \frac{1}{N_{i}}\sum_{s=N_{\textrm{bi}}+1}^{N_{i}}\indicator{k\text{ points in } (i,j)}{\mat{\Phi}^{(s)}}.
\end{equation}
\textit{Remark: If more detailed posterior statistics are needed, it is possible to fix the dimensionality of the problem using the estimate $\mat{\hat{\Phi}} $ and run a fixed dimensional sampler for additional $N_{i}$ iterations (see  \cref{APP: posterior statistics}).}

Many samplers capable of exploring different model dimensions, i.e., different numbers of points, are available in the point process literature (a complete summary can be found in \cite[Chapter~9]{brooks2011handbook}). The continuous birth-death chain method builds a continuous-time Markov chain that converges to the posterior distribution of interest. Alternatively, perfect sampling approaches generates samples using a rejection sampling scheme, which incurs in a bigger computational load. Finally, the Reversible jump Markov chain Monte Carlo (RJ-MCMC) sampler, introduced by Green in \cite{green1995reversible}, constructs a discrete time Markov chain, where moves between different dimensions are proposed and accepted or rejected in order to converge to the posterior distribution of interest. In this work, we choose an RJ-MCMC sampler, as this option allows us to design application-specific proposals that speed up the convergence rate.

In addition, we propose a data augmentation scheme to sample the background levels. This technique introduces extra auxiliary (latent) variables $\myvec{u}$ and generates samples in this augmented model space $(\mat{B}^{(s)},\myvec{u}^{(s)})\sim p(\mat{B},\myvec{u}|\mat{Z},\mat{\Phi},\alpha_{B})$, which is easier than sampling the marginal distribution $p(\mat{B}|\mat{Z},\mat{\Phi},\alpha_{B})$. The resulting samples $\mat{B}^{(s)}$ are distributed according to the desired marginal density (detailed theory and applications of data augmentation can be found in \cite[Chapter~10]{brooks2011handbook}). 

\subsection{Reversible jump Markov chain Monte Carlo}
RJ-MCMC can be seen as a natural extension of the Metropolis-Hastings algorithm for problems with an unknown \emph{a priori} dimensionality. Given the actual state of the chain $\myvec{\theta} =\{\mat{\Phi} , \mat{B} \}$ of model order $N_{\Phi }$, a random vector of auxiliary variables $\myvec{u}$ is generated to create a new state  $\myvec{\theta}'=\{\mat{\Phi}',\mat{B'}\}$ of model order $N_{\Phi'}$, according to an appropriate deterministic function $\myvec{\theta}'=g(\myvec{\theta} ,\myvec{u})$. To ensure reversibility, an inverse mapping with auxiliary random variables $\myvec{u'}$ has to exist such that $\myvec{\theta} =g^{-1}(\myvec{\theta}',\myvec{u'})$. The move $\myvec{\theta} \rightarrow\myvec{\theta}'$ is accepted or rejected with probability $\rho=\min\{1,r\left(\myvec{\theta} ,\myvec{\theta}'\right)\}$, where $r(\cdot,\cdot)$ satisfies the so-called dimension balancing condition
\begin{equation}
\label{EQ:RevJump}
r\left(\myvec{\theta},\myvec{\theta'}\right) = \frac{p(\myvec{\theta'}|\mat{Z},\Psi)K(\myvec{\theta}|\myvec{\theta'})p(\myvec{u'})}{p(\myvec{\theta}|\mat{Z},\Psi)K(\myvec{\theta'}|\myvec{\theta})p(\myvec{u})}\left|\frac{\partial g(\myvec{\theta},\myvec{u})}{\partial(\myvec{\theta},\myvec{u})}\right|
\end{equation}
where $K(\myvec{\theta'}|\myvec{\theta})$ is the probability of proposing the move $\myvec{\theta}\rightarrow\myvec{\theta'}$, $p(\myvec{u})$ is the probability distribution of the random vector $\myvec{u}$, and $\left|\frac{\partial g(\myvec{\theta},\myvec{u})}{\partial(\myvec{\theta},\myvec{u})}\right|$ is the Jacobian of the mapping $g(\cdot)$. All the terms involved in \cref{EQ:RevJump} have a complexity that depends only on the size of the neighbourhood, except the prior distribution of the \ans{intensity} values defined in \cref{EQ:a_joint}. Note that \cref{EQ:RevJump} involves the computation of the ratio of determinants of the precision matrices $\mat{P}$ and $\mat{P}'$, which have a global dependency on all the points in $\Phi_r$. To keep the computational complexity low, we address this difficulty by only considering a block diagonal approximation of $\mat{P}$, which includes only points in local neighbourhoods (see  \cref{APP:approx} for more details). \textcolor{black}{The RJ-MCMC algorithm performs birth, death, dilation, erosion, spatial shift, mark shift, split and merge moves with probabilities $p_{\textrm{birth}}$, $p_{\textrm{death}}$, $p_{\textrm{dilation}}$, $p_{\textrm{erosion}}$, $p_{\textrm{shift}}$, $p_{\textrm{mark}}$, $p_{\textrm{split}}$ and $p_{\textrm{merge}}$.} These moves are detailed in the following subsections. For ease of reading we summarize the key aspects of each move, without specifying the full acceptance rate expression of \cref{EQ:RevJump}, which can be found in  \cref{APP:acc_ratio}.

\subsubsection{Birth and death moves}
The birth move proposes a new point $(\myvec{c}_{N_\Phi+1},r_{N_\Phi+1})$ uniformly at random in $\mathcal{T}$. 
The \ans{intensity} of the new point is computed according to the following scheme
\begin{equation}
\left\{
\begin{aligned}
\textcolor{black}{u} & \sim \mathcal{U}(0,1),  b_{i,j}' = \textcolor{black}{u}b_{i,j}\\
e^{m_{N_\Phi+1}} &= (1-\textcolor{black}{u})b_{i,j}\frac{T}{\sum_{t=1}^{T}h(t)} 
\end{aligned}
\right.
\end{equation}
This mapping preserves the total posterior intensity of the pixel, since 
\begin{eqnarray}
e^{m_{N_\Phi+1}}\sum_{t=1}^{T}h(t) + b_{i,j}'T = b_{i,j}T,
\end{eqnarray}
thus yielding a relatively high acceptance probability. Its reversible pair, the death move, proposes to remove one point randomly. In this case, the inverse mapping is given by
\begin{equation}
b_{i,j}' = b_{i,j} +e^{m_{N_\Phi+1}}\frac{\sum_{t=1}^{T}h(t)}{T},
\end{equation}
The acceptance ratio for the birth move reduces to $\rho=\min\{1,C_1\}$ with $C_1$ given by \cref{EQ:RevJump}, where the posterior ratio is computed according to \cref{EQ:posterior}, $K(\myvec{\theta'}|\myvec{\theta})=p_{\textrm{birth}}$, $K(\myvec{\theta}|\myvec{\theta'})=p_{\textrm{death}}$, \textcolor{black}{$p(\myvec{u})=\frac{\lambda(\cdot)}{\lambda(\mathcal{T})}$ and $p(\myvec{u}')=\frac{1}{N_\Phi+1}$  and a Jacobian equal to
\begin{equation}
\left|\frac{\partial g(\myvec{\theta},\myvec{u})}{\partial(\myvec{\theta},\myvec{u})}\right|=\frac{1}{1-u}.
\end{equation}
The death move is accepted or rejected with probability $\rho = \min\{1,C_1^{-1}\}$, modifying $p(\myvec{u})$ accordingly (i.e., changing $\frac{1}{N_\Phi+1}$ to $\frac{1}{N_\Phi}$). }

\subsubsection{Dilation and erosion moves}
Standard birth and death moves yield low acceptance rates, because the probability of proposing a point in a likely position is relatively low, as the detected surfaces only occupy a small subset of the full 3D volume $\mathcal{T}$. To overcome this problem, we propose new RJ-MCMC moves that explore the target distribution by dilating and eroding existing surfaces. The dilation move randomly picks a point $\myvec{c}_{n}$ that has less than 8 neighbours, and then proposes a new neighbour $\myvec{c}_{N_\Phi+1}$ with uniform probability across all possible pixel positions (where a point can be added). 
The new \ans{intensity} can be sampled from the Gaussian prior, taking into account the available information from the neighbours, i.e., $u$ is sampled from the conditional distribution specified in \cref{EQ:a_n_cond} and $m_{N_\Phi+1} = u$. The background level is adjusted to keep the total intensity of the pixel unmodified
\begin{equation}
\label{EQ:background_update_dilero}
b'_{i,j} = b_{i,j} - e^{m_{N_\Phi+1}}\frac{\sum_{t=1}^{T}h(t)}{T}.
\end{equation}
If the resulting background level in \cref{EQ:background_update_dilero} is negative, the move is rejected. The complementary move (named erosion) proposes to remove a point $\myvec{c}_{n}$ with one or more neighbours. 
In a similar fashion to the birth move, a dilation is accepted with probability $\rho=\min\{1,C_2\}$, with $C_2$ computed according to \cref{EQ:RevJump}. In this case, $p(\myvec{u})=p(u_1)p(u_2)$ with
\begin{equation}
p(u_1) = \frac{1}{N_\Phi(2N_b+1)}\sum_{m\in\mathcal{M}_{pp}(\myvec{c}_{N_\Phi+1})}  \card\mathcal{M}_{pp}(\myvec{c}_{m})
\end{equation}
where $0\le\card\mathcal{M}_{pp}(\myvec{c}_{m})\le8$ denotes the number of neighbouring points of $\myvec{c}_m$. The expression of $p(u_2)$ is given by the conditional distribution defined in \cref{EQ:a_n_cond} and the Jacobian term equals $1$. The probability of $u'$ is given by
\begin{equation}
p(u')=\frac{1}{\sum_{m=1}^{N_\Phi+1} \indicator{\mathbb{Z}_{+}}{\card\mathcal{M}_{pp}(\myvec{c}_{m})}}
\end{equation}
and the transition probabilities are $K(\myvec{\theta'}|\myvec{\theta})=p_{\textrm{dilation}}$ and $K(\myvec{\theta}|\myvec{\theta'})=p_{\textrm{erosion}}$.
An erosion move is accepted with probability $\rho=\min\{1,C_2^{-1}\}$.

\subsubsection{Shift move}
The shift move modifies the position of a given point. The point is chosen uniformly at random and a new position inside the same pixel is proposed using a random walk Metropolis proposal \textcolor{black}{defined as
\begin{equation}\label{EQ:shift_prop}
 u \sim \mathcal{N}\left(t_n ,\delta_{t}\right).
\end{equation}
and $t'_n=u$.} The resulting acceptance ratio is $\rho=\min\{1,C_3\}$, with $C_3$ computed according to \cref{EQ:RevJump}, where $K(\myvec{\theta'}|\myvec{\theta})=K(\myvec{\theta}|\myvec{\theta'})=p_{\textrm{shift}}$, $p(u)=p(u')$ given by the Gaussian distribution of \cref{EQ:shift_prop}  and a Jacobian equal to $1$. The value of $\delta_{t}$ is set to $(\frac{N_{b}}{3})^2$ to obtain an acceptance ratio close to $41\%$, which is the optimal value, as explained in \cite[Chapter~4]{brooks2011handbook}.
\subsubsection{Mark move}
Similarly to the shift move, the mark move refines the \ans{intensity} value of a randomly chosen point. The corresponding proposal is a Gaussian distribution with variance $\delta_{m}$
\begin{equation}\label{EQ:mark_prop}
u \sim \mathcal{N}\left(m_{n} ,\delta_{m}\right).
\end{equation}
and $m'_{n}=u$. In this move, the acceptance ratio is $\rho=\min\{1,r(\myvec{\theta},\myvec{\theta'})\}$, where $K(\myvec{\theta'}|\myvec{\theta})=K(\myvec{\theta}|\myvec{\theta'})=p_{\textrm{mark}}$, $p(u)=p(u')$ given by \cref{EQ:mark_prop} and a Jacobian equal to $1$. As in the shift move, we set the value of $\delta_{m}$ to $(0.5)^2$ to obtain an acceptance ratio close to $41\%$. 
\subsubsection{Split and merge moves}
\begin{figure}[!h]
	\centering
	\includegraphics[width=0.6\textwidth]{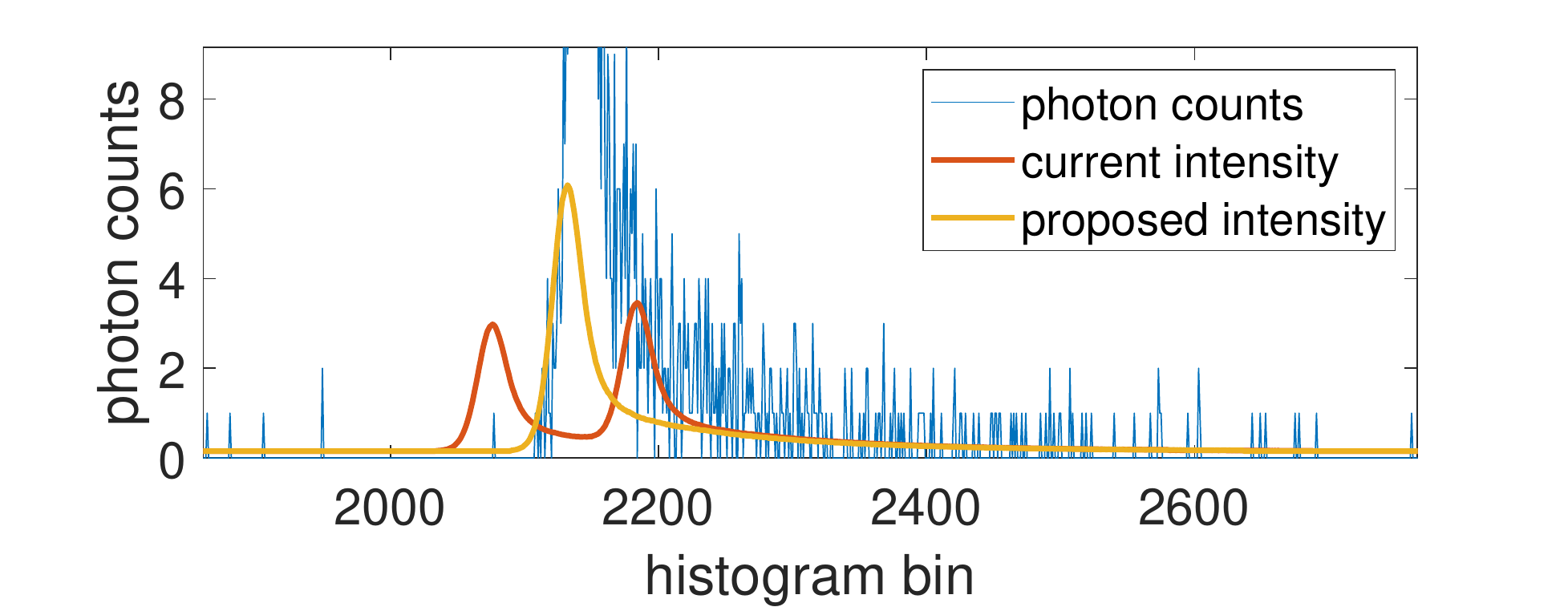}
	\caption{In scenarios where the sampler proposes two points (red line) instead of one (yellow line), the probability of killing one of them and shifting the other is very low. However, accepting a merge move has high probability.}
	\label{FIG:merge_split_example}
\end{figure}

In Lidar histograms with many photon counts per pixel, the likelihood function becomes very peaky and the non-convexity of the problem becomes more difficult to handle. This non-convexity is related to the discrete nature of the point process, similar to problems where the $l_0$ pseudo-norm regularization is used, as discussed in \cite{zhou2009non}. In such cases, when one true \ans{surface} is associated with two points, as illustrated in  \cref{FIG:merge_split_example}, the probability of performing a death move followed by a shift move is very low. To alleviate this problem, we propose a merge move and its complement, the split move. 
A merge move is performed by randomly choosing two points $\myvec{c}_{k_1}$ and $\myvec{c}_{k_2}$ inside the same pixel ($x_{k_1}=x_{k_2}$ and $y_{k_1}=y_{k_2}$) that satisfy the condition 
\begin{equation}
\label{EQ: merge_condition}
d_{\min}<\left|t_{k_1}-t_{k_2}\right| \le \text{attack}_{h(t)}+\text{decay}_{h(t)}
\end{equation}
where $\text{attack}_{h(t)}$ is the length of the impulse response until the maxima and $\text{decay}_{h(t)}$ is the length after the maxima until the value where $h(t)$ is negligible. The merged point $(\myvec{c'}_n,r'_n)$ is finally obtained by the mapping
\begin{equation}
\left\{
\begin{aligned}
\label{EQ:merge_mapping}
e^{m'_{n}} &= e^{m_{k_1}} + e^{m_{k_2}}  \\
t'_{n} &= t_{k_1} \frac{e^{m_{k_1}}}{e^{m_{k_1}} + e^{m_{k_2}}} +t_{k_2} \frac{e^{m_{k_2}}}{e^{m_{k_1}} + e^{m_{k_2}}}
\end{aligned}
\right.
\end{equation}
that preserves the total pixel intensity and weights the spatial shift of each peak according to its relative amplitude. For instance, if two peaks of significantly different amplitudes are merged, the resulting peak will be closer to the original peak which presents the highest amplitude. The split move randomly picks a point $(\myvec{c'}_n,r'_n)$ and proposes two new points, $(\myvec{c}_{k_1},r_{k_1})$ and $(\myvec{c}_{k_2},r_{k_2})$, following the inverse mapping 
\begin{equation}
\left\{
\begin{aligned}
u &\sim \mathcal{U}(0,1) \\
\Delta &\sim \mathcal{U}(d_{\min},\text{attack}_{h(t)}+\text{decay}_{h(t)}) \\
m_{k_1} &= m'_{n} + \log(u) \\
m_{k_2} &= m'_{n} + \log(1-u)  \\
t_{k_1} &= t'_{n} - (1-u)\Delta \\
t_{k_2} &= t'_{n} + u\Delta
\end{aligned}
\right.
\end{equation}
which is based on the auxiliary variables $u$ and $\Delta$. This proposal verifies \cref{EQ:merge_mapping}, ensuring reversibility. \textcolor{black}{The acceptance ratio for the split move is 
$\rho=\min\{1,C_4\}$, with $C_4$ computed according to \cref{EQ:RevJump}, where the Jacobian is $1/u(1-u)$, $K(\myvec{\theta'}|\myvec{\theta})=p_{\textrm{shift}}$, $K(\myvec{\theta}|\myvec{\theta'})=p_{\textrm{merge}}$, $p(u)=\frac{1}{N_\Phi}(d_{\min}+\text{attack}_{h(t)}+\text{decay}_{h(t)})^{-1}$ and $p(u')$ is the inverse of the number of points in $\mat{\Phi}$ that verify \cref{EQ: merge_condition}. The acceptance probability of the merge move is simply $\rho=\min\{1,C_3^{-1}\}$.}

\subsection{Sampling the background}
In the presence of at least one peak in a given pixel, Gibbs updates cannot be directly applied to obtain background samples, as the linear combination between the \ans{objects} and the background level in \cref{EQ:likelihood} cancels the conjugacy between the Poisson likelihood and the gamma prior. However, this problem can be overcome by introducing auxiliary variables in a data augmentation scheme. In a similar fashion to \cite{zhou2012beta}, we propose to augment \cref{EQ:likelihood} as
\begin{align*}
z_{i,j,t} =& \sum_{\substack{n:(x_n,y_n)=(i,j)}} \tilde{z}_{i,j,t,n} + \tilde{z}_{i,j,t,b} \\
\tilde{z}_{i,j,t,b} \sim&  \mathcal{P}(g_{i,j} b_{i,j}) \\
\tilde{z}_{i,j,t,n} \sim&  \mathcal{P}(g_{i,j} r_n h(t-t_n)) 
\end{align*}
where $\tilde{z}_{i,j,t,n}$ are the photons in bin \textrm{$\# t$} associated with the $k$th \ans{surface} and $\tilde{z}_{i,j,t,b}$ are the ones associated with the background. If we also add the auxiliary variables $w_{i,j}$ of the gamma Markov random field (as explained in \cref{SEC:backreg}), we can construct the following Gibbs sampler
\begin{equation}
\label{EQ:B_scheme}
\left\{
\begin{aligned}
\tilde{z}_{i,j,t,b} &\sim \mathcal{B}\left(z_{i,j,t},\frac{b_{i,j}}{\sum_{\substack{n:(x_n,y_n)=(i,j)}}\exp(m_{n})h(t-t_{n})}\right) 
\\ w_{i,j} &\sim  \mathcal{IG}(\alpha_{B},\alpha_{B}\overline{w}_{i,j})
\\ b_{i,j} & \sim \mathcal{G} \left(\alpha_{B}+\sum_{t=1}^{T}\tilde{z}_{i,j,t,b},\frac{1}{T+\frac{\alpha_{B}}{\overline{b}_{i,j}}}\right)
\end{aligned}
\right.
\end{equation}
where $\mathcal{B}(\cdot)$ denotes the Binomial distribution, $\overline{w}_{i,j}$ and $\overline{b}_{i,j}$ are defined according to \cref{EQ:u_bar} and \cref{EQ:b_bar} respectively. The transition kernel defined by \cref{EQ:B_scheme} produces samples of $b_{i,j}$ distributed according to the marginal distribution of \cref{EQ:marginalized GMRF}. In practice, we use only one iteration of this kernel. 

\subsection{Full algorithm}
The RJ-MCMC algorithm alternates between birth, death, dilation, erosion, shift, mark, split and merge moves with probabilities as reported in \cref{TAB:RJMCMC params}. A complete background update is done every $N_{B}=N_rN_c$ iterations. After each accepted update, we compute the difference in the posterior density $\delta_{\textrm{map}}$ in order to keep track of the maximum density $\textrm{map}_{\max}$. After $N_{\textrm{bi}}=N_{i}/2$ burn-in iterations, we save the set of parameters $\mat{\Phi}$ that yield the highest posterior density and we also accumulate the samples of $\mat{B}$ to compute \cref{EQ:B_estimator}. \Cref{ALG: ManiPoP} shows a pseudo-code of the resulting RJ-MCMC sampler.

\begin{table}
	\centering
	\begin{tabular}{| c | c | c | c | c | c| c | c|}
		\hline	$p_{\textrm{birth}}$ & $1/24$ & $p_{\textrm{death}}$ & $1/24$ & $p_{\textrm{dilation}}$  & $5/24$ & $p_{\textrm{erosion}}$  &  $5/24$ \\  \hline
		$p_{\textrm{shift}}$ &  $5/24$ & 	$p_{\textrm{mark}}$ &   $5/24$ & 	$p_{\textrm{split}}$ & $1/24$ & 	$p_{\textrm{merge}}$ & $1/24$\\ \hline
	\end{tabular}
	\caption{Move probabilities used in the RJ-MCMC sampler.}
	\label{TAB:RJMCMC params}
\end{table}

\begin{algorithm}
	\caption{ManiPoP}\label{ALG: ManiPoP}
	\begin{algorithmic}[1]
		\STATE \textbf{Input:} Lidar waveforms $\mat{Z}$, initial estimate 
		$(\mat{\Phi}^{(0)},\mat{B}^{(0)})$ and hyperparameters $\Psi$
		\STATE \textbf{Initialization:} 
		\STATE $(\mat{\Phi},\mat{B}) \gets (\mat{\Phi}^{(0)},\mat{B}^{(0)})$
		\STATE $s \gets 0$ 
		\STATE \textbf{Main loop:} 
		\WHILE{$s < N_{i}$}
		\IF {$rem(s,N_{B})==0$}
		\STATE  $(\mat{\Phi},\mat{B},\delta_{\textrm{map}}) \gets$ sample $\mat{B}$ using \cref{EQ:B_scheme} 
		\ENDIF
		\STATE $\textrm{move} \sim \text{Discrete}(p_{\textrm{birth}},\dots,p_{\textrm{merge}})$ 
		\STATE  $(\mat{\Phi},\mat{B},\delta_{\textrm{map}}) \gets$ perform selected move 
		\STATE $\textrm{map} \gets \textrm{map} + \delta_{\textrm{map}}$
		\IF {$s \ge N_{\textrm{bi}}$}
		\STATE $\mat{\hat{B}} \gets \mat{\hat{B}} + \mat{B}$
		\IF {$\textrm{map}>\textrm{map}_{\max}$}
		\STATE $\mat{\hat{\Phi}} \gets \mat{\Phi}$
		\STATE $\textrm{map}_{\max} \gets \textrm{map}$
		\ENDIF
		\ENDIF
		\STATE $s \gets s+1$
		\ENDWHILE
		\STATE $\mat{\hat{B}} \gets \mat{\hat{B}}/(N_{i}-N_{\textrm{bi}})$
		\STATE \textbf{Output:} Final estimates $(\mat{\hat{\Phi}},\mat{\hat{B}})$
	\end{algorithmic}
\end{algorithm}

\section{Efficient implementation}\label{SEC: implementation}
In order to achieve a computational performance similar to other optimization-based approaches, while allowing a more complex modelling of the input data, we have considered the following implementation aspects
\begin{enumerate}
	\item Recently, the algorithm reported in \cite{azzari2016variance} showed that state-of-the-art denoising of images corrupted with Poisson noise can be obtained by starting from a coarser scale and progressively refining the estimates in finer scales. We propose a similar multiscale approach to achieve faster processing times and better scalability with the total data size. The proposed sequential procedure is detailed in \cref{SUBSEC:MR approach}.
	\item In the photon-starved regime considered in this work, the recorded histograms are generally extremely sparse, meaning that more than $95\%$ of the time bins are empty. Therefore, a histogram representation is inefficient, both in terms of likelihood evaluation and memory requirements. In \cite{shin2015photon}, the authors replaced the histograms by modelling directly each detected photon. Similarly, we represent the Lidar data by using an ordered list of bins and photon counts, only considering bins with at least one count (see  \cref{APP:log-likelihood evaluation} for more details).
	\item In order to avoid finding neighbours of a point to be updated at each iteration, we store and update an adjacency list for each point. This list allows the neighbour search only during the creation or shift of a point.
	\item To reduce the search space, we add a preprocessing step that computes the matched-filter response at the coarsest resolution. The time bins whose values are below a threshold (equal to \ans{$\frac{0.05}{T}\sum_{t=1}^{T}z_{i,j,t}\sum_{t=1}^{T}\log h(t)$) are assigned zero intensity in the point process prior, i.e., $\lambda(\cdot) = 0$. In this way, the search includes with high probability objects in pixels with signal-to-background ratio higher than 0.05 (see \cref{APP:threshold} for a more detailed explanation). }
\item When the number of photons per pixel is very high, the binomial sampling step of \cref{EQ:B_scheme} is replaced by using a Poisson approximation, i.e., \\ $\sum_{t=1}^{T}\tilde{z}_{i,j,t,b}\sim \mathcal{P}(\sum_{t=1}^{T}\frac{b_{i,j} z_{i,j,t}}{\sum_{\substack{n : (x_n,y_n)=(i,j)}} r_{n}h(t-t_{n})+b_{i,j}})$.
\end{enumerate}

\subsection{Multiresolution approach} \label{SUBSEC:MR approach}

\begin{algorithm}
	\caption{Multiresolution ManiPoP}
	\label{ALG: MR_approach}
	\begin{algorithmic}[h]
		\STATE \textbf{Input:} Lidar scene $\mat{Z}$, hyperparameters $\Psi$,  window size $N_{binning}$ and number of scales $K$
		\STATE \textbf{Initialization:} 
		\STATE $\mat{\Phi}_1^{(0)} \gets \emptyset $
		\STATE $ \mat{B}_1^{(0)} \sim \prod\mathcal{U}(10^{-2},10^{-3}) $
		\STATE \textbf{Main loop:} 
		\FOR{$k = 1,\dots,K$}
		\IF {$k==1$}
		\STATE $(\mat{\Phi}_k^{(0)},\mat{B}_k^{(0)}) \gets$ upsample$(\hat{\mat{\Phi}}_{k-1},\mat{\hat{B}}_{k-1})$
		\ENDIF
		\STATE $(\mat{\hat{\Phi}}_k,\mat{\hat{B}}_k) \gets$ManiPoP$(\mat{Z}_k,(\mat{\Phi}_k^{(0)},\mat{B}_k^{(0)}),\Psi)$ 
		\ENDFOR
		\STATE \textbf{Output:}  $(\hat{\mat{\Phi}}_K,\mat{\hat{B}}_K)$
	\end{algorithmic}

\end{algorithm}

We downsample the input 3D data by summing the contents over $N_{\textrm{bin}} \times N_{\textrm{bin}}$ windows. This aggregation results in a smaller Lidar image that keeps the same Poisson statistics, where each bin can present an intensity $N_{\textrm{bin}}^2$ bigger (on average). Hence, a Lidar data cube with higher signal-to-noise ratio, approximately $N_{\textrm{bin}}^2$ less points to infer and a similar observational model (if the broadening of the impulse response can be neglected) is obtained. In this way, we run \cref{ALG: ManiPoP} on the downsampled data to get an initial coarse estimate of the 3D scene. This estimate is then upsampled and used as the initial condition for the finer resolution data. The point cloud $\mat{\Phi}$ is upsampled using a linear interpolator for fast computation. \ans{Following the connected-surface structure of ManiPoP, each of the estimated surfaces is upsampled independently of the rest}. However, more elaborate algorithms can be also used, such as moving least squares (MLS), as detailed in \cite{lancaster1981surfaces}. These two steps can be performed in $K$ scales, whereby, for each scale, the Lidar data $\mat{Z}_k$ is obtained by aggregating $\mat{Z}_{k+1}$. \Cref{ALG: MR_approach} summarizes the proposed sequential multiscale approach. 

\section{Experiments}\label{SEC:experiments}
The proposed method was evaluated with synthetic and real Lidar data. In all experiments, we denote the bin length as $\Delta_{b} = \frac{T_{b}c}{2}$, where $c$ is the speed of light in the scene medium and $T_b$ is the bin width used in the TCSPC timing histogram. We also indicate the mean number of photons per pixel as $\bar{\lambda}_{p}$, which is proportional to the per pixel acquisition time. Our method is compared with the classical log-matched filtering solution and two recent algorithms. The first one  is referred to as SPISTA \cite{shin2016computational} and considers an $\ell_1$ regularization to promote sparsity in the recovered peaks. \textcolor{black}{The second algorithm is the method presented in \cite{halimi2017} and is referred to as $\ell_{21}$+TV. It considers an $\ell_{21}$ and total variation regularizations to promote smoothness between points in neighbouring pixels}. The RJ-MCMC algorithm proposed in \cite{hernandez2008multilayered} was not considered in this work as its computational complexity is hardly compatible with large images (for a scene of $N_r=100=N_c=100$ pixels and $T=4500$ bins, the algorithm takes more than a day of computation). The log-matched filtering solution is the depth maximum likelihood estimator when the background is negligible and in the presence of a single peak, i.e., $\hat{t}_{i,j} = \argmax_{t_{i,j}\in[1,T]} \sum_{t=1}^{T} z_{i,j,t} \log [h(t-t_{i,j})] $. The \ans{intensity} estimator can then be obtained as $\hat{r}_{i,j} = \sum_{t=1}^{T} z_{i,j,t}/ (g_{i,j}\sum_{t=1}^{T} h(t)).
$ In order to infer the background levels, we constrain the \ans{intensity} estimate to the support of $h(t)$ leading to
$\tilde{r}_{i,j}  = \sum_{t=\hat{t}_{i,j}-\textrm{attack}}^{\hat{t}_{i,j}+\textrm{decay}} z_{i,j,t} / \left(g_{i,j}\sum_{t=1}^{T} h(t)\right).$
The background components can then be computed using the residual photons as $
\hat{b}_{i,j}  = \sum_{t=1}^{T}z_{i,j,t}  \indicator{h(t-\hat{t}^k_{i,j})=0}{t}/ \left(g_{i,j}\sum_{t=1}^{T}  \indicator{h(t-\hat{t}^k_{i,j})=0}{t}\right)$.
The corrected \ans{intensity} estimate is finally computed as $\hat{r}_{i,j}  = \min\{\tilde{r}_{i,j}-\hat{b}_{i,j},0\}$. 
In the experiments, we used only 2 scales, a coarse one using a binning window of $N_{\textrm{bin}}=3$ pixels and the full resolution. The hyperparameters were adjusted with the following considerations
\begin{itemize}
	\item  The cuboid length $N_{\textrm{bin}}$ should be fixed according to the relative scale between the bin width and the pixel resolution. In our real data experiments, we set $N_{\textrm{bin}}$ to $8\Delta_{p}/\Delta_{b}$.
	\item The minimum distance between two points in the same pixel can be set as $d_{\min}=2N_{\textrm{bin}}+1$, thus verifying the condition $d_{\min}>2N_{\textrm{bin}}$.
	\item  The parameters controlling the number of points and the spatial correlation were set by cross-validation using many Lidar data sets leading to $\gamma_{a}=e^{2}$ and $\lambda_{a}=(N_rN_c)^{1.5}$.
	\item For each scale, we scaled the impulse response $h'(t)=h(t)\frac{\bar{\lambda}_{p}}{5\sum_{t}h(t)}$, where $h(t)$ is the unit gain impulse response, such that all \ans{intensity} values lie approximately in the interval $[0,10]$. The regularization parameters were then fixed to $\sigma^2=0.6^2$ and $\beta=\sigma^2/100$ by cross-validation in order to obtain smooth estimates. 
	\item The hyperparameter controlling the smoothness in the background image $\mat{B}$ was also adjusted by cross-validation yielding $\alpha_{B}=2$. 
\end{itemize}
\Cref{TAB:hyperparameters} summarizes the different hyperparameter values for the coarse and fine scales. All the experiments were performed using $N_{i}=25N_rN_c$ iterations in the coarse scale and finest scale.
\begin{table}
	\centering
\begin{tabular}{|l|l|l|l|l|l|l|l|}
	\hline
	Hyperparameter & $\gamma_{a}$ & $\lambda_{a}$            & $N_\textrm{bin}$                         & $d_{\textrm{min}}$    & $\sigma^2$ & $\beta$        & $\alpha_B$ \\ \hline
	Coarse scale   & $e^2$        & $(N_rN_r/N_{b}^2)^{1.5}$ & $3N_{b}\Delta_{p}/\Delta_{\textrm{bin}}$ & $2N_{\textrm{bin}}+1$ & $0.6^2$    & $\sigma^2/100$ & $2$        \\ \hline
	Fine scale     & $e^3$        & $(N_rN_r)^{1.5}$         & $3\Delta_{p}/\Delta_{\textrm{bin}}$      & $2N_{\textrm{bin}}+1$ & $0.6^2/3$  & $\sigma^2/100$ & $2$        \\ \hline
\end{tabular}
	\caption{Hyperparameters values.}
	\label{TAB:hyperparameters}
\end{table}
\subsection{Error metrics}
Three different error metrics are used to evaluate the performance of the proposed algorithm. We compare the percentage of true detections   $F_{\textrm{true}}(\tau)$ as a function of the distance $\tau$, considering an estimated point as a true detection if there is another point in the ground truth/reference point cloud in the same pixel ($x^{\textrm{true}}_n=x^{\textrm{est}}_{n'}$ and $y^{\textrm{true}}_n=y^{\textrm{est}}_{n'}$) such that $|t^{\textrm{true}}_{n}-t^{\textrm{est}}_{n'}|\le\tau$. We also consider the number of points that were falsely created denoted as $F_{\textrm{false}}(\tau)$ (i.e., the estimated points that cannot be assigned to any true point at a distance of $\tau$). Regarding the \ans{intensity} estimates, we focus on target-wise comparison, by gating the 3D reconstruction between the ranges where a specific target can be found, keeping only the point with biggest \ans{intensity} and assigning zero \ans{intensity} to the empty pixels. We computed the normalized mean squared error of the resulting 2D \ans{intensity} image as 
\begin{equation}
\textrm{NMSE}_{\textrm{target}} = \frac{\sum_{i=1}^{N_r}\sum_{j=1}^{N_r} (r^{\textrm{true}}_{i,j}-\hat{r}_{i,j})^2}{\sum_{i=1}^{N_r} \sum_{j=1}^{N_r} {\left(r^{\textrm{true}}_{i,j} \right)^2}}.
\end{equation}
Finally, we consider the NMSE metric for the background image
\begin{equation}
\textrm{NMSE}_{\mat{B}} = \frac{\sum_{i=1}^{N_r}\sum_{j=1}^{N_r} (b^{\textrm{true}}_{i,j}-\hat{b}_{i,j})^2}{\sum_{i=1}^{N_r}\sum_{j=1}^{N_r} {\left(b^{\textrm{true}}_{i,j} \right)}^2}.
\end{equation}

\subsection{Synthetic data}
\begin{figure}[t!]
	\centering
		\centering
		\includegraphics[width=.5\textwidth]{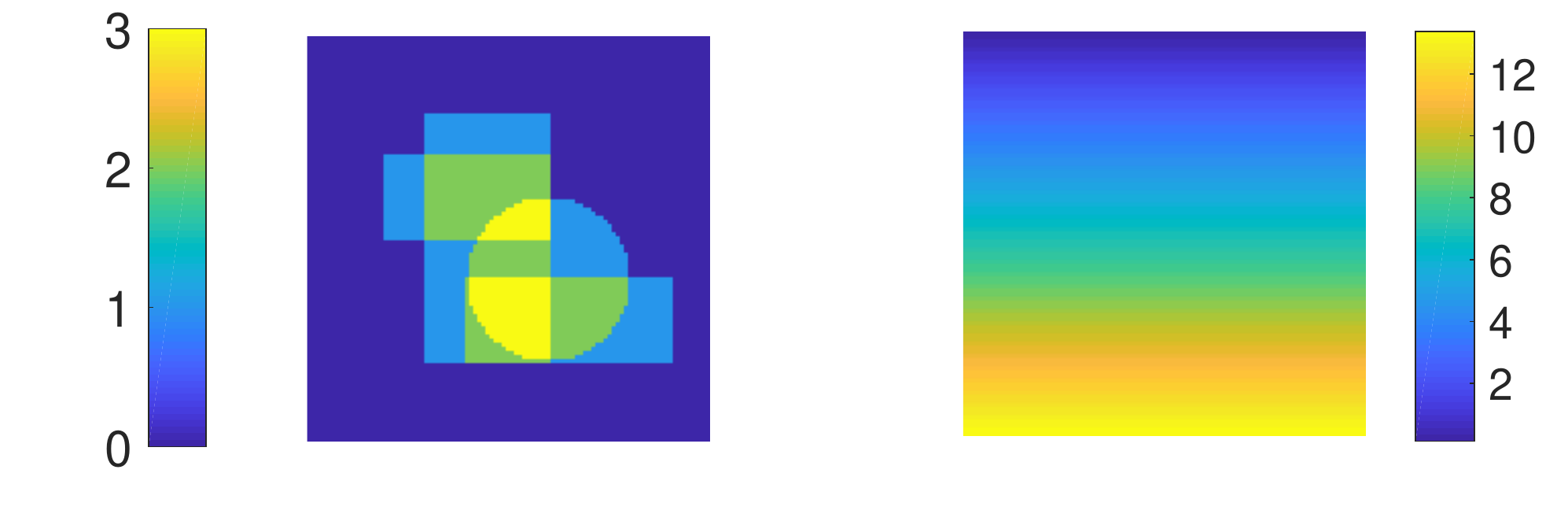}
	\caption{The 3D scene depicted in \cref{FIG:example} consists in 3 plates with different sizes and orientations and one paraboloid shaped object. Left: Number of \ans{objects} per pixel. Right: Mean background photon count $T\mat{B}$.}
	\label{FIG:synthetic}
\end{figure}

\begin{figure}[t!]
	\centering
	\begin{subfigure}[t]{0.55\textwidth}
		\centering
		\includegraphics[width=1\textwidth]{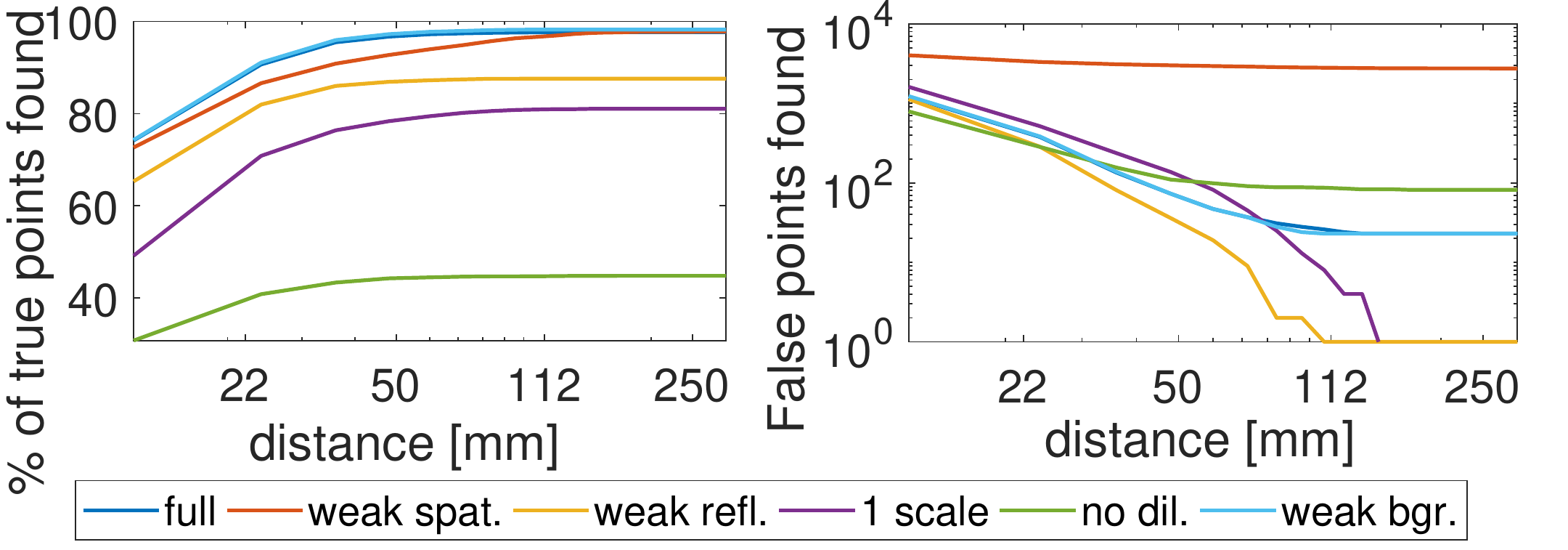}
		\caption{}
	\end{subfigure}
	\begin{subfigure}[t]{0.38\textwidth}
	\includegraphics[width=1\textwidth]{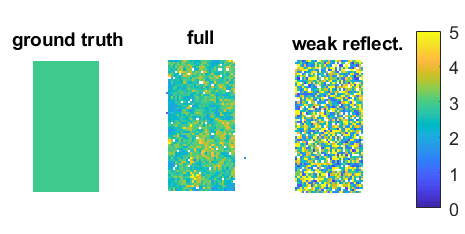}
	\caption{}
	\label{FIG:target_synth}
	\end{subfigure}
	\caption{Left: Percentage of true detections for different priors as a function of maximum distance $\tau$, $F_{true}(\tau)$. Right: Number of false detections, $F_{true}(\tau)$. Ground truth (left), estimates with $\sigma^2=0.6^2$ (center) and $\sigma^2=100^2$ (right).}
	\label{FIG:detections_synthetic}
\end{figure}

\ans{We evaluated the algorithm in two synthetic datasets: A simple one, containing basic geometric shapes and a complex one, based on a scene from the Middlebury dataset \cite{altmann2016lidar}. Both scenes present multiple surfaces per pixel.} The first scene, shown in \cref{FIG:synthetic}, has dimensions $N_r=N_r=99$, $T=4500$, $\Delta_{b}=1.2$ mm and $\Delta_{p} \approx 8.5$ mm. The impulse response used in our experiments was obtained from real Lidar measurements, with $\text{\textrm{attack}}=58$ bins and $\text{decay}=460$ bins. The background was created using a linear intensity profile, as shown in \cref{FIG:synthetic}. The resulting mean intensity per pixel was $\bar{\lambda}_{p} = 11$, meaning that $99.75\%$ of the bins are empty and approximately $4$ photons per pixel are due to 3D objects.
First we evaluated the performance with and without the proposed priors to show their effect on the final estimates. The algorithm was tested in the following conditions 
\begin{enumerate}
	\item With all the priors as reported in \cref{TAB:hyperparameters}.
	\item Without spatial regularization ($\gamma_{a}=1$).
	\item With a weak \ans{intensity} regularization ($\sigma^2=100^2$).
	\item With a softer spatial regularization for the background levels ($\alpha_{B}=1$).
	\item Without erosion and dilation moves.
	\item Only using the finest scale, adjusting the number of iterations to yield the same computing time.
\end{enumerate}
The total execution time for all cases was approximately $120$ seconds. \Cref{FIG:detections_synthetic} shows $F_{\textrm{true}}(\tau)$ and $F_{\textrm{false}}(\tau)$ for all the configurations. The number of false points increases dramatically when the area interaction process is not considered, as the sampler tends to create many points of low \ans{intensity}, mistaking background counts as false \ans{surfaces}.  The background regularization does not affect the detected points significantly, but yields a better estimation of $\mat{B}$, leading to $\textrm{NMSE}=0.107$ for $\alpha_{B}=1$ and $\textrm{NMSE}=0.0912$ for $\alpha_{B}=2$. The number of true points detected without dilation and erosion moves or using only one scale decreases dramatically to $44\%$ and $80\%$ respectively. 
\Cref{FIG:target_synth} compares the estimated intensity of the biggest plate with different values of $\sigma^2$. The NMSE obtained with $\sigma^2=0.6^2$ is 0.058, compared to 0.399 in the absence of correlation (i.e., when $\sigma^2=100^2$). \ans{\Cref{APP:SBR} shows the performance of ManiPoP for different signal-to-background ratios and mean photons per pixels for this specific synthetic scene.}

\begin{figure}[t!]
	\centering
	\includegraphics[width=1\textwidth]{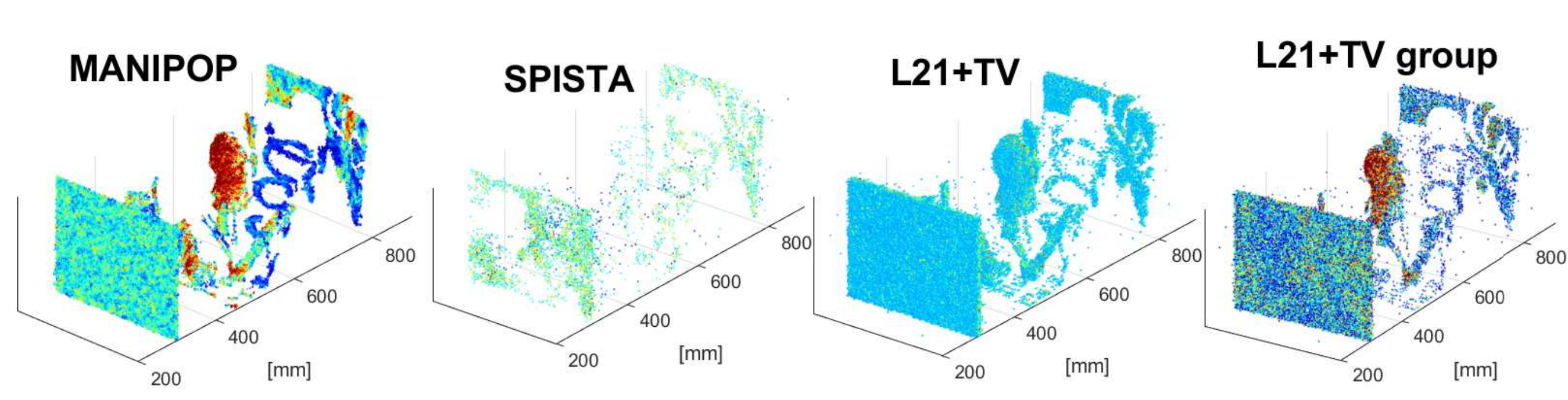}
	\caption{Estimated 3D point cloud by the proposed algorithm, SPISTA and $\ell_{21}$+TV and $\ell_{21}$+TV with grouping.}
	\label{FIG:synthetic_MIT}
\end{figure}

\begin{figure}[t!]
	\centering
	\includegraphics[width=.7\textwidth]{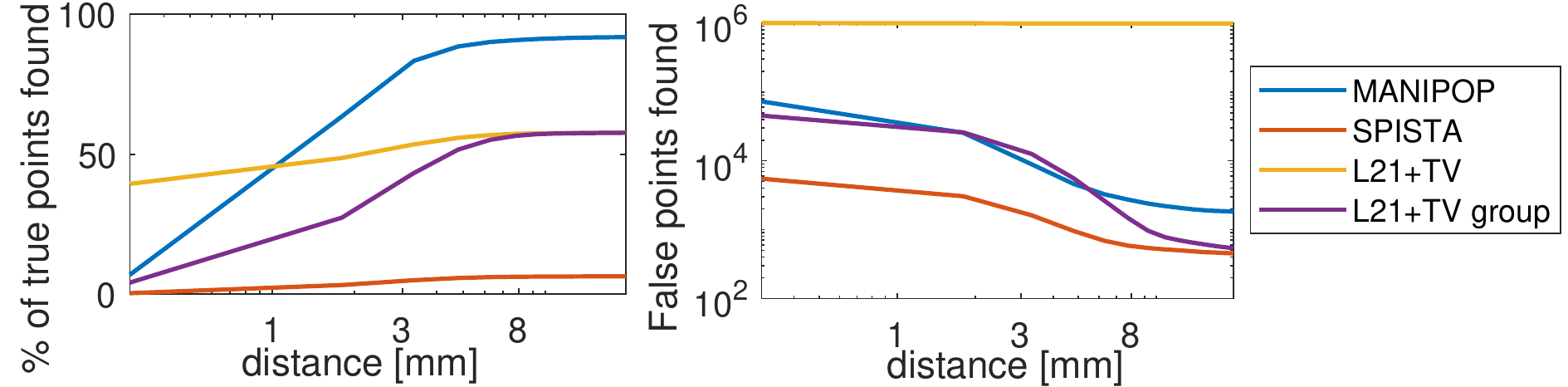}
	\caption{Upper row: Percentage of true detections for different algorithms as a function of maximum distance $\tau$, $F_{\textrm{true}}(\tau)$. Bottom row: Number of false detections, $F_{\textrm{true}}(\tau)$.}
	\label{FIG:points_SPISTA}
\end{figure}

\begin{figure}[!h]
	\centering
	\includegraphics[width=0.9\textwidth]{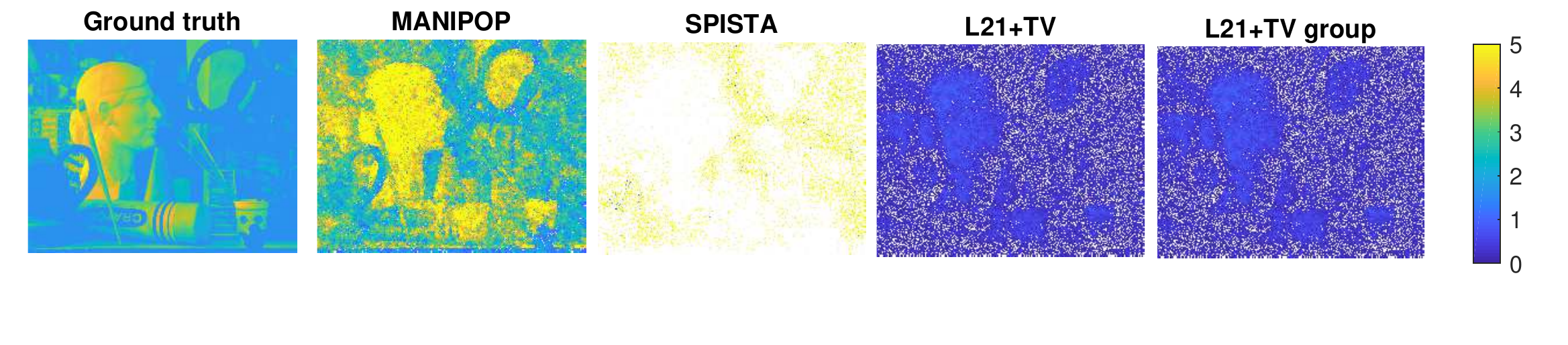}
	\caption{Intensity estimates of the surfaces behind the semi-transparent object. From left to right: Ground truth, proposed method, SPISTA, $\ell_{21}$+TV and $\ell_{21}$+TV with grouping.}
	\label{FIG:target_mit_synth}
\end{figure}

The second dataset was created with the 'Art' scene from \cite{scharstein2007learning}. In order to have multiple surfaces per pixel, we added a semi-transparent plane in front of the scene. We simulated the Lidar measurements, as if they were taken by the system described in \cite{McCarthy:09}. The scene consists in $N_r=183$, $N_c=231$ pixels $T=4500$ histogram bins. The bin width is $\Delta_b=0.3$ mm and the pixel size is $\Delta_p\approx1.2$ mm. In this complex scene, we compared the proposed method with the optimization algorithms SPISTA and $\ell_{21}$+TV. SPISTA relies on the specification of a background level that was set to the true background value. It is important to note that this information is not available in real Lidar applications, as the background levels depend on the imaged scene.  We also show the results for the regularization parameter giving best results, which was found to equal $50$ after many trials (the empirical rule for setting this parameter provided in \cite{shin2016computational} achieved worse results). Similarly, we adjusted the 2 regularization parameters of $\ell_{21}$+TV in order to obtain the best results. The $\ell_{21}$+TV algorithm also relies on a thresholding step on the final estimates, as the output of the optimization method is not sparse. Again, the thresholding constant was adjusted to achieve the best results. To further improve the results of $\ell_{21}$+TV, we included a grouping step (the same procedure as described in \cite{shin2016computational}), which reduces the number of false detections by pairing similar ones in the same pixel. \Cref{FIG:synthetic_MIT} shows the 3D point clouds obtained for each algorithm whereas \cref{FIG:points_SPISTA} shows $F_{\textrm{true}}(\tau)$ and $F_{\textrm{false}}(\tau)$. SPISTA finds 7\% of the true \ans{points} and around 452 false detections. $\ell_{21}$+TV improves the detection rate to 57\%, but also increases the false \ans{detections} to $10^6$. The grouping technique improves the results provided by $\ell_{21}$+TV, reducing the false detections by a factor of 200. The proposed method obtains the best results, finding 92\% of all the true points and 1852 false detections. As shown in \cref{TAB:results_synth}, the proposed algorithm yields the best \ans{intensity} estimates with the lowest execution time. 
\begin{table}
	\centering
	\begin{tabular}{| c | c | c | }
		\hline
		Method 	& Total time [seconds] & NMSE intensity\\ \hline 
		SPISTA \cite{shin2016computational} & 3147 & $>1$ \\\hline
		$\ell_{21}$+TV \cite{halimi2017}	& 2453  & $0.75$ \\ \hline
		$\ell_{21}$+TV group &  2455 & $0.75$ \\ \hline
		ManiPoP  & \textbf{630} & \textbf{0.29} \\ \hline
	\end{tabular}
	\caption{Performance of the proposed method, SPISTA and $\ell_{21}$+TV on the synthetic data.}
	\label{TAB:results_synth}
\end{table}
\Cref{FIG:target_mit_synth} shows the \ans{intensity} estimate of the last vertical plate for each algorithm.
\subsection{Real Lidar data}
We assessed the proposed algorithm using two different Lidar datasets: The multi-layered scene provided in \cite{shin2016computational,mit_data} recorded in the Massachusetts Institute of Technology and the polystyrene target imaged in Heriot-Watt University \cite{altmann2016lidar}. 
\subsubsection{Mannequin behind a scattering object}
\begin{figure}[t!]
	\centering
	\includegraphics[width=0.8\textwidth]{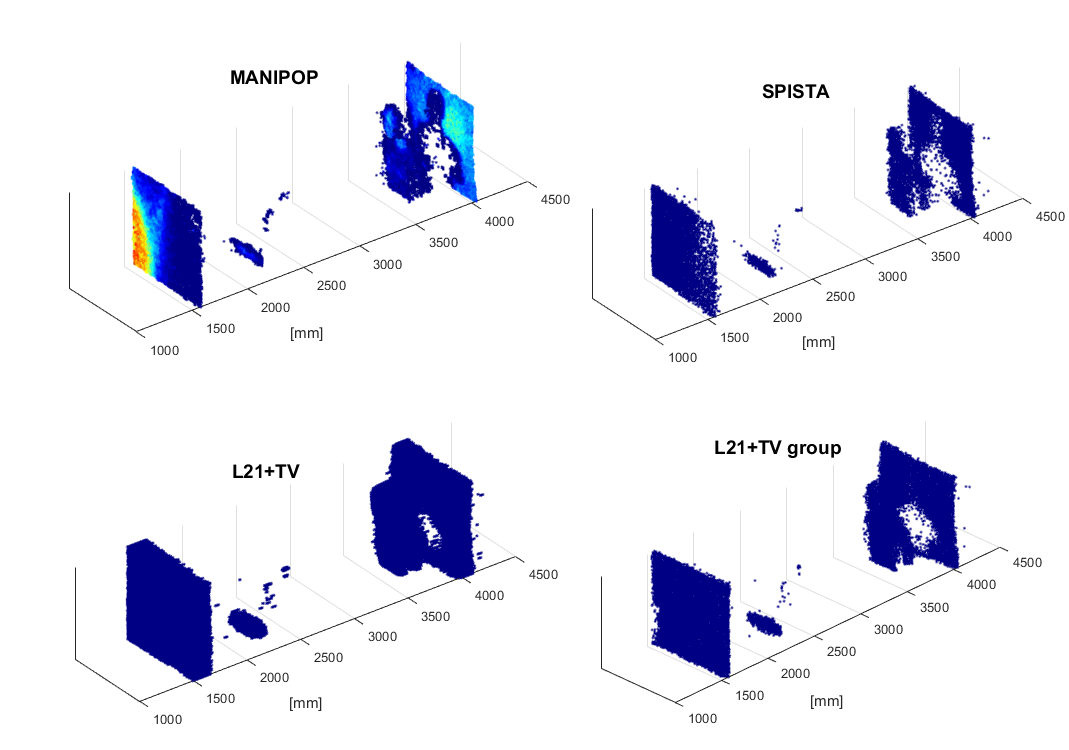}
	\caption{Estimated 3D point cloud by ManiPoP, SPISTA, $\ell_{21}$+TV and $\ell_{21}$+TV with grouping.}
	\label{FIG:data_MIT}
\end{figure}

\begin{figure}[t!]
	\centering
	\includegraphics[width=0.8\textwidth]{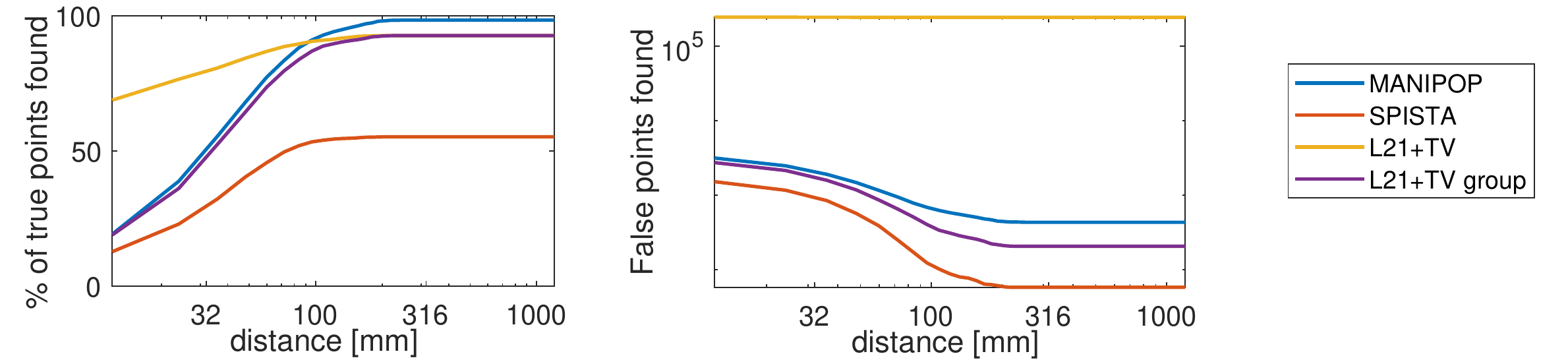}
	\caption{Percentage of true detections at a maximum distance $\tau$, $F_{\textrm{true}}(\tau)$, for ManiPoP, SPISTA, $\ell_{21}$+TV and $\ell_{21}$+TV with grouping. The number of false detections, $F_{\textrm{false}}(\tau)$, is shown in (b).}
	\label{FIG:points_mannequin}
\end{figure}

\begin{figure}[!h]
	\centering
	\includegraphics[width=1\textwidth]{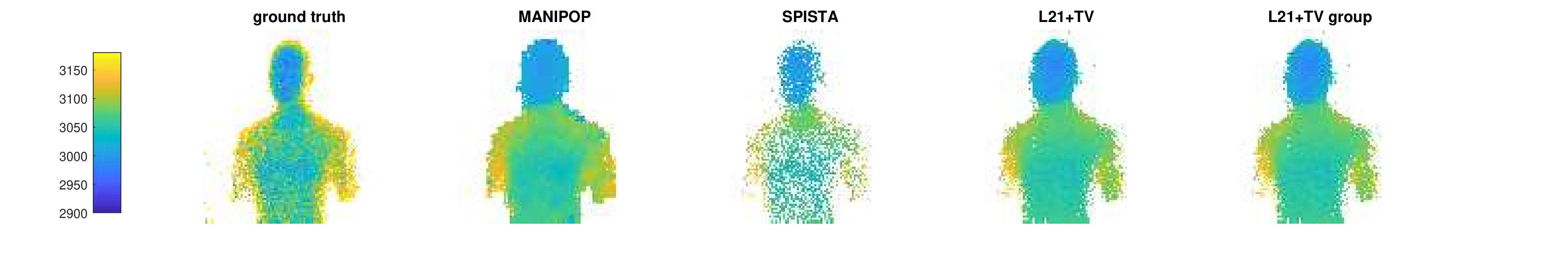}
	\caption{Depth estimates of the mannequin. From left to right: Long acquisition reference, ManiPoP, SPISTA, $\ell_{21}$+TV and $\ell_{21}$+TV with grouping estimates.} 
	\label{FIG:mannequin_depth_mit}
\end{figure}

\begin{figure}[!h]
	\centering
	\includegraphics[width=.85\textwidth]{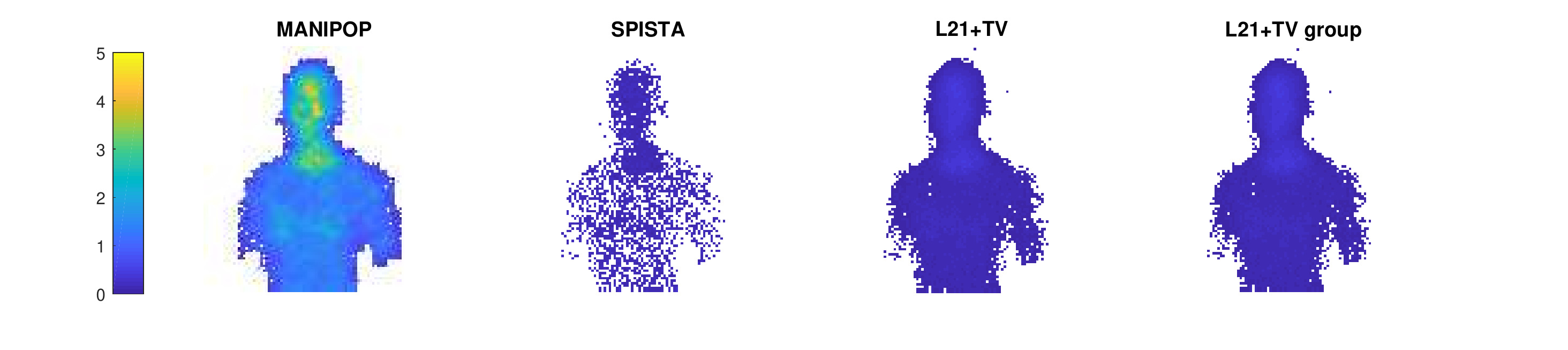}
	\caption{From left to right: Mannequin intensity estimates by Manipop, SPISTA, $\ell_{21}$+TV and $\ell_{21}$+TV with grouping.} 
	\label{FIG:target_mit}
\end{figure}

The first scene consists of a mannequin located $4$ meters behind a partially scattering object, with $N_r=N_c=100$ pixels and $T=4000$ bins. This Lidar scene is publicly available online \cite{mit_data}. The mean photon-count per pixel is $\bar{\lambda}_{p}=45$ and the dimensions are $\Delta_{p}\approx8.4mm$ and $\Delta_{b}=1.2 mm$. In \cite{shin2016computational}, a Gaussian shaped impulse response is suggested. However, we used a data-retrieved impulse response that yields better results (see \cref{APP: impulse_mit} for a detailed explanation). \Cref{FIG:data_MIT} shows the reconstructed point clouds for each algorithm. ManiPoP achieves a sparse and smooth solution, whereas the estimate of SPISTA presents more random scattering of points. The $\ell_{21}$+TV output presents more spatial structure than SPISTA, but also fails to find the the border of the mannequin.
The dataset contains a reference depth of the mannequin obtained using a long acquisition time. This reference was computed using the log-matched filtering solution of a cropped Lidar cuboid where only the mannequin is present. \textcolor{black}{\Cref{FIG:mannequin_depth_mit} shows the ground truth depth and the estimates obtained by ManiPoP, SPISTA and $\ell_{21}$+TV. The proposed method outperforms the SPISTA and $\ell_{21}$+TV outputs, finding 97.9\% of the reference detections, whereas SPISTA only detects 55.3\% of these detections and $\ell_{21}$+TV finds 92.7\%, as shown in \cref{FIG:points_mannequin}. The SPISTA and $\ell_{21}$+TV with grouping algorithms detect 218 and 206 false points respectively, compared to the 341 points found by ManiPoP. This increase in false detections can be attributed to the scattering object that was (probably) removed when the reference dataset was obtained.} The scattering effect can be also seen in \cref{FIG:data_MIT}, as it is possible to find \ans{some parts of the low intensity surface} behind the mannequin. Despite not having a reference for reflectivity values of the target, we can say that the proposed method attains significantly better visual results, as shown in  \cref{FIG:target_mit}. The total execution time of ManiPoP (146 seconds) was around $5$ times less than SPISTA (712 seconds) and slightly shorter than $\ell_{21}$+TV (202 seconds). 


\subsubsection{Polystyrene head}
\begin{figure}[!h]
	\centering
	\begin{subfigure}[t]{0.2\textwidth}
		\centering
		\includegraphics[width=1\textwidth]{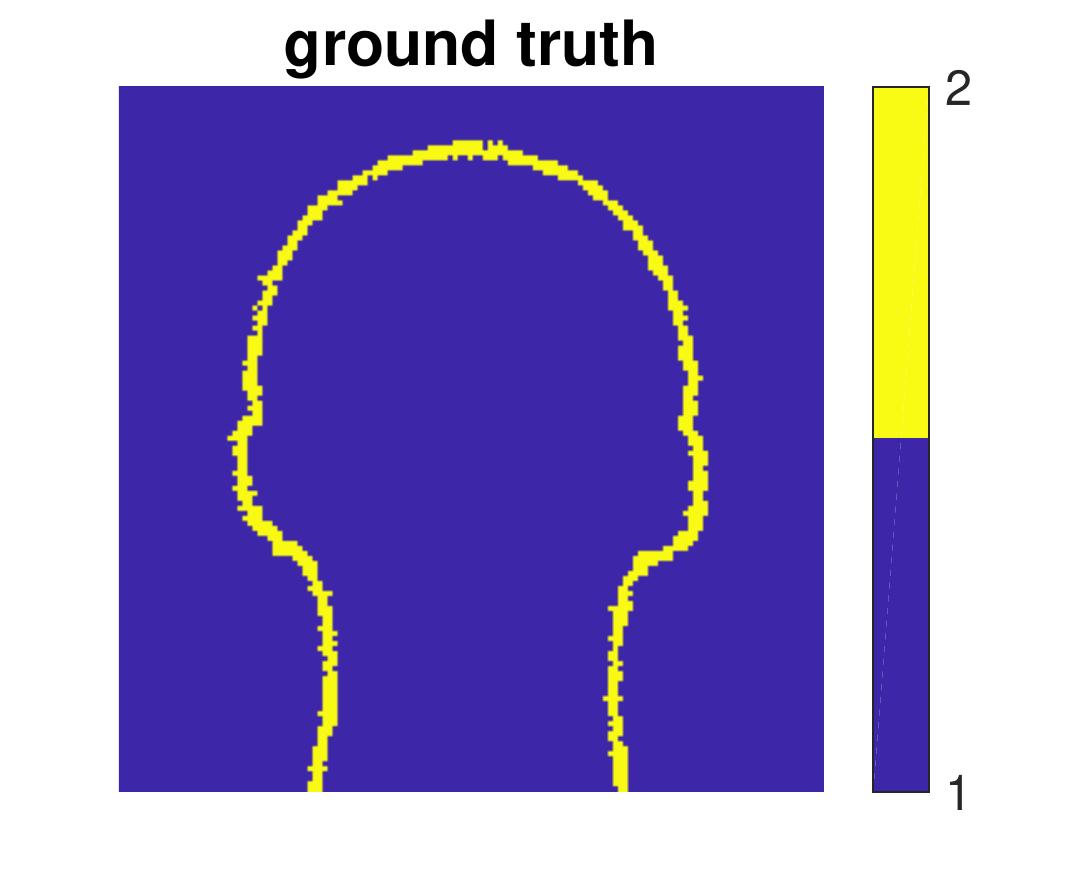}
		\caption{}
	\end{subfigure}
	\begin{subfigure}[t]{0.6\textwidth}
		\centering
		\includegraphics[width=1\textwidth]{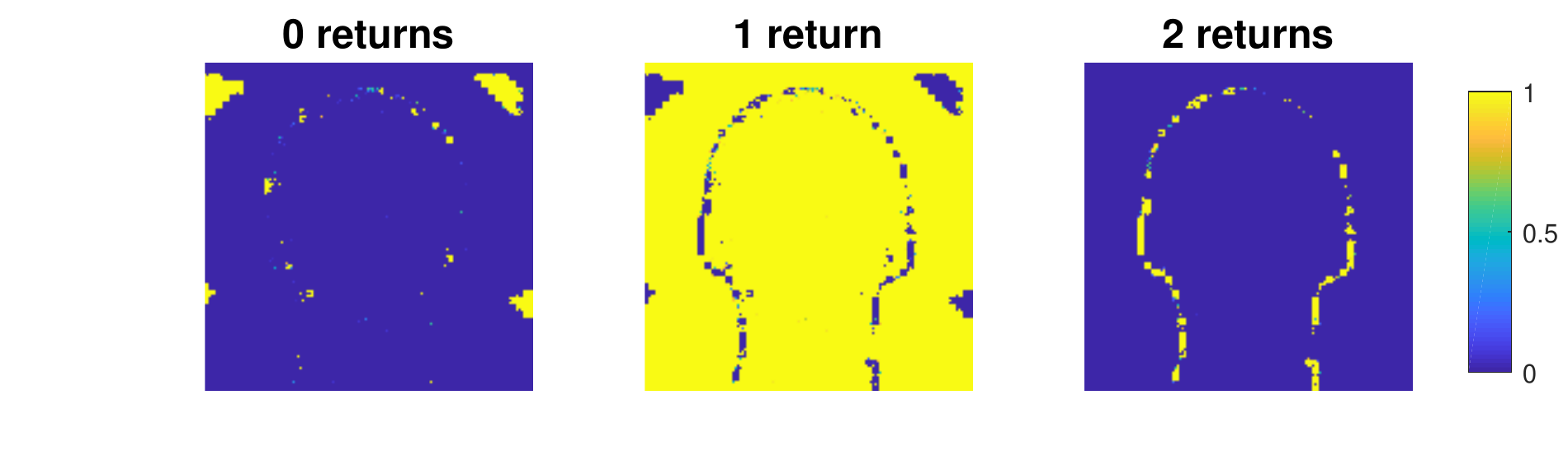}
		\caption{}
	\end{subfigure}
	\caption{ \ans{(a) shows the true number of surfaces per pixel.} The figures in (b) show the probability of having $k=0,1,2$ \ans{objects} per pixel for an acquisition time of $1$ ms.} 
	\label{FIG:returns_head}
\end{figure}

\begin{figure}[t!]
	\centering
	\includegraphics[width=1\textwidth]{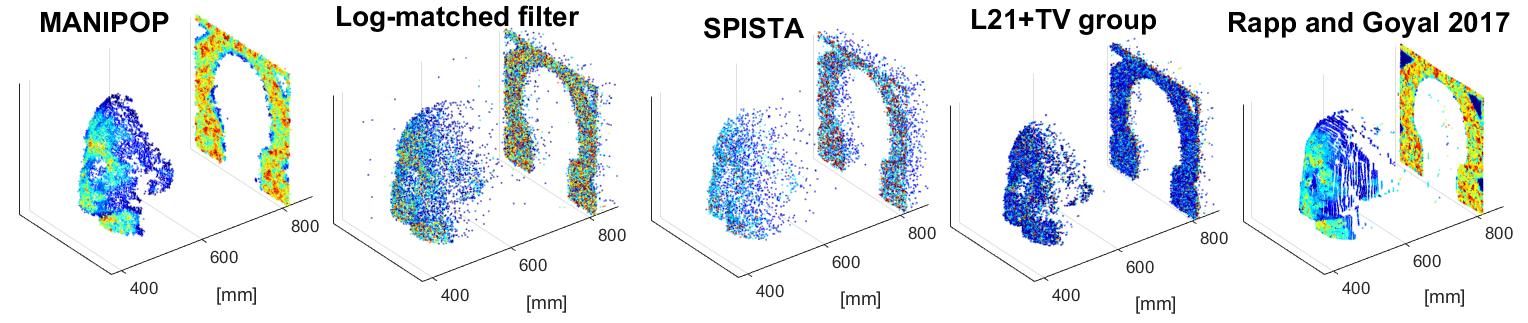}
	\caption{Estimated 3D point clouds using the polystyrene head dataset with an acquisition time of 1 ms.}
	\label{FIG:data_head}
\end{figure}

\begin{figure}[t!]
	\centering
	\includegraphics[width=.8\textwidth]{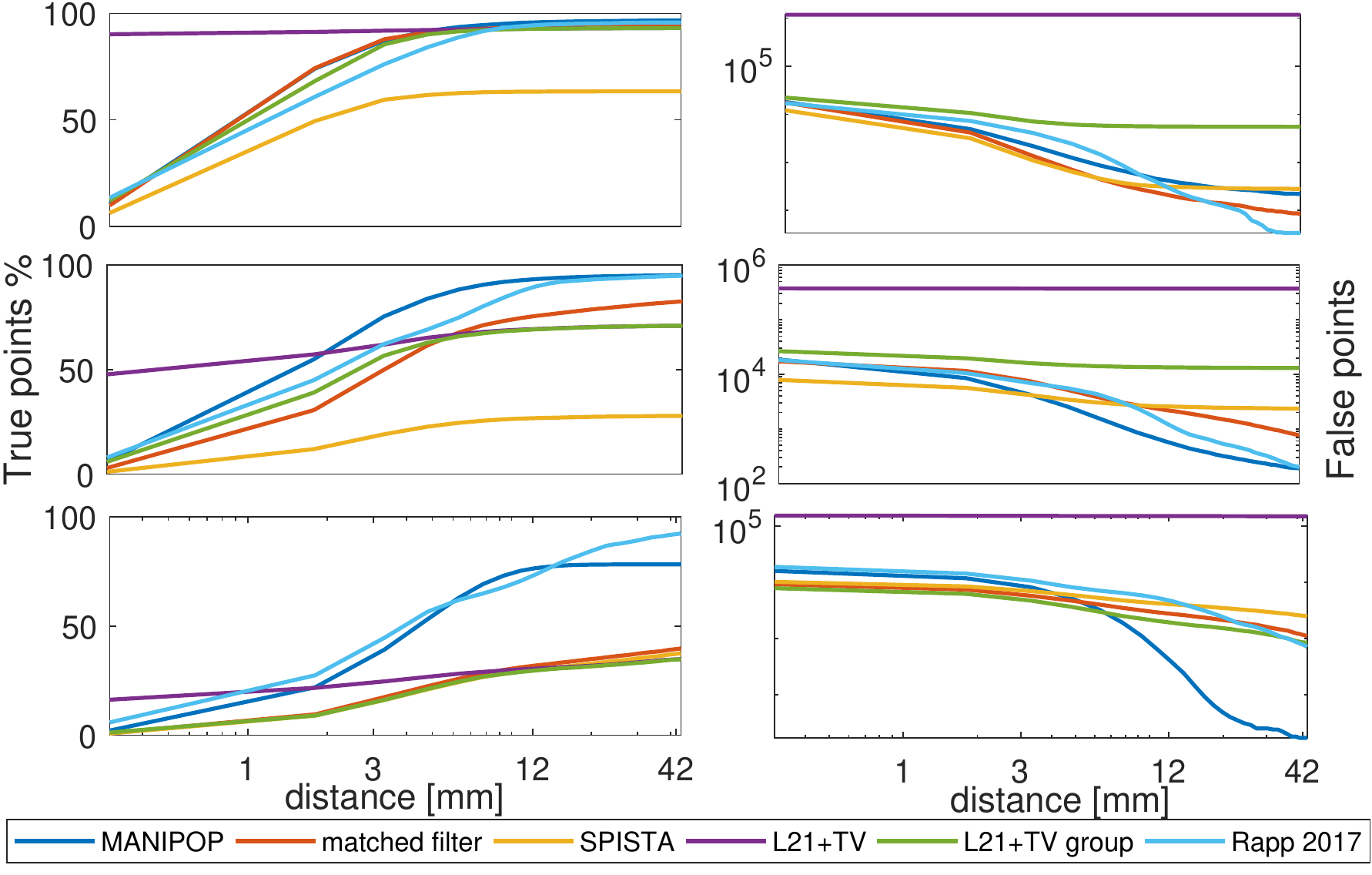}
	\caption{$F_{\textrm{true}}(\tau)$ and  $F_{\textrm{false}}(\tau)$ for the polystyrene head using acquisition times of 10 ms (top), 1 ms (middle) and 0.2 ms (bottom).}
	\label{FIG:points_head}
\end{figure}

The second dataset was obtained in Heriot-Watt University and consists of a life-sized polystyrene head at 40 meters from the imaging device (an image can be found in \cite{altmann2016lidar}). The data cuboid has size $N_r=N_c=141$ pixels and $T=4613$ bins. The physical dimensions are $\Delta_{p}\approx2.1$ mm and $\Delta_{\textrm{bin}}=0.3$ mm. A total acquisition time of $100$ milliseconds was used for each pixel, yielding $\lambda_{p}=337$ with approximately $23$ background photons per pixel. The scene consists mainly in one \ans{object} per pixel, only with $2$ \ans{surfaces per pixel} around the borders of the head. We compare the proposed method with the log-matched filtering solution and the SPISTA algorithm for different acquisition times, i.e., many values of $\lambda_{p}$. As no ground truth is available, we used as reference \ans{the log-matched filter solution, manually dividing the Lidar cube into segments with only one surface, using the largest acquisition time} ($100$ milliseconds). Although the dataset seems to have only one \ans{active depth} per pixel, \ans{two surfaces per pixel} can be found in the borders of the head, as shown in \cref{FIG:returns_head}. \ans{As only a few pixels contain two surfaces, we also compared with \cite{rappfew}, which is a state-of-the-art 3D reconstruction algorithm under a single-surface per pixel assumption.}
\Cref{FIG:data_head} shows the reconstructed 3D point clouds for an acquisition time of $1$ ms whereas \cref{FIG:points_head} shows $F_{\textrm{true}}(\tau)$ and $F_{\textrm{false}}(\tau)$  for acquisition times of 10, 1 \ans{and 0.2 ms}. In the 10 and 1 ms cases, ManiPoP outperforms the other methods, finding almost all true points and providing relatively few false estimates. \ans{The log-matched filter solution (of the complete Lidar cube)} shows a significant error in depth estimates and fails to find $10\%$ of true points, as it is only capable of finding one \ans{object} per pixel. We noticed that SPISTA provides large errors in the \ans{intensity} estimates, as the gradient of the Poisson likelihood is not Lipschitz continuous and the gradient step iterations may diverge in very low-photon scenarios \cite{combettes2011proximal}. \ans{In the 0.2 ms case, there are only $\lambda_{p}=0.7$ photons per pixel on average. Thus, the best performing algorithm is \cite{rappfew}, as the single-surface assumption plays a fundamental role to inpaint the missing depth information. ManiPoP performs in second place, finding $14\%$ less true points than \cite{rappfew}.}

The fastest algorithm is the log-matched filtering solution with less than $20$ seconds in all cases. However, ManiPoP still requires less computing time than SPISTA and $\ell_{21}$+TV. It is worth noticing that the $\ell_{21}$+TV algorithm has a memory requirement proportional to 6 times the whole data cube due to the ADMM algorithm, which can be prohibitively large when the Lidar cube is relatively big. The sparse nature of the ManiPoP algorithm only requires an amount of memory proportional to the number of bins with one photon or more plus the number of 3D points to infer. 

\begin{table}
	\centering
	\begin{tabular}{| c | c | c | c | c |}
		\hline
		& 100 ms & 10 ms  & 1 ms & \ans{0.2 ms}\\ 
		Algo./Acq. time		& \ans{($\lambda_{p}=337$)} & \ans{($\lambda_{p}=33.7$)}  &  \ans{($\lambda_{p}=3.4$)} 
		& \ans{($\lambda_{p}=0.7$)} \\  \hline 
		SPISTA \cite{shin2016computational} & 2029 & 1813 & 1939 & \ans{1809} \\ \hline
		$\ell_{21}$+TV \cite{halimi2017} & 792  & 697 & 704 & \ans{535} \\ \hline
		$\ell_{21}$+TV group & 793 & 697  & 705 & \ans{535.4}\\ \hline
		ManiPoP    & 322 & 229  & 201 & \ans{173.4} \\ \hline
		Log-matched filter &  \textbf{18} &  \textbf{11} & \textbf{7.8} & \textbf{5.6} \\ \hline
		\ans{Rapp and Goyal 2017 \cite{rappfew}} &  \ans{196.87} &  \ans{40} & \ans{37} & \ans{38.4} \\ \hline
	\end{tabular}
	\caption{Computing time of the proposed method, SPISTA, $\ell_{21}$+TV, log-matched filtering and \cite{rappfew} on the polystyrene head dataset.}
	\label{TAB:results_head}
\end{table}

\begin{figure}[t!]
	\centering
	\begin{subfigure}[t]{.35\textwidth}
		\centering
		\includegraphics[width=1\textwidth]{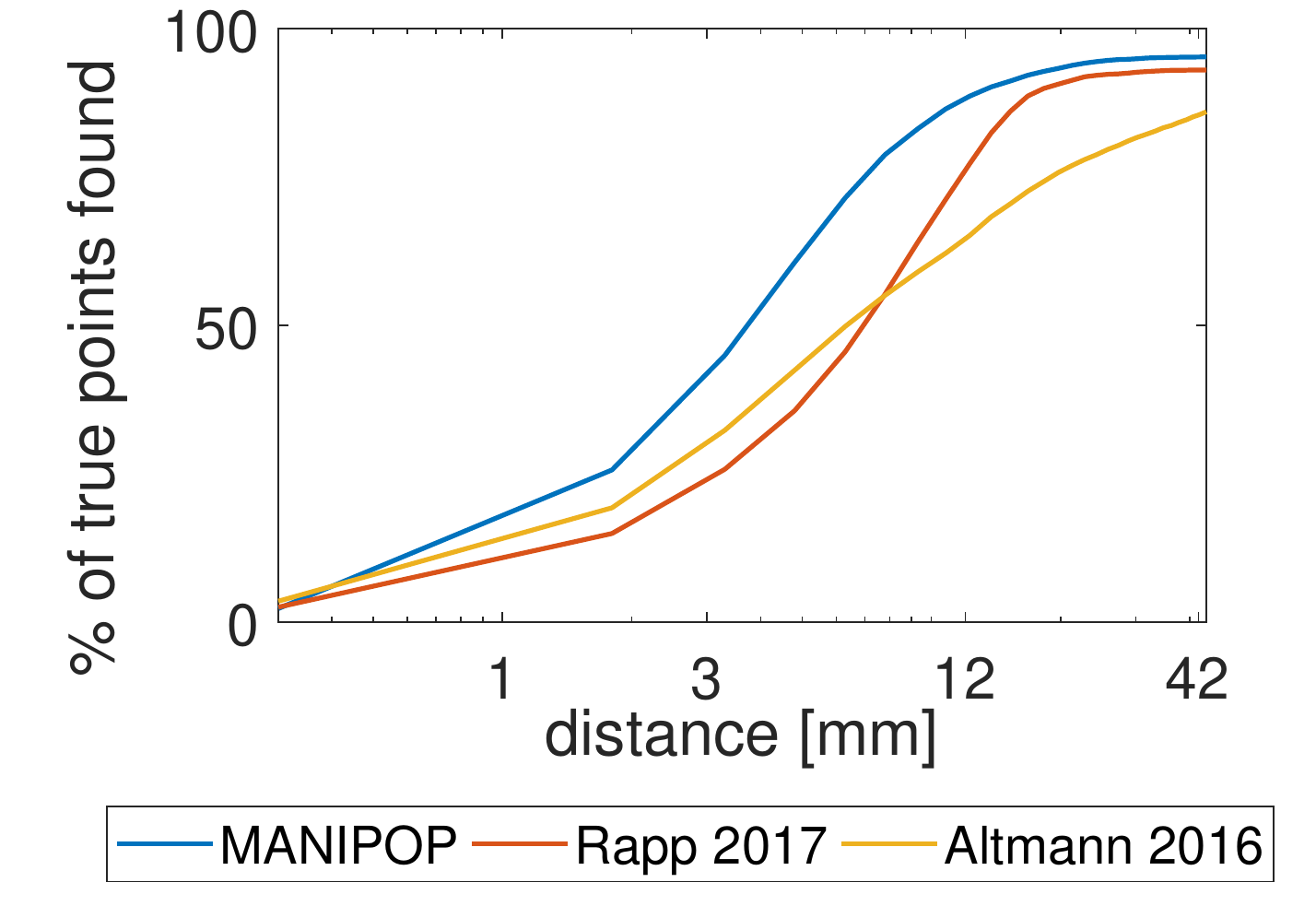}
		\caption{}
	\end{subfigure}%
	\begin{subfigure}[t]{.65\textwidth}
		\centering
		\includegraphics[width=1\textwidth]{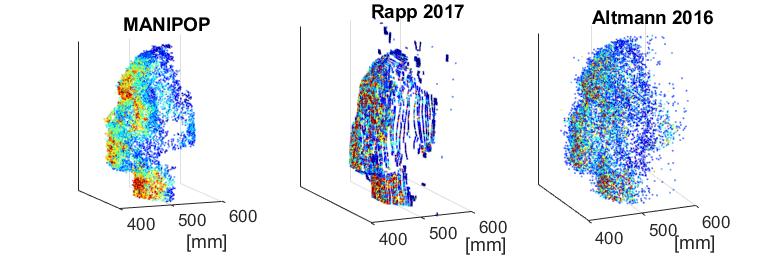}
		\caption{}
		\label{FIG:3Dpoints_target_head}
	\end{subfigure}
	\caption{\ans{(a) $F_{\textrm{true}}(\tau)$ and  $F_{\textrm{false}}(\tau)$ for the polystyrene head without backplane using an acquisition time of 1 ms. The 3D reconstructions are shown in (b).}}
	\label{FIG:points_target_head}
\end{figure}

\ans{To further demonstrate the generality of the proposed method, we studied the case where only one surface is present per pixel, but not all the pixels contain surfaces, which occurs in most outdoor measurements. If a single-surface per pixel algorithm is used, such as \cite{kirmani2014first,altmann2016lidar,shin2015photon,halimi2016restoration,rappfew}, a non-trivial post-processing step is necessary to discriminate which pixels have active depths. We also included the results obtained by the Bayesian target detection algorithm \cite{altmanndetect2016}, which assumes at most one surface per pixel. To recreate this case using the polystyrene head dataset, we removed the backplane from the 1ms dataset, obtaining a new 3D Lidar cube that only contains the polystyrene head. \cref{FIG:points_target_head} shows the results obtained using ManiPoP, \cite{altmanndetect2016} and \cite{rappfew}. In the latter, we applied a global thresholding based on the recovered reflectivity values, such that only the target would be present in the final results. The value of the threshold was manually chosen to obtain the best results. ManiPoP obtains the best results, finding 95.2\% of the points with only 21 false detections, whereas \cite{rappfew} finds 95.3\% of the points and 542 false detections and \cite{altmanndetect2016} obtains 94.5\% of the points and 4 false detections. As shown in \cref{FIG:3Dpoints_target_head}, the estimates of \cite{rappfew} degrade significantly towards the borders of the target, as the single-surface assumption imposes a false correlation with the background photons in neighbouring pixels where no surface is present. While \cite{altmanndetect2016} performs similarly in terms of true and false point detections than ManiPoP, the depth and reflectivity estimates are worse. This result can be attributed to the lack of prior spatial correlation for the depth and reflectivity values in \cite{altmanndetect2016}}.

\begin{figure}[!t]
	\centering
	\includegraphics[width=0.85\textwidth]{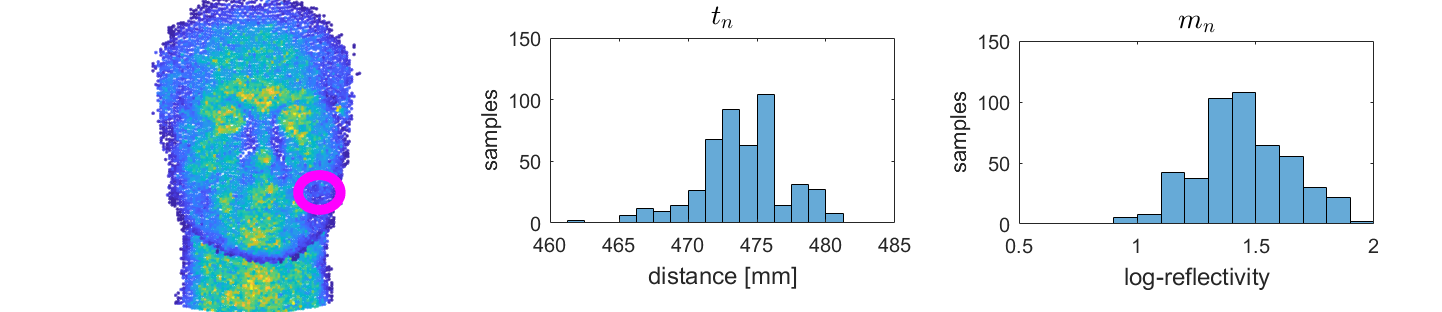}
	\caption{The center and right plots show the position and log-intensity histograms for the point encircled in violet in the left plot, using an acquisition time of 1 ms.}
	\label{FIG:histograms}
\end{figure}

Note that the samples generated by the proposed RJ-MCMC method are asymptotically distributed according to the posterior \cref{EQ:posterior} and can thus be used to compute various uncertainty measures. For instance,  \cref{FIG:returns_head} shows the probability of having $k={0,1,2}$ peaks for  an acquisition time of $1$ ms, computed according to \cref{EQ:prob_k_returns}. Another example is displayed in  \cref{FIG:histograms}, which shows the position and log-intensity histograms that were computed using the samples from additional $N_i=400N_rN_c$ iterations in a fixed dimension (only allowing mark and shift moves).

\subsubsection{Human behind camouflage}
\begin{figure}[!t]
	\centering
	\begin{subfigure}[t]{0.6\textwidth}
		\centering
		\includegraphics[width=1\textwidth]{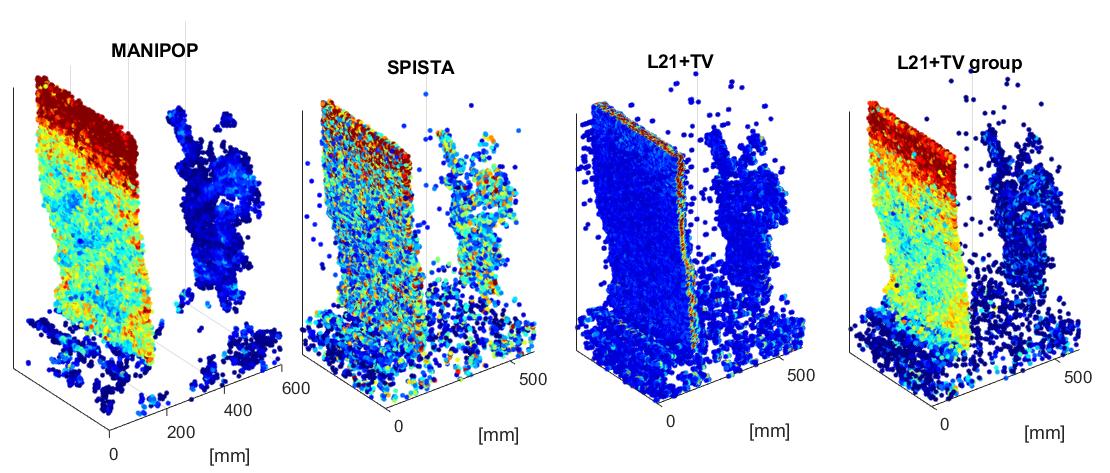}
		\caption{}
	\end{subfigure}
	\begin{subfigure}[t]{0.6\textwidth}
		\centering
		\includegraphics[width=1\textwidth]{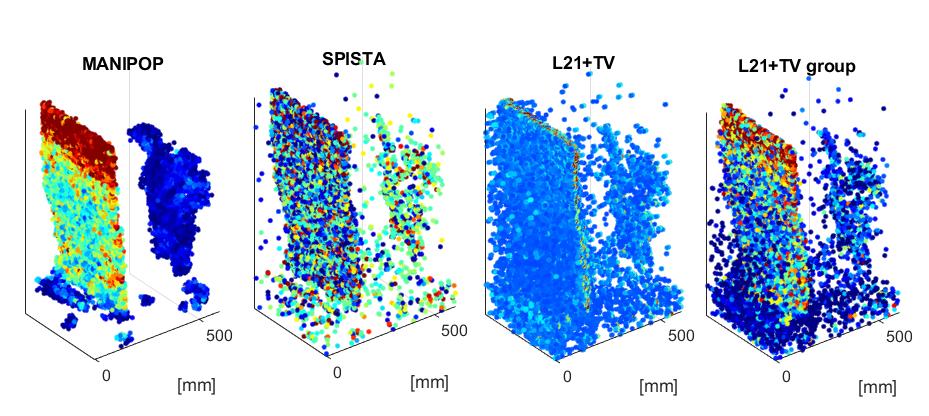}
		\caption{}
	\end{subfigure}
	\caption{Estimated 3D point clouds using the camouflage dataset for per-pixel acquisition times of 3.2 ms (a) and 0.32 ms (b).}
	\label{FIG:camouflage}
\end{figure}

\ans{The last dataset consists of a man standing behind camouflage at a stand-off distance of $230$ meters from the Lidar system. An in-depth description of the scene can be found in \cite{tobin2017long,halimi2017}. An acquisition time of $3.2$ milliseconds was used for each pixel, obtaining $\lambda_{p}=44.6$ photons per pixel on average, where approximately $13.3$ photons correspond to background levels. The Lidar cube has $N_r=159$ and $N_c=78$ pixels and $T=550$ histogram bins. The physical dimensions are $\Delta_{p}\approx2.1$ mm and $\Delta_{\textrm{bin}}=5.6$ mm. We evaluated the performance of the algorithms for the per-pixel acquisition times of $3.2$ ms and $0.32$ ms.  \cref{FIG:camouflage} shows the reconstructions obtained by ManiPoP, SPISTA, TV+$\ell_{21}$ and TV+$\ell_{21}$ with depth grouping. In both cases, ManiPoP obtains a more structured reconstruction, without spurious detections and more dense reconstructions in the regions where the target is present.}

\section{Conclusions and future work}\label{SEC:Conclusions}
\begin{figure}[H]
	\centering
	\includegraphics[width=.85\textwidth]{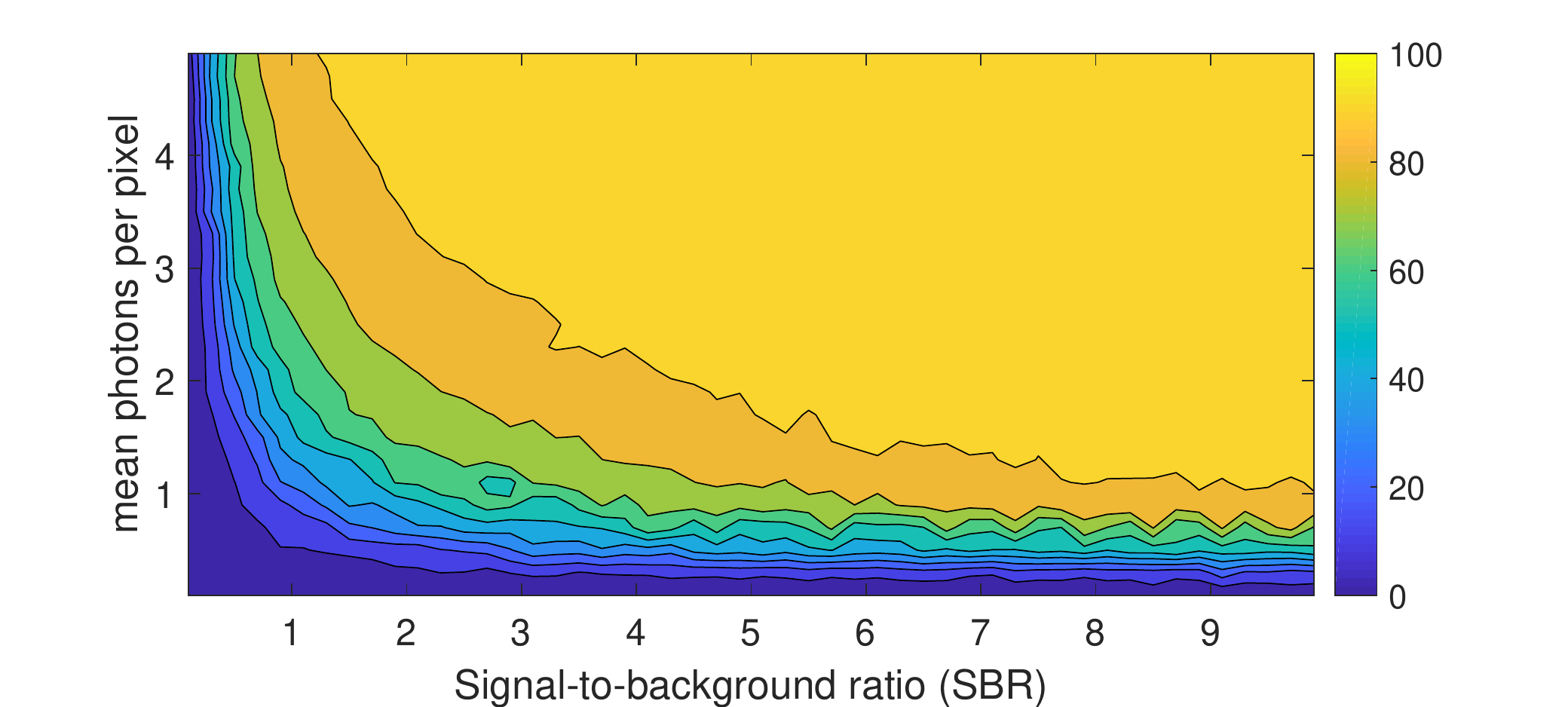}
	\caption{\ans{$F_{\textrm{true}}(\tau)$ for different signal-to-background ratio (SBR) and mean photons per pixel ($\lambda_{p}$).} The results were obtained using the synthetic scene of \cref{FIG:example}.}
	\label{FIG:results_SBR}
\end{figure}

In this paper, we proposed a new Bayesian spatial point process model for describing single-photon depth images. This model promotes spatially correlated and sparse structures, which can be interpreted as a structured $l_0$ pseudo-norm regularization. From a compressive sensing viewpoint, structured sparsity priors can yield the lowest number of necessary measurements to reconstruct a signal \cite{baraniuk2010model}. Finding the MAP estimate of the proposed model is an NP-hard problem \cite{natarajan1995sparse}. We overcame this problem by developing a stochastic RJ-MCMC algorithm with new moves that find a solution relatively fast. In addition, a multiresolution approach improved the estimates and reduced the execution time. The proposed method yielded good 3D reconstructions, with better depth and \ans{intensity} estimates. In our experiments, we noted that for each dataset, a different set of hyperparameters and thresholding values is needed both for SPISTA and $\ell_{21}$+TV, thus making user supervision compulsory, whereas the proposed algorithm uses the same set of hyperparameters across all datasets. \ans{In extremely low-photon cases, i.e., less than one photon per pixel on average, ManiPoP might fail to recover the surface, thus performing worse than other single-surface 3D reconstruction algorithms \cite{rappfew}. As shown in \cref{FIG:results_SBR}, the algorithm performs relatively well for a signal-to-background ratio higher than 1 and more than one photon per pixel. Excluding the aforementioned extremely low-photon or extremely low SBR cases, ManiPoP generalizes other single-surface per pixel and target detection algorithms, as it can provide accurate estimates in scenes with only one surface per pixel and scenes where the target is present in a subset of pixels.}

The algorithm requires less execution time when compared to other optimization \cite{shin2016computational,halimi2017} and RJ-MCMC approaches \cite{hernandez2007bayesian,hernandez2008multilayered}. 
Although there is a significant increase in computational time with respect to the classic log-matched filtering solution, a C++ implementation with efficient handling of the connected-surface structure would reduce the computing time considerably. A profiling analysis of the present code shows that around $70\%$ of the total computational time is due to these computations. In addition, the Markovian structure of the algorithm could be further exploited to perform multiple parallel moves. 

\ans{Scenes containing scattering media may present a broadening of the impulse response. Moreover, surfaces with normals that have a significant angle with respect to the laser beam might also show a broadening of $h(t)$. In such cases, the proposed method might have a reduced performance. Future work will be devoted to estimating the degree of broadening of each point. Moreover, the hard constraint on the minimum distance between two surfaces within a pixel \cref{EQ:strauss} may not apply in some scenes, such as dense foliage or scenes with extremely close objects. In this setting, the hard constraint Strauss process should be modified for a soft constraint one \cite{van2000markov}.}

\ans{Another} important direction of future work is the extension of the proposed model to handle multispectral Lidar (MSL) data, by considering jointly $L>1$ bands and classifying the 3D point cloud according to different materials. Our model is easily extendible to this configuration, as we can add a mark to each point that labels the spectral signature of the \ans{object}. We note that the presented model can be also used for single-photon imaging \cite{kirmani2014first,altmanncamsap}. Finally, we note that ManiPoP can be used as a first processing step to recover denoised 3D point clouds from raw Lidar data to then perform other higher-level computer vision tasks. 

\section*{Acknowledgment}
The authors thank Prof. V. Goyal for providing experimental data used in \cref{SEC:experiments} and the codes of \cite{shin2016computational,rappfew} and Dr A. Halimi for providing codes of \cite{halimi2017}. \ans{We would like to thank Bradley Schilling of the US Army RDECOM CERDEC NVESD and his team for their assistance with the camouflage field trial measurements described in this paper. We also thank Rachael Tobin (Heriot-Watt) and Dr Ken McEwan (DSTL) for their help with the camouflage field trial work.}.

\bibliographystyle{ieeetr}

\begin{thebibliography}{10}
	
	\bibitem{mit_data}
	https://github.com/photon-efficient-imaging/full-waveform/tree/master/fcns.
	
	\bibitem{altmann2017bayesian}
	{\sc Y.~Altmann, R.~Aspden, M.~Padgett, and S.~McLaughlin}, {\em A {B}ayesian
		approach to denoising of single-photon binary images}, IEEE Trans. Comput.
	Imaging, 3 (2017), pp.~460--471.
	
	\bibitem{altmann2017robust}
	{\sc Y.~Altmann, A.~Maccarone, A.~McCarthy, G.~Newstadt, G.~Buller,
		S.~McLaughlin, and A.~Hero}, {\em Robust spectral unmixing of sparse
		multispectral lidar waveforms using gamma {M}arkov random fields}, IEEE
	Trans. Comput. Imaging, 3 (2017), pp.~658--670.
	
	\bibitem{altmanncamsap}
	{\sc Y.~Altmann, S.~McLaughlin, and M.~Padgett}, {\em Unsupervised restoration
		of subsampled images constructed from geometric and binomial data}, in 2017
	IEEE 7th International Workshop on Computational Advances in Multi-Sensor
	Adaptive Processing (CAMSAP), Dec 2017, pp.~1--5,
	\url{https://doi.org/10.1109/CAMSAP.2017.8313187}.
	
	\bibitem{altmann2016lidar}
	{\sc Y.~Altmann, X.~Ren, A.~McCarthy, G.~S. Buller, and S.~McLaughlin}, {\em
		Lidar waveform-based analysis of depth images constructed using sparse
		single-photon data}, IEEE Trans. Image Process., 25 (2016), pp.~1935--1946.
	
	\bibitem{altmanndetect2016}
	{\sc Y.~Altmann, X.~Ren, A.~McCarthy, G.~S. Buller, and S.~McLaughlin}, {\em
		Robust bayesian target detection algorithm for depth imaging from sparse
		single-photon data}, IEEE Transactions on Computational Imaging, 2 (2016),
	pp.~456--467, \url{https://doi.org/10.1109/TCI.2016.2618323}.
	
	\bibitem{altmann2016target}
	{\sc Y.~Altmann, X.~Ren, A.~McCarthy, G.~S. Buller, and S.~McLaughlin}, {\em
		Target detection for depth imaging using sparse single-photon data}, in 2016
	IEEE International Conference on Acoustics, Speech and Signal Processing
	(ICASSP), March 2016, pp.~3256--3260,
	\url{https://doi.org/10.1109/ICASSP.2016.7472279}.
	
	\bibitem{azzari2016variance}
	{\sc L.~Azzari and A.~Foi}, {\em Variance stabilization for noisy+ estimate
		combination in iterative poisson denoising}, IEEE Signal Process. Lett., 23
	(2016), pp.~1086--1090.
	
	\bibitem{baddeley1995area}
	{\sc A.~J. Baddeley and M.~Van~Lieshout}, {\em Area-interaction point
		processes}, Annals of the Institute of Statistical Mathematics, 47 (1995),
	pp.~601--619.
	
	\bibitem{baraniuk2010model}
	{\sc R.~G. Baraniuk, V.~Cevher, M.~F. Duarte, and C.~Hegde}, {\em Model-based
		compressive sensing}, IEEE Trans. Inf. Theory, 56 (2010), pp.~1982--2001.
	
	\bibitem{brooks2011handbook}
	{\sc S.~Brooks, A.~Gelman, G.~Jones, and X.-L. Meng}, {\em Handbook of {M}arkov
		chain {M}onte {C}arlo}, CRC press, 2011.
	
	\bibitem{combettes2011proximal}
	{\sc P.~L. Combettes and J.-C. Pesquet}, {\em Proximal splitting methods in
		signal processing}, in Fixed-point algorithms for inverse problems in science
	and engineering, Springer, 2011, pp.~185--212.
	
	\bibitem{dikmen2010gamma}
	{\sc O.~Dikmen and A.~T. Cemgil}, {\em Gamma {M}arkov random fields for audio
		source modeling}, IEEE Trans. Audio, Speech, Language Process., 18 (2010),
	pp.~589--601.
	
	\bibitem{Ferstl_2013_ICCV}
	{\sc D.~Ferstl, C.~Reinbacher, R.~Ranftl, M.~Ruether, and H.~Bischof}, {\em
		Image guided depth upsampling using anisotropic total generalized variation},
	in The IEEE International Conference on Computer Vision (ICCV), December
	2013.
	
	\bibitem{gao2011research}
	{\sc J.~Gao, J.~Sun, J.~Wei, and Q.~Wang}, {\em Research of underwater target
		detection using a slit streak tube imaging lidar}, in Proc. Academic
	International Symposium on Optoelectronics and Microelectronics Technology
	(AISOMT), Harbin, China, Mar. 2012, pp.~240--243.
	
	\bibitem{green1995reversible}
	{\sc P.~J. Green}, {\em Reversible jump {M}arkov chain {M}onte {C}arlo
		computation and {B}ayesian model determination}, Biometrika, 82 (1995),
	pp.~711--732.
	
	\bibitem{hakala2012full}
	{\sc T.~Hakala, J.~Suomalainen, S.~Kaasalainen, and Y.~Chen}, {\em Full
		waveform hyperspectral lidar for terrestrial laser scanning}, Optics express,
	20 (2012), pp.~7119--7127.
	
	\bibitem{halimi2016restoration}
	{\sc A.~Halimi, Y.~Altmann, A.~McCarthy, X.~Ren, R.~Tobin, G.~S. Buller, and
		S.~McLaughlin}, {\em Restoration of intensity and depth images constructed
		using sparse single-photon data}, in Proc. Signal Processing Conference
	(EUSIPCO), Budapest-Hungary, Sept. 2016, pp.~86--90.
	
	\bibitem{halimiwater}
	{\sc A.~Halimi, A.~Maccarone, A.~McCarthy, S.~McLaughlin, and G.~S. Buller},
	{\em Object depth profile and reflectivity restoration from sparse
		single-photon data acquired in underwater environments}, IEEE Transactions on
	Computational Imaging, 3 (2017), pp.~472--484,
	\url{https://doi.org/10.1109/TCI.2017.2669867}.
	
	\bibitem{halimi2017}
	{\sc A.~Halimi, R.~Tobin, A.~McCarthy, S.~McLaughlin, and G.~S. Buller}, {\em
		Restoration of multilayered single-photon 3d lidar images}, in 2017 25th
	European Signal Processing Conference (EUSIPCO), Aug 2017, pp.~708--712,
	\url{https://doi.org/10.23919/EUSIPCO.2017.8081299}.
	
	\bibitem{hammersley1971markov}
	{\sc J.~M. Hammersley and P.~Clifford}, {\em Markov fields on finite graphs and
		lattices},  (1971).
	
	\bibitem{hartley2003multiple}
	{\sc R.~Hartley and A.~Zisserman}, {\em Multiple view geometry in computer
		vision}, Cambridge university press, 2003.
	
	\bibitem{hernandez2007bayesian}
	{\sc S.~Hernandez-Marin, A.~M. Wallace, and G.~J. Gibson}, {\em Bayesian
		analysis of lidar signals with multiple returns}, IEEE Trans. Pattern Anal.
	Mach. Intell., 29 (2007), pp.~2170--2180.
	
	\bibitem{hernandez2008multilayered}
	{\sc S.~Hernandez-Marin, A.~M. Wallace, and G.~J. Gibson}, {\em Multilayered
		3{D} lidar image construction using spatial models in a {B}ayesian
		framework}, IEEE Trans. Pattern Anal. Mach. Intell., 30 (2008),
	pp.~1028--1040.
	
	\bibitem{kirmani2014first}
	{\sc A.~Kirmani, D.~Venkatraman, D.~Shin, A.~Cola{\c{c}}o, F.~N. Wong, J.~H.
		Shapiro, and V.~K. Goyal}, {\em First-photon imaging}, Science, 343 (2014),
	pp.~58--61.
	
	\bibitem{lancaster1981surfaces}
	{\sc P.~Lancaster and K.~Salkauskas}, {\em Surfaces generated by moving least
		squares methods}, Mathematics of computation, 37 (1981), pp.~141--158.
	
	\bibitem{maccarone2015underwater}
	{\sc A.~Maccarone, A.~McCarthy, X.~Ren, R.~E. Warburton, A.~M. Wallace,
		J.~Moffat, Y.~Petillot, and G.~S. Buller}, {\em Underwater depth imaging
		using time-correlated single-photon counting}, Optics express, 23 (2015),
	pp.~33911--33926.
	
	\bibitem{mallet2009full}
	{\sc C.~Mallet and F.~Bretar}, {\em Full-waveform topographic lidar:
		State-of-the-art}, ISPRS Journal of photogrammetry and remote sensing, 64
	(2009), pp.~1--16.
	
	\bibitem{mallet2010marked}
	{\sc C.~Mallet, F.~Lafarge, M.~Roux, U.~Soergel, F.~Bretar, and C.~Heipke},
	{\em A marked point process for modeling lidar waveforms}, IEEE Trans. Image
	Process., 19 (2010), pp.~3204--3221.
	
	\bibitem{McCarthy:09}
	{\sc A.~McCarthy, R.~J. Collins, N.~J. Krichel, V.~Fern\'{a}ndez, A.~M.
		Wallace, and G.~S. Buller}, {\em Long-range time-of-flight scanning sensor
		based on high-speed time-correlated single-photon counting}, Appl. Opt., 48
	(2009), pp.~6241--6251, \url{https://doi.org/10.1364/AO.48.006241},
	\url{http://ao.osa.org/abstract.cfm?URI=ao-48-32-6241}.
	
	\bibitem{mccool2016robust}
	{\sc P.~McCool, Y.~Altmann, A.~Perperidis, and S.~McLaughlin}, {\em Robust
		{M}arkov random field outlier detection and removal in subsampled images}, in
	Proc. Statistical Signal Processing Workshop (SSP), Palma de Mallorca, Spain,
	June 2016, pp.~1--5.
	
	\bibitem{moller2007modern}
	{\sc J.~M{\o}ller and R.~P. Waagepetersen}, {\em Modern statistics for spatial
		point processes}, Scandinavian Journal of Statistics, 34 (2007),
	pp.~643--684.
	
	\bibitem{murray2012mcmc}
	{\sc I.~Murray, Z.~Ghahramani, and D.~MacKay}, {\em {MCMC} for
		doubly-intractable distributions}, arXiv preprint arXiv:1206.6848,  (2012).
	
	\bibitem{natarajan1995sparse}
	{\sc B.~K. Natarajan}, {\em Sparse approximate solutions to linear systems},
	SIAM journal on computing, 24 (1995), pp.~227--234.
	
	\bibitem{nilsson1996estimation}
	{\sc M.~Nilsson}, {\em Estimation of tree heights and stand volume using an
		airborne lidar system}, Remote Sensing of Environment, 56 (1996), pp.~1--7.
	
	\bibitem{ogawa2006lane}
	{\sc T.~Ogawa and K.~Takagi}, {\em Lane recognition using on-vehicle lidar}, in
	Proc. Intelligent Vehicles Symposium, Tokyo, Japan, June 2006, pp.~540--545.
	
	\bibitem{pereyra2014maximum}
	{\sc M.~Pereyra, N.~Whiteley, C.~Andrieu, and J.-Y. Tourneret}, {\em Maximum
		marginal likelihood estimation of the granularity coefficient of a
		potts-{M}arkov random field within an {MCMC} algorithm}, in Proc. IEEE
	Workshop on Statistical Signal Processing (SSP), Gold Coast, Australia, June
	2014, pp.~121--124.
	
	\bibitem{rappfew}
	{\sc J.~Rapp and V.~K. Goyal}, {\em A few photons among many: Unmixing signal
		and noise for photon-efficient active imaging}, IEEE Transactions on
	Computational Imaging, 3 (2017), pp.~445--459,
	\url{https://doi.org/10.1109/TCI.2017.2706028}.
	
	\bibitem{rue2005gaussian}
	{\sc H.~Rue and L.~Held}, {\em Gaussian Markov random fields: theory and
		applications}, CRC press, 2005.
	
	\bibitem{scharstein2007learning}
	{\sc D.~Scharstein and C.~Pal}, {\em Learning conditional random fields for
		stereo}, in Computer Vision and Pattern Recognition, 2007. CVPR'07. IEEE
	Conference on, IEEE, 2007, pp.~1--8.
	
	\bibitem{schwarz2010lidar}
	{\sc B.~Schwarz}, {\em Lidar: Mapping the world in 3d}, Nature Photonics, 4
	(2010), pp.~429--430.
	
	\bibitem{shin2015photon}
	{\sc D.~Shin, A.~Kirmani, V.~K. Goyal, and J.~H. Shapiro}, {\em
		Photon-efficient computational 3-{D} and reflectivity imaging with
		single-photon detectors}, IEEE Trans. Comput. Imaging, 1 (2015),
	pp.~112--125.
	
	\bibitem{shin2016computational}
	{\sc D.~Shin, F.~Xu, F.~N. Wong, J.~H. Shapiro, and V.~K. Goyal}, {\em
		Computational multi-depth single-photon imaging}, Optics express, 24 (2016),
	pp.~1873--1888.
	
	\bibitem{snyder2012random}
	{\sc D.~L. Snyder and M.~I. Miller}, {\em Random point processes in time and
		space}, Springer Science \& Business Media, 2012.
	
	\bibitem{Tenenbaum2319}
	{\sc J.~B. Tenenbaum, V.~d. Silva, and J.~C. Langford}, {\em A global geometric
		framework for nonlinear dimensionality reduction}, Science, 290 (2000),
	pp.~2319--2323, \url{https://doi.org/10.1126/science.290.5500.2319},
	\url{http://science.sciencemag.org/content/290/5500/2319},
	\url{https://arxiv.org/abs/http://science.sciencemag.org/content/290/5500/2319.full.pdf}.
	
	\bibitem{tobin2017long}
	{\sc R.~Tobin, A.~Halimi, A.~McCarthy, X.~Ren, K.~J. McEwan, S.~McLaughlin, and
		G.~S. Buller}, {\em Long-range depth profiling of camouflaged targets using
		single-photon detection}, Optical Engineering, 57 (2017), p.~031303.
	
	\bibitem{van2000markov}
	{\sc M.~Van~Lieshout}, {\em Markov point processes and their applications},
	World Scientific, 2000.
	
	\bibitem{wallace2014design}
	{\sc A.~M. Wallace, A.~McCarthy, C.~J. Nichol, X.~Ren, S.~Morak,
		D.~Martinez-Ramirez, I.~H. Woodhouse, and G.~S. Buller}, {\em Design and
		evaluation of multispectral lidar for the recovery of arboreal parameters},
	IEEE Transactions on Geoscience and Remote Sensing, 52 (2014),
	pp.~4942--4954.
	
	\bibitem{zhou2009non}
	{\sc M.~Zhou, H.~Chen, L.~Ren, G.~Sapiro, L.~Carin, and J.~W. Paisley}, {\em
		Non-parametric bayesian dictionary learning for sparse image
		representations}, in Proc. Advances in neural information processing systems,
	Vancouver, B.C., Canada, Dec. 2009, pp.~2295--2303.
	
	\bibitem{zhou2012beta}
	{\sc M.~Zhou, L.~Hannah, D.~B. Dunson, and L.~Carin}, {\em Beta-negative
		binomial process and poisson factor analysis.}, in Proc. AISTATS, vol.~22, La
	Palma, Canary Islands, Apr. 2012, pp.~1462--1471.
	
\end{thebibliography}

\section{Point process definitions}\label{APP:point_process_theory}
\ans{Following \cite{van2000markov}, the space of all point configurations can be defined as 
	\begin{equation}
	\Omega = \bigcup_{N_\Phi\in \mathbb{Z}_{+}} \Omega_{N_\Phi} 
	\end{equation}
	where $\Omega_{N_\Phi}$ denotes the space of configurations containing exactly $N_\Phi$ points and $\mathbb{Z}_{+}$ is the set of positive integers. A Poisson point process is the basic building block for more elaborate point processes \cite{van2000markov}. Points $\mat{\Phi}_c$ of a Poisson process with intensity $\lambda(\cdot)$ are independently distributed in $\mathcal{T}$. The number of points found in a Borel subset $B$ of $\mathcal{T}$ is a random variable distributed according to a Poisson distribution with mean $\lambda(B)$. Moreover, the numbers of points in $K$ disjoint Borel subsets $B_1,\dots,B_K$ are mutually independent. The probability measure $\pi(\cdot)$ associated with a Poisson process on a subset $A$ of the configuration space 
	$\Omega$ is
	\begin{equation}\label{EQ:poisson_ref_measure}
	\pi(d\mat{\Phi}_c) = e^{-\lambda(\mathcal{T})}\sum_{N_{\Phi_c}\in \mathbb{Z}_{+}} \indicator{\Omega_{N_\Phi}}{\mat{\Phi}_c} \lambda(d\myvec{c}_1) \dots \lambda(d\myvec{c}_{N_{\Phi_c}})
	\end{equation}
	where $\lambda(\mathcal{T})$ is the expected total number of points, $\mathds{1}_{A}(\cdot)$ is the indicator function defined in $A$.
	Interactions between points can be characterized using a normalized density $f: \Omega \rightarrow \mathbb{R}_+$, defined with respect to the Poisson reference measure $\pi(\cdot)$ such that 
	\begin{equation*}
	\int_{\Omega} f(\mat{\Phi}_c) \pi(d\mat{\Phi}_c) = 1 .
	\end{equation*}
}

\section{Resulting point process when $\gamma_{a}=1$}\label{APP:lambda_ai_equal_1}
If only an area interaction process density is considered with respect to the Poisson reference measure \cref{EQ:poisson_ref_measure} and no spatial correlations are considered, i.e. $\gamma_{a}=1$, the resulting point process is Poisson with density $\lambda_{a}\lambda(\cdot)$. This can be shown by noticing that the resulting density is
\begin{align*}
f(\mat{\Phi})\pi(d\mat{\Phi}) &\propto e^{-\lambda(\mathcal{T})}\sum_{N_{\Phi_c}\in \mathbb{Z}_+} \lambda_{a}^{N_\Phi}  \indicator{\Omega_{N_\Phi}}{\mat{\Phi}_c} \lambda(d\myvec{c}_1) \dots \lambda(d\myvec{c}_{N_{\Phi}}) \\
&\propto e^{-\lambda(\mathcal{T})}\sum_{N_{\Phi_c}\in \mathbb{Z}_+} \indicator{\Omega_{N_\Phi}}{\mat{\Phi}_c} \lambda_{a}\lambda(d\myvec{c}_1) \dots \lambda_{a}\lambda(d\myvec{c}_{N_{\Phi}}) ,
\end{align*}
which is equivalent to a Poisson point process with intensity $\lambda_{a}\lambda(\cdot)$.


\section{Marginal density of a gamma Markov random field}\label{APP: GMRF marginal}
The marginal gamma Markov field joint density, $p(\mat{B}|\alpha_{\mat{B}})$, can be derived by integrating out the auxiliary variables $w_{i,j}$ from the complete joint density $p(U,\mat{B}|\alpha_{\mat{B}})$, that is
\begin{align}
\label{EQ:integrate_out}
p(\mat{B}|\alpha_{\mat{B}}) = \int_{\mathcal{\mat{W}}} p(\mat{W},\mat{B}|\alpha_{B})dU
\end{align}
For notation simplicity we replace the indices $i=1,\dots,N_r$ and $j=1,\dots,N_r$ for a unique linear index $n=1,\dots,N_rN_r$. The density $p(\mat{W},\mat{B}|\alpha_{B})$ can be expressed using Hammersley and Clifford theorem \cite{hammersley1971markov} as
\begin{equation*}
p(\mat{B},\mat{W}|\alpha_{B}) = \frac{1}{Z}\exp(\sum_{n=1}^{N} -(\alpha_{B}+1)\log(w_n) 
+(\alpha_{B}-1)\log(b_n)-\sum_{n'\in \mathcal{M}_B(w_{n})} \frac{4}{\alpha_{B}} \frac{b_n'}{w_n})
\end{equation*}
where $Z$ is an intractable normalizing constant. Then \cref{EQ:integrate_out} can be expressed as
\begin{equation*}
p(\mat{B}|\alpha_{B}) = \int_{\mathbb{R}^N_{+}}p(\mat{B},\mat{W}|\alpha_{B}) d\mat{W} 
\end{equation*}
\begin{align*}
&\propto  \prod_{n=1}^{N_rN_c}  \left( \int_{0}^{\infty} w_n^{-(\alpha_{B}+1)} \prod_{n'\in \mathcal{M}_B(w_{n})}e^{- \frac{4}{\alpha_{B}} \frac{b_n'}{w_{n}}} dw_{n}\right) b_n^{\alpha_{B}-1}.
\end{align*}
Considering that each integral has the following analytical solution
\begin{equation*}
\int_{0}^{\infty} w_n^{-(\alpha_{B}+1)} \prod_{n'\in \mathcal{M}_B(w_{n})}e^{- \frac{4}{\alpha_{B}} \frac{b_n'}{w_n}} dw_n  =  \frac{\Gamma(\alpha_{B})}{\left( \frac{4}{\alpha_{B}}  \sum_{n'\in \mathcal{M}_B(w_{n})} b_n'\right)^{\alpha_{B}}}
\end{equation*}
then
\begin{align}
p(\mat{B}|\alpha_{B}) &\propto \prod_{n=1}^{N_rN_c} b_n^{\alpha_{B}-1} \frac{\Gamma(\alpha_{B})}{\left( \frac{4}{\alpha_{B}}  \sum_{n'\in \mathcal{M}_B(w_{n})} b_n'\right)^{\alpha_{B}}} \\
&\propto \prod_{n=1}^{N_rN_c} \frac {b_n^{\alpha_{B}-1}} { \tilde{b}_{n}^{\alpha_{B}}}
\end{align}
where $\tilde{b}_n = \frac{4}{\alpha_{B}}\sum_{n'\in \mathcal{M}_B(w_{n})} b_n'$ is a low-pass filtered version of $b_n$. 

\section{Sampling the prior distribution}\label{APP:sampling_prior}
\begin{figure}[!h]
	\centering
	\begin{subfigure}{0.5\textwidth}
		\centering
		\includegraphics[width=1\textwidth]{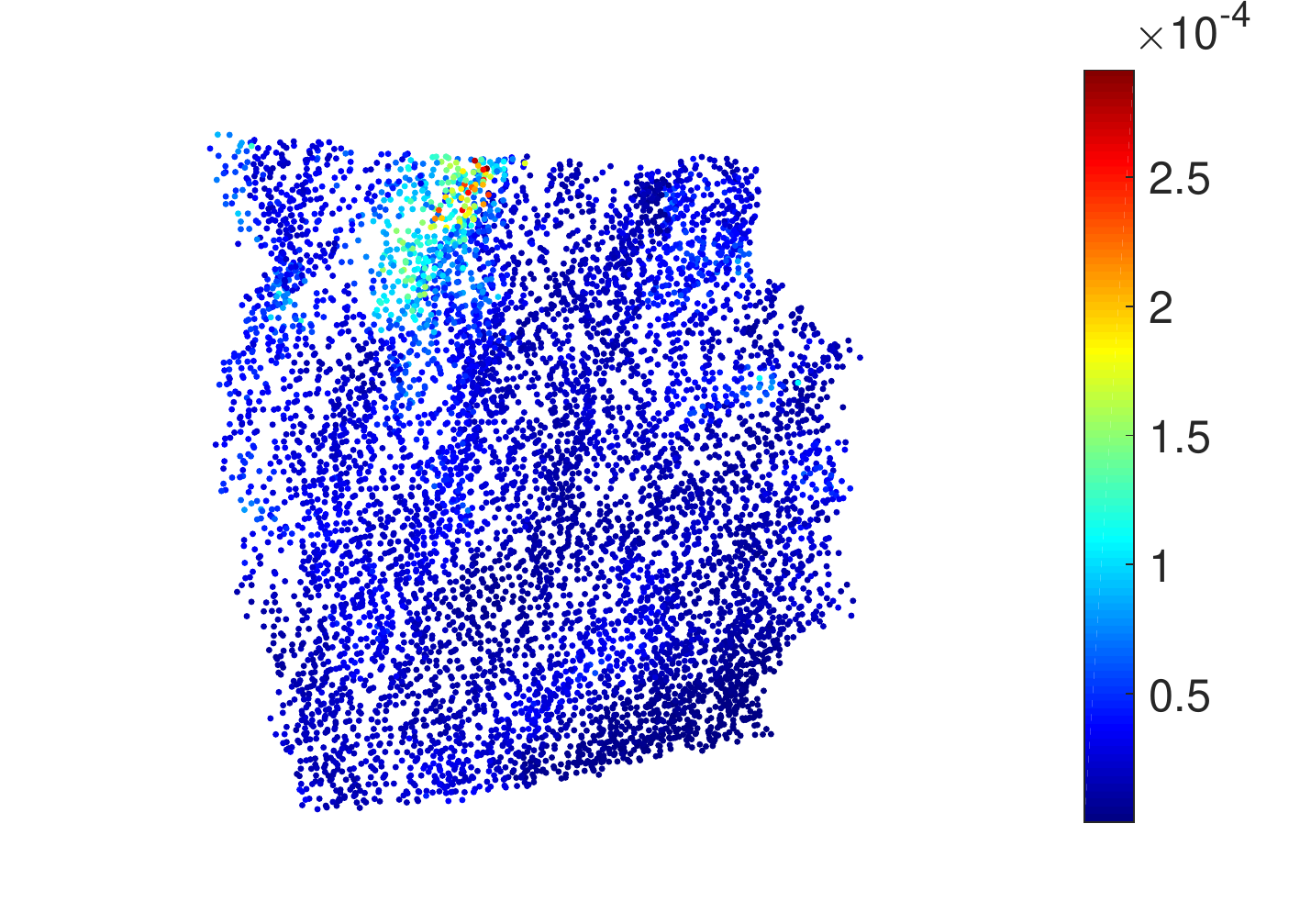}
		\caption{}\label{SUBFIG:prior3D}
	\end{subfigure}%
	\begin{subfigure}{0.5\textwidth}
		\centering
		\includegraphics[width=1\textwidth]{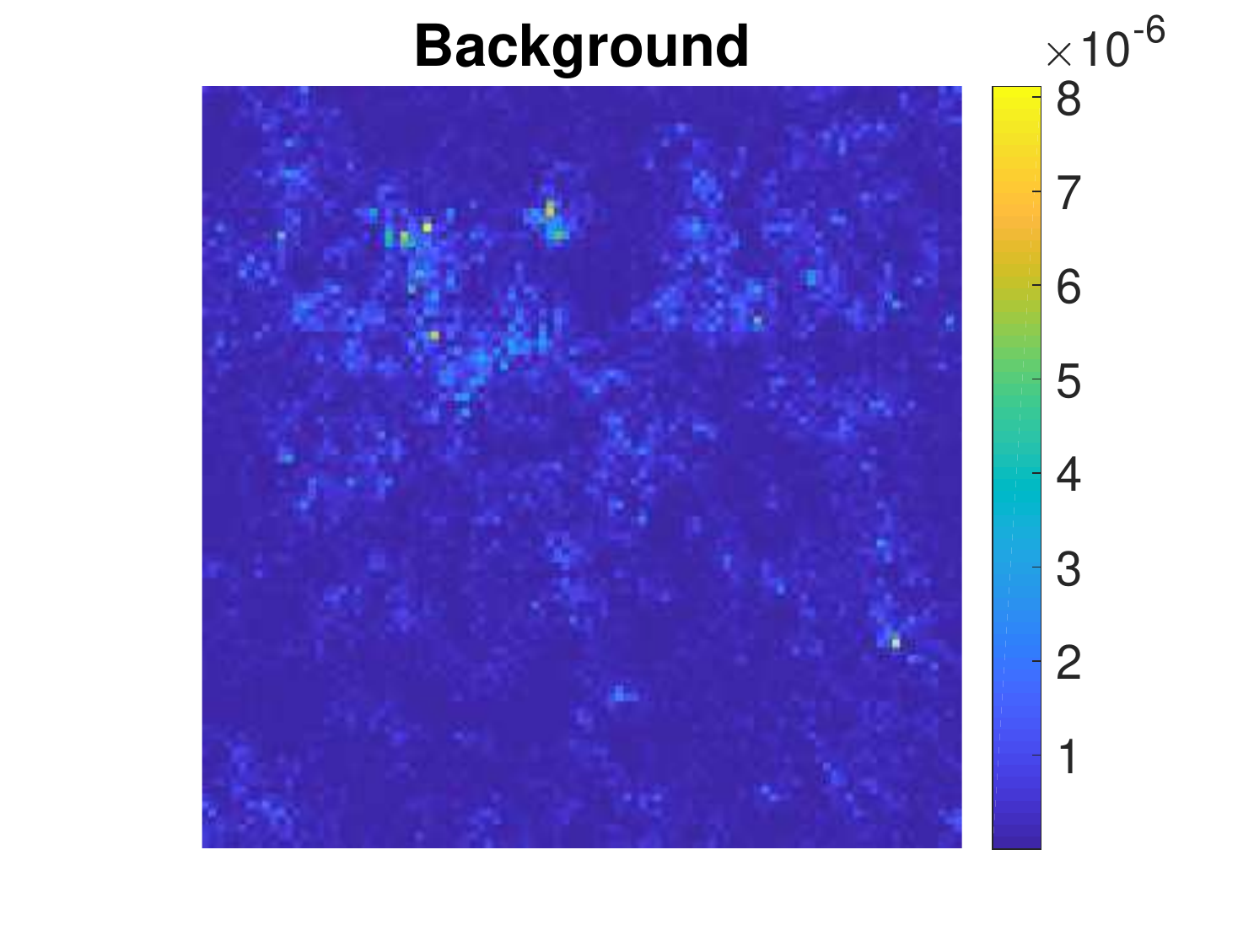}
		\caption{}\label{SUBFIG:prior_bkg}
	\end{subfigure}%
	\caption{\ans{One realization of the ManiPoP point process. Figure (a) illustrates the 3D point cloud and (b) shows the background levels.}}
	\label{FIG:prior_sample}
\end{figure}

\ans{To illustrate the prior distribution proposed in this work, we show one realization of the point process and background process. The sample was obtained using the RJ-MCMC algorithm described in \cref{SEC:estimation_strategy} in a 3D cube of $N_r=N_c=99$ pixels and $T=4500$ histogram bins. The hyperparameters were chosen according to \cref{TAB:hyperparameters}, except for $\gamma_{a}=e^{3.5}$ and $\lambda_{a}=(N_rN_r)^{2.2}$. The points in \cref{SUBFIG:prior3D} are connected (with respect the definition in \cref{SUBSEC:point process model}) and define a single surface. Note that the reflectivity of this surface also shows spatial correlation. \Cref{SUBFIG:prior_bkg} illustrates the smoothing effect of the gamma Markov random field, as the background levels show spatial correlation.}

\section{Other posterior statistics of interest}\label{APP: posterior statistics}
After $N_{iter}$ RJMCMC iterations, we can fix the model dimension by only keeping the background update and the shift and mark moves. The sampler can continue obtaining samples in this configuration for $N_{iter2}$ iterations. The position and amplitude of each point can be then estimated using the empirical mean of the posterior samples:

\begin{equation}
\hat{t}_{n}=\frac{1}{N_{iter2}}\sum_{s=1}^{N_{iter2}}{t^{(s)}_{n}}  \quad \forall n=1,\dots,N_\Phi
\end{equation}

\begin{equation}
\hat{r}_{n}=\frac{1}{N_{iter2}}\sum_{s=1}^{N_{iter2}}{e^{m^{(s)}_{n}}}  \quad \forall n=1,\dots,N_\Phi
\end{equation}

We can also estimate the \emph{a posteriori} standard deviation of each point and its reflectivity:

\begin{equation}
\text{var}(t_n) \approx \frac{1}{N_{iter2}}\sum_{s=1}^{N_{iter2}}{ t^{(s)}_{n}} -\hat{t}_{n}^2  \quad \forall n=1,\dots,N_\Phi
\end{equation}

\begin{equation}
\text{var}(r_n) \approx \frac{1}{N_{iter2}}\sum_{s=1}^{N_{iter2}}{e^{2m_n^{(s)}}}-\hat{r}_{n}^2  \quad \forall n=1,\dots,N_\Phi 
\end{equation}

\section{Efficient likelihood evaluation} \label{APP:log-likelihood evaluation}
We can rearrange the log-likelihood expression \cref{EQ:likelihood} for one pixel carefully in the following way
\begin{multline}
\label{EQ:simplified_log_likelihood}
\log(p(z_{i,j,t} | \dots)) \propto \\ \sum_{z_{i,j,t} \ge 1} z_{i,j,t} \log\left( g_{i,j}  \sum_{\substack{n:x_n=i\\ y_n=j \\ h(t-t_n)>\epsilon}}r_{n}h(t-t_{n})+g_{i,j} b_{i,j} \right) \\ -g_{i,j} \left( \sum_{\substack{n:x_n=i\\ y_n=j}}r_{n}\sum_{t=1}^{T}h(t)+b_{i,j} T \right).
\end{multline}
Note that
\begin{itemize}
	\item We only need to keep account of the bins where $z_{i,j,t}\ge1$, that in most Lidar images represent less than $3\%$ of the total Lidar 3D cube. A list of bin positions plus number of returns is considerably more efficient than using a full 3D matrix representation of the Lidar data.  
	\item The intensity $g_{i,j}  \sum r_{n}h(t-t_{n})+g_{i,j} b_{i,j}$ only has to be evaluated for bins with one photon or more, thus the computational load is proportional to the total photon counts. Moreover, if we bound the number of non-zero values of $h(t)$ with the threshold $\epsilon$, only the returns with $h(t-t_{n})$ have to be evaluated for $p(z_{i,j,t} | \dots)$.
	\item The third line of \cref{EQ:simplified_log_likelihood}, only involves multiplications and sums proportional the total number of returns in pixel $(i,j)$, considering that $\sum_{t=1}^{T}h(t)$ can be precomputed.
\end{itemize}
Regarding the sampling strategy for the background image $\mat{B}$, we only have to perform the binomial sampling step,  \cref{EQ:B_scheme}, when the $z_{i,j,t}\ge1$ there is at least one return active, i.e. $h(t-t_{n})>\epsilon$. If $z_{i,j,t}=0$, then $\tilde{y}_{i,j,t,b}=0$ and if $y_{i,j,t,b}\ge1$ and $h(t-t_{n})<\epsilon$ for all the returns in pixel $(i,j)$, then $y_{i,j,t,b}=z_{i,j,t}$.
\section{Local approximation of $\mat{P'}/\mat{P}$}\label{APP:approx}
\begin{figure}[!h]	 
	\centering
	\includegraphics[width=1\textwidth]{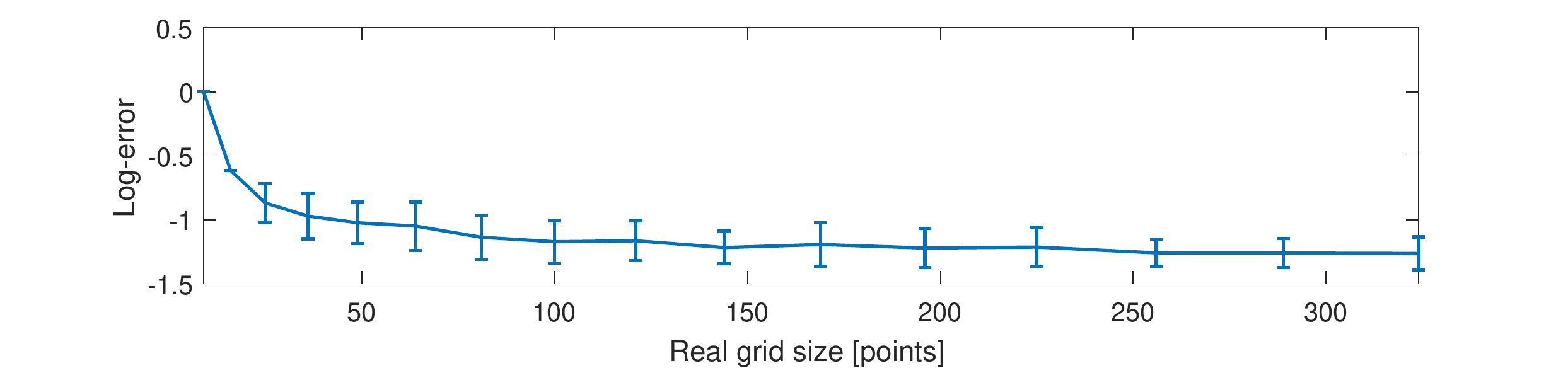}
	\caption{$log(|\mat{P'}|/|\mat{P}|)_{exact}-log(|\mat{P'}|/|\mat{P}|)_{approx}$ as a function of number of total connected points. The graph shows the mean and standard deviation over 100 independent random realizations.} 
	\label{FIG:p_approx}
\end{figure}

In all variable dimension moves, the acceptance ratio of  \cref{EQ:RevJump} involves the computation of $\mat{P'}/\mat{P}$ where one  matrix is of size $N_{\Phi}\times N_{\Phi}$ and the other is of size $(N_{\Phi}+1 )\times (N_{\Phi}+1)$. The exact value of this ratio depends all the $N_{\Phi}+1$, which involves a prohibitive computational time for each update. However, since we only modify one point of $N_{\Phi}+1$ points, the ratio will depend mostly on the local neighbourhood of the point and will have little dependence on the global structure. In this work, we used a local approximation of $\mat{P'}$ and $\mat{P}$, by only considering a reduced $N_{p}^2-1$ neighbouring points and computing the entries of $|\mat{P}|$ and $|\mat{P'}|$ according to  \cref{EQ:P}. We demonstrate empirically that the mean log-error, i.e. $log(|\mat{P'}|/|\mat{P}|)_{exact}-log(|\mat{P'}|/|\mat{P}|)_{approx}$ is always smaller than $-1.5$, as shown in figure \cref{FIG:p_approx}.

\section{Green ratio expressions}\label{APP:acc_ratio}
The birth move of point $(\myvec{c}_{N_\Phi+1},r_{N_\Phi+1})$ has an acceptance ratio given by $\rho=\min\{1,r\left(\myvec{\theta},\myvec{\theta'}\right)\}$ with
\begin{equation*}
r\left(\myvec{\theta},\myvec{\theta'}\right) = \left\{
\begin{array}{ll}
C_1  & \mbox{if } |t_{N_\Phi+1}-t_{n}| > d_{max} \quad \forall n\ne{N_\Phi+1}:\\
&  x_n=x_{N_\Phi+1} \text{ and } y_n=y_{N_\Phi+1} \\
0 & \mbox{otherwise}
\end{array}
\right.
\end{equation*}
where $C_1$ is 
\begin{multline*}
C_1 = \prod_{t=1}^{T}\left(\frac{  \sum_{\substack{n:x_n=i\\ y_n=j \\}}r'_{n}h(t-t'_{n})+ b'_{i,j}  }{ \sum_{\substack{n:x_n=i\\ y_n=j \\}}r_{n}h(t-t_{n})+ b_{i,j}   } \right)^{z_{i,j,t}} \frac{p_{\textrm{death}}}{p_{\textrm{birth}}}
\\  \lambda_{a}\gamma_{a}^{-m\left(S(\myvec{c}_{N_\Phi+1}) \setminus  \bigcup_{n'\in \mathcal{M}_{pp}(\myvec{c}_{N_\Phi+1})}S(\myvec{c}_{n'}) \right)}
\frac{1}{N_{\Phi}+1}  
\\
\exp\left(-\frac{1}{2\sigma^2}\left(\sum_{n'\in\mathcal{M}_{pp}(\myvec{c}_n)}\frac{(m_{N_\Phi+1}-m_{n'})^2}{d(\myvec{c}_{N_\Phi+1};\myvec{c}_{n'})}+{m_{N_\Phi+1}}^2\beta\right)\right) \sqrt{\frac{|\mat{P'}|}{|\mat{P}|}}  \sqrt{\frac{1}{2\pi\sigma^2}}
\\ 
\prod_{(i,j)\in \mathcal{M}_B(b_{i,j})} \left(\frac{{b'}_{i,j}}{{b}_{i,j}}\right)^{\alpha_{B}-1} \left(\frac{\sum_{(i',j')\in \mathcal{M}_B(b_{i,j})} {b}_{i',j'}}{\sum_{(i',j')\in \mathcal{M}_B(b_{i,j})} {b'}_{i',j'}}\right)^{\alpha_{B}} \frac{1}{1-u}.
\end{multline*}
Similarly, the death move is accepted with probability $\rho=\min\{1,C_1^{-1}\}$, where the term $\frac{1}{N_{\Phi}+1}$ in the second line is replaced by $\frac{1}{N_{\Phi}}$. The dilation move of point $(\myvec{c}_{N_\Phi+1},r_{N_\Phi+1})$ is accepted with probability 
$\rho=\min\{1,r\left(\myvec{\theta},\myvec{\theta'}\right)\}$ with
\begin{equation*}
r\left(\myvec{\theta},\myvec{\theta'}\right) = \left\{
\begin{array}{ll}
C_2  & \mbox{if } |t_{N_\Phi+1}-t_{n}| > d_{max} \quad \forall n\ne{N_\Phi+1}:\\
&  x_n=x_{N_\Phi+1} \text{ and } y_n=y_{N_\Phi+1} \\
0 & \mbox{otherwise}
\end{array}
\right.
\end{equation*}
where $C_2$ is 
\begin{multline*}
C_2 = \prod_{t=1}^{T}\left(\frac{  \sum_{\substack{n:x_n=i\\ y_n=j \\}}r'_{n}h(t-t'_{n})+ b'_{i,j}  }{ \sum_{\substack{n:x_n=i\\ y_n=j \\}}r_{n}h(t-t_{n})+ b_{i,j}   } \right)^{z_{i,j,t}}\frac{p_{\textrm{erosion}}}{p_{\textrm{dilation}}}
\\  \lambda_{a}\gamma_{a}^{-m\left(S(\myvec{c}_{N_\Phi+1}) \setminus  \bigcup_{n'\in \mathcal{M}_{pp}(\myvec{c}_{N_\Phi+1})}S(\myvec{c}_{n'}) \right)}
\\
\frac{N_\Phi(2N_b+1)}{\sum_{m\in\mathcal{M}_{pp}(\myvec{c}_{N_\Phi+1})}  \card\mathcal{M}_{pp}(\myvec{c}_{m})} \times
\frac{1}{\sum_{m=1}^{N_\Phi+1} \indicator{\mathbb{Z}_{+}}{\card\mathcal{M}_{pp}(\myvec{c}_{m})}}
\\
\exp\left(-\frac{1}{2\sigma^2}\left(\sum_{n'\in\mathcal{M}_{pp}(\myvec{c}_n)}\frac{(m_{N_\Phi+1}-m_{n'})^2}{d(\myvec{c}_{N_\Phi+1};\myvec{c}_{n'})}+{m_{N_\Phi+1}}^2\beta\right)\right) \sqrt{\frac{|\mat{P'}|}{|\mat{P}|}}  \sqrt{\frac{1}{2\pi\sigma^2}}
\\ 
\prod_{(i,j)\in \mathcal{M}_B(b_{i,j})} \left(\frac{{b'}_{i,j}}{{b}_{i,j}}\right)^{\alpha_{B}-1} \left(\frac{\sum_{(i',j')\in \mathcal{M}_B(b_{i,j})} {b}_{i',j'}}{\sum_{(i',j')\in \mathcal{M}_B(b_{i,j})} {b'}_{i',j'}}\right)^{\alpha_{B}}.
\end{multline*}
A shift of the point $(\myvec{c}_n,r_n)$ to the new position $\myvec{c'}_n=(x_n,y_n,t'_n)^T$, has an acceptance probability of $\rho=\min\{1,r\left(\myvec{\theta},\myvec{\theta'}\right)\}$ with
\begin{equation*}
r\left(\myvec{\theta},\myvec{\theta'}\right) = \left\{
\begin{array}{ll}
C_3  & \mbox{if }  |t'_{n}-t_{m}| > d_{max} \quad \forall n\ne m: \\
& x_m=x_n \text{ and } y_m=y_n   \\
0 & \mbox{otherwise}
\end{array}
\right.
\end{equation*}
where  
\begin{multline*}
C_3 = \prod_{t=1}^{T}\left(\frac{  \sum_{\substack{n:x_n=i\\ y_n=j \\}}r'_{n}h(t-t'_{n})+ b'_{i,j}  }{ \sum_{\substack{n:x_n=i\\ y_n=j \\}}r_{n}h(t-t_{n})+ b_{i,j}   } \right)^{z_{i,j,t}}
\\
\exp\left(-\frac{1}{2\sigma^2}\left(\sum_{n'\in\mathcal{M}_{pp}(\myvec{c'}_n)}\frac{(m_{n}-m_{n'})^2}{d(\myvec{c'}_{n};\myvec{c}_{n'})}\right)\right) 
\\
\exp\left(\frac{1}{2\sigma^2}\left(\sum_{n'\in\mathcal{M}_{pp}(\myvec{c}_n)}\frac{(m_{n}-m_{n'})^2}{d(\myvec{c}_{n};\myvec{c}_{n'})}\right)\right) 
\sqrt{\frac{|\mat{P'}|}{|\mat{P}|}}  
\\
\gamma_{a}^{-m\left(S(\myvec{c'}_{n}) \setminus  \bigcup_{n'\in \mathcal{M}_{pp}(\myvec{c'}_{n})}S(\myvec{c}_{n'}) \right)+m\left(S(\myvec{c}_{n}) \setminus  \bigcup_{n'\in \mathcal{M}_{pp}(\myvec{c}_{n})}S(\myvec{c}_{n'}) \right)}.
\end{multline*}
A mark update of point $(\myvec{c}_n,r_n)$ to a new reflectivity $r'_n=\log(m'_n)$, is accepted with probability $\rho=\min\{1,C_4\}$, where
\begin{multline*}
C_4 =  \prod_{t=1}^{T}\left(\frac{  \sum_{\substack{n:x_n=i\\ y_n=j \\}}r'_{n}h(t-t'_{n})+ b'_{i,j}  }{ \sum_{\substack{n:x_n=i\\ y_n=j \\}}r_{n}h(t-t_{n})+ b_{i,j}   } \right)^{z_{i,j,t}}
\\
\exp\left(-\frac{1}{2\sigma^2}\left(\sum_{n'\in\mathcal{M}_{pp}(\myvec{c'}_n)}\frac{(m'_{n}-m_{n'})^2}{d(\myvec{c'}_{n};\myvec{c}_{n'})}+{m'_n}^2\beta\right)\right) 
\\
\exp\left(\frac{1}{2\sigma^2}\left(\sum_{n'\in\mathcal{M}_{pp}(\myvec{c}_n)}\frac{(m_{n}-m_{n'})^2}{d(\myvec{c}_{n};\myvec{c}_{n'})}+m_n^2\beta\right)\right) .
\end{multline*}
The split move from $(\myvec{c}_n=(x_n,y_n,t_{n})^T,r_n)$ to $(\myvec{c'}_{k_1}=(x_n,y_n,t'_{k_1})^T,r'_{k_1})$ and $(\myvec{c'}_{k_2}=(x_n,y_n,t'_{k_2})^T,r'_{k_2})$ is accepted with probability $\rho=\min\{1,r\left(\myvec{\theta},\myvec{\theta'}\right)\}$, where
\begin{equation*}
r\left(\myvec{\theta},\myvec{\theta'}\right) = \left\{
\begin{array}{ll}
C_5  & \mbox{if }  |t'_{n}-t_{m}| > d_{max} \quad \forall n\ne m: \\
& x_m=x_n \text{ and } y_m=y_n   \\
0 & \mbox{otherwise}
\end{array}
\right.
\end{equation*} 
and
\begin{multline*}
C_5 = \prod_{t=1}^{T}\left(\frac{  \sum_{\substack{n:x_n=i\\ y_n=j \\}}r'_{n}h(t-t'_{n})+ b'_{i,j}  }{ \sum_{\substack{n:x_n=i\\ y_n=j \\}}r_{n}h(t-t_{n})+ b_{i,j}   } \right)^{z_{i,j,t}} \frac{p_{\textrm{merge}}}{p_{\textrm{split}}}
\\
\frac{1}{u(1-u)} N_\Phi (\card\text{points in } \mat{\Phi} \text{ that verify  \cref{EQ: merge_condition}})^{-1} 
\\
\exp\left(-\frac{1}{2\sigma^2}\left(\sum_{n'\in\mathcal{M}_{pp}(\myvec{c'}_{k_1})}\frac{(m_{k_1}-m_{n'})^2}{d(\myvec{c'}_{k_1};\myvec{c}_{n'})}\right)\right) 
\\
\exp\left(-\frac{1}{2\sigma^2}\left(\sum_{n'\in\mathcal{M}_{pp}(\myvec{c'}_{k2})}\frac{(m_{k_1}-m_{n'})^2}{d(\myvec{c'}_{k_2};\myvec{c}_{n'})}\right)\right) 
\\
\exp\left(\frac{1}{2\sigma^2}\left(\sum_{n'\in\mathcal{M}_{pp}(\myvec{c}_n)}\frac{(m_{n}-m_{n'})^2}{d(\myvec{c}_{n};\myvec{c}_{n'})}\right)\right) 
\sqrt{\frac{|\mat{P'}|}{|\mat{P}|}}  
\\
\gamma_{a}^{-m\left(S(\myvec{c'}_{k_1}) \setminus  \bigcup_{n'\in \mathcal{M}_{pp}(\myvec{c'}_{k_1})}S(\myvec{c}_{n'}) \right)+m\left(S(\myvec{c}_{n}) \setminus  \bigcup_{n'\in \mathcal{M}_{pp}(\myvec{c}_{n})}S(\myvec{c}_{n'}) \right)}.
\\
\lambda_{a}\gamma_{a}^{-m\left(S(\myvec{c'}_{k_2}) \setminus  \bigcup_{n'\in \mathcal{M}_{pp}(\myvec{c'}_{k_2})}S(\myvec{c}_{n'}) \right)}(d_{max}+\text{attack}_{h(t)}+\text{decay}_{h(t)}).
\end{multline*}
Finally, the merge move is accepted with probability $\rho=\min\{1,C_5^{-1}\}$.


\section{Impulse response election for mannequin data}\label{APP: impulse_mit}
\begin{figure}[!h]
	\centering
	\includegraphics[width=0.7\textwidth]{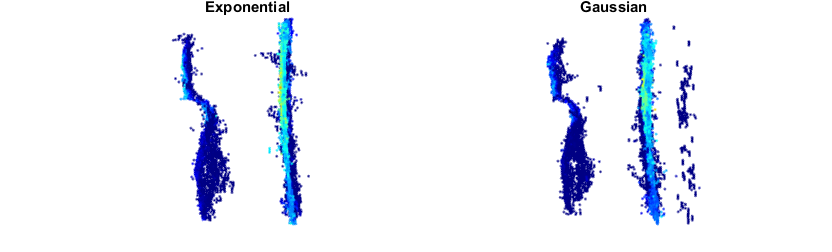}
	\caption{Recovered mannequin and back wall using a Gaussian $h(t)$ (right) and an exponential $h(t)$ (left). The mismatch between the true impulse response and the Gaussian one produces false returns behind de back wall.} 
	\label{FIG:exp_vs_gauss}
\end{figure}

We evaluated the ManiPoP algorithm with the proposed Gaussian shape and also with the exponential decay retrieved directly from the data. As shown in \cref{FIG:exp_vs_gauss}, the Gaussian-shaped instrumental response generally fits two points per each real point due to the mismatch between the true impulse response and the Gaussian curve (similar results are obtained using SPISTA with both impulse responses). Hence, the rest of the experiments with this dataset were performed using the data-driven exponential impulse response.
\begin{figure}[!h]
	\centering
	\includegraphics[width=0.85\textwidth]{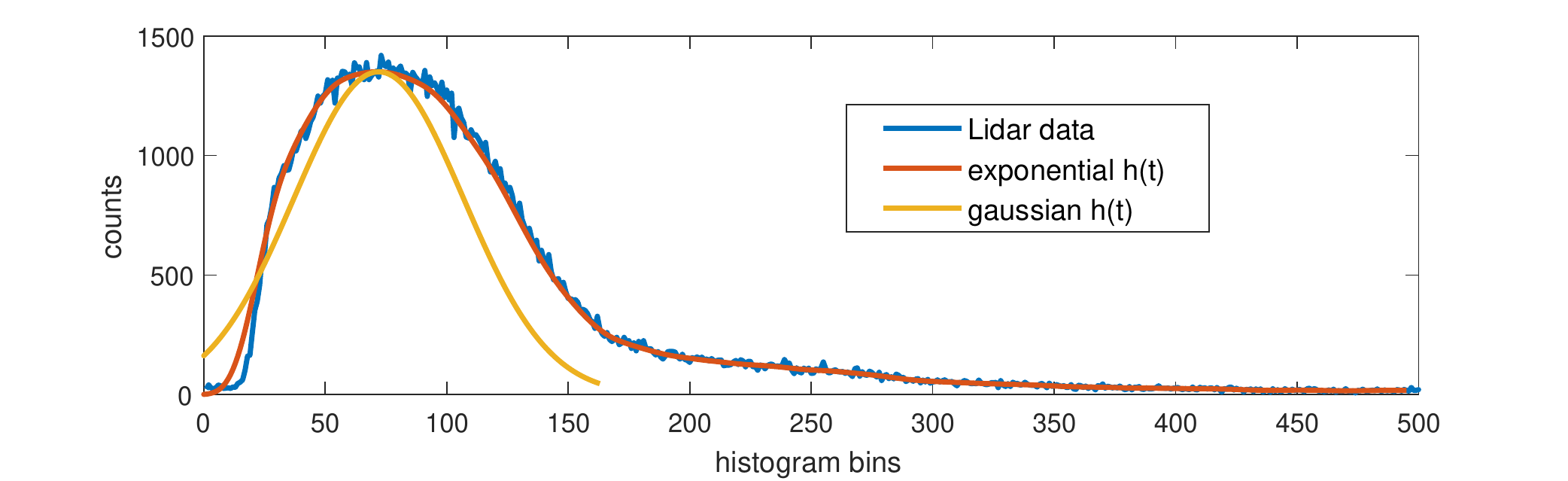}
	\caption{The integrated response of the backplane is shown in blue, the exponential impulse response in red and the Gaussian one in yellow.} 
	\label{FIG:correct_impulse}
\end{figure}

\section{Election of threshold value}\label{APP:threshold}
\begin{figure}[!h]
	\centering
	\includegraphics[width=0.7\textwidth]{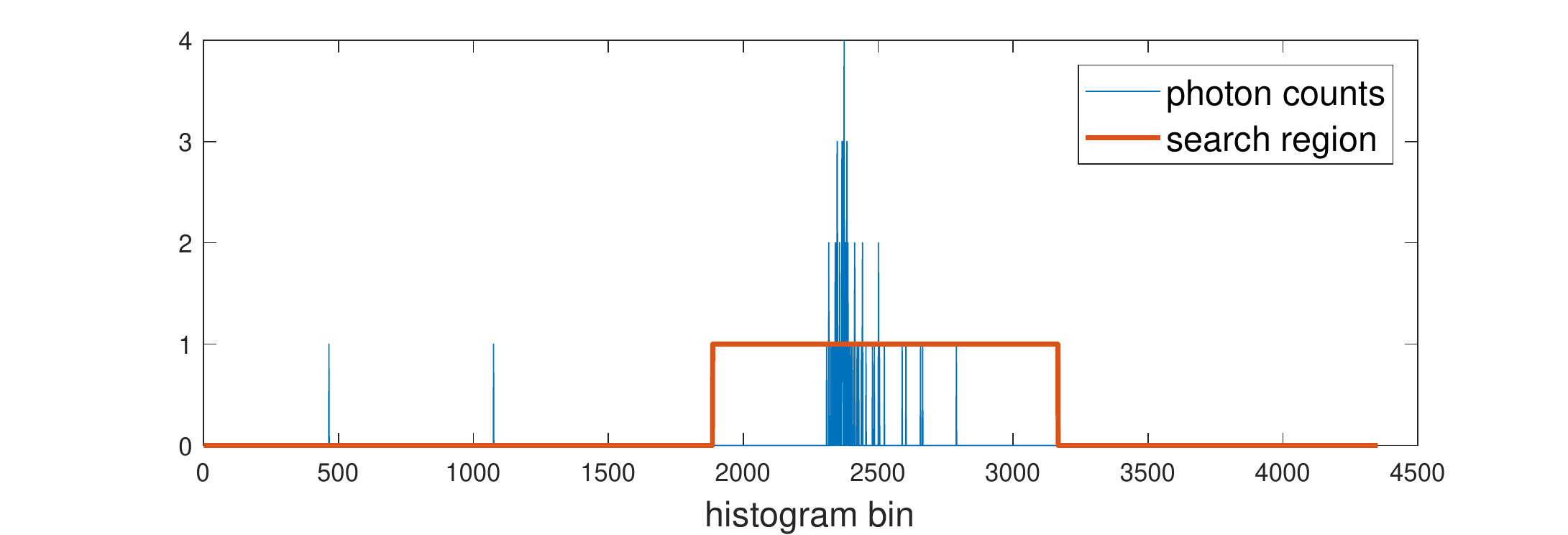}
	\caption{Illustration of the search space reduction (by a factor up to $70\%$) when adding a preprocessing step that discards histogram regions with zero or very low photon counts.} 
	\label{FIG:threshold}
\end{figure}

\begin{figure}[H]
	\centering
	\includegraphics[width=0.5\textwidth]{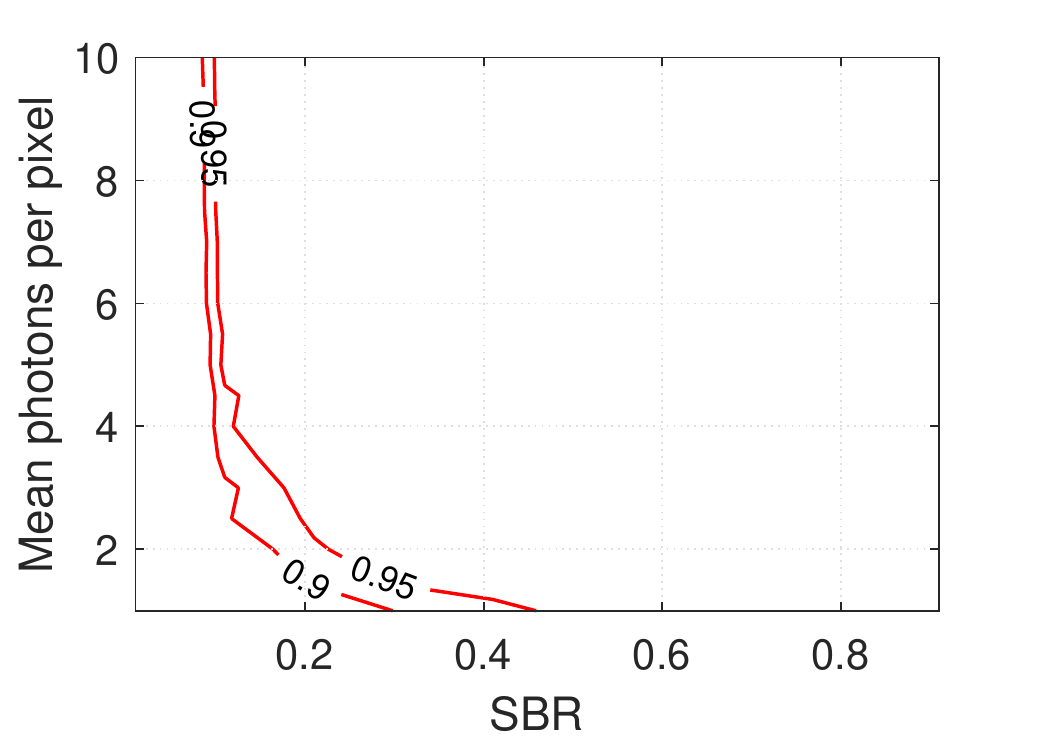}
	\caption{Level curves of 0.9 and 0.95 probability of not discarding the signal for different values of signal-to-background ratio and mean number of photon counts per pixel.}	\label{FIG:threshold_montecarlo}
\end{figure}

A preprocessing step reduces the 3D search space by applying a log-matched filter to the data and assigning zero probability to the occurrence of points where the filtered output is below $\frac{0.05}{T}\sum_{t=1}^{T}z_{i,j,t}\sum_{t=1}^{T}\log h(t)$.  An example is shown in \cref{FIG:threshold}. 
Using this threshold, we have a high probability (0.9 or higher) of not discarding signals up to a signal-to-background ratio (SBR) of 0.05. The Monte Carlo simulations shown in figure \cref{FIG:threshold_montecarlo} demonstrate that the statement holds for very low mean photon counts. Note that the thresholding operation is applied on the coarse scale data, increasing the probability of including the signal in the search space, even when a given pixel might not have any signal photons in the fine scale.

\section{Performance as a function of signal-to-background (SBR) and total number of photons} \label{APP:SBR}
We have added an analysis of the reconstruction quality as a function of background-to-signal ratio and mean number of photons per pixel in the synthetic image. As explained in \cite{rappfew}, this characterization avoids the ambiguity of having a high number of photon counts (i.e., high SNR), but no signal photons. The algorithm performs well for a signal-to-background ratio higher than 1 and more than 1 photon per pixel.

\end{document}